\documentclass[aps,amsmath,amssymb,showpacs,pre]{revtex4-1}
\usepackage{graphicx,color}
 \graphicspath{{./}{./figures/}}
\usepackage{ams math}
\usepackage{enumerate}
\usepackage{mathbbol}
\usepackage{amsfonts}
\usepackage{natbib}

\usepackage{bm}

\usepackage{color}

\draft 

\newcommand{\nn}{\nonumber}

\newcommand{\be}{\begin{equation}}
\newcommand{\ee}{\end{equation}}
\newcommand{\bea}{\begin{eqnarray}}
\newcommand{\eea}{\end{eqnarray}}
\newcommand{\beq}{\begin{eqnarray}}
\newcommand{\eeq}{\end{eqnarray}}

\newcommand{\Tr}{{\rm Tr}}

\newcommand{\B}{{\sf B}}

\newlength{\bilderlength}
\newcommand{\bilderscale}{0.25}

\newcommand{\bilderskip}{\hspace*{0.8ex}}

\newcommand{\diagram}[1]{%
\settowidth{\bilderlength}{\bilderskip%
\includegraphics[scale=\bilderscale]{./#1}\bilderskip}%
\parbox{\bilderlength}{\bilderskip%
\includegraphics[scale=\bilderscale]{./#1}\bilderskip}}

\begin{document}

\title{Non-interacting fermions at finite temperature in a $d$-dimensional trap: universal correlations} 



\author{David S. \surname{Dean}}
\affiliation{Univ. Bordeaux and CNRS, Laboratoire Ondes et Mati\`ere  d'Aquitaine
(LOMA), UMR 5798, F-33400 Talence, France}
\author{Pierre Le Doussal}
\affiliation{CNRS-Laboratoire de Physique Th\'eorique de l'Ecole Normale Sup\'erieure, 24 rue Lhomond, 75231 Paris Cedex, France}
\author{Satya N. \surname{Majumdar}}
\affiliation{LPTMS, CNRS, Univ. Paris-Sud, Universit\'e Paris-Saclay, 91405 Orsay, France}
\author{Gr\'egory \surname{Schehr}}
\affiliation{LPTMS, CNRS, Univ. Paris-Sud,  Universit\'e Paris-Saclay, 91405 Orsay, France}



\date{\today}

\begin{abstract} 
We study a system of $N$ non-interacting spin-less fermions trapped in a confining potential, in arbitrary dimensions $d$ and
arbitrary temperature $T$. The presence of the confining trap breaks the translational invariance and introduces {\it an edge} where the average 
density of fermions vanishes. Far from the edge, near the center of the trap (the so called ``bulk regime''), where the fermions do not feel the curvature of the trap, physical properties of the fermions have traditionally been understood using the Local Density (or Thomas Fermi) Approximation. However, these approximations drastically fail near the edge where the density vanishes {and thermal and quantum fluctuations are thus enhanced}. The main goal of this paper is to show that, even near the edge, novel
universal properties emerge, independently of the details of the shape of the confining potential. We present a unified framework to 
investigate both the bulk and the edge properties of the fermions.   
We show that for large $N$, these fermions in a confining trap, in arbitrary dimensions and at finite temperature, form a determinantal point process. 
As a result, any $n$-point correlation function, including the average density profile, can be expressed as an $n \times n$ determinant whose entry is called the kernel, a central object for such processes. Near the edge, we derive the large $N$ scaling form of the kernels, parametrized by $d$ and $T$. 
In $d=1$ and $T=0$, this reduces to the so called Airy kernel, that appears in the Gaussian Unitary Ensemble (GUE) of random matrix theory. In $d=1$ and $T>0$ we show a remarkable connection between our kernel and the one appearing in the $1+1$-dimensional Kardar-Parisi-Zhang equation at finite time.  
Consequently our result provides a finite $T$ generalization of the Tracy-Widom distribution, that describes the fluctuations of the rightmost fermion at $T=0$. In $d>1$ and $T \geq 0$, while the connection to GUE no longer holds, the process is still determinantal whose analysis provides a new class of kernels, generalizing the $1d$ Airy kernel at $T=0$ obtained in random matrix theory.  Some of our finite temperature results should be testable in present-day cold atom experiments, most notably our detailed predictions for the temperature dependence of the fluctuations near the edge.
\end{abstract}


\maketitle 

\tableofcontents

\section{Introduction}


\subsection{Overview}

Over the past few years, experimental developments in trapping and cooling of dilute Bose and
Fermi gases have led to spectacular progress in the study of many-body quantum systems \cite{BDZ08,GPS08}. Even in
the absence of interactions bosons and fermions display collective many-body effects emerging purely from the quantum
statistics \cite{Mahan, Castin, Castin2}. For non-interacting fermions, which we focus on here, the Pauli exclusion principle induces
highly non-trivial spatial (and temporal) correlations between the particles. Remarkably, the recent development of 
Fermi quantum microscopes \cite{Cheuk:2015,Haller:2015,Parsons:2015} provides a direct access to these spatial correlations, via a direct 
in situ imaging of the individual fermions, with a resolution comparable to the inter-particle spacing. It is thus
important to have a precise theoretical description of these correlations in such fermionic
gases.  

In many experimental setups, including the aforementioned Fermi quantum microscopes, the fermions are trapped by an external
potential. This trapping potential breaks the translational invariance and generically induces {\it an edge} of the Fermi
gas. Indeed, beyond a certain distance from the center of the trap, the average density of fermions vanishes.   
Far from the edge, i.e., close to the center of the trap, the properties of the Fermi gas are well described by standard
tools of many-body quantum physics, like the local 
density (or Thomas-Fermi) approximation (LDA) \cite{Castin,butts}. However, this approximation breaks down close to the edge where 
the density  vanishes, and where the fluctuations (both quantum and thermal) are large \cite{Kohn}. The purpose of this
paper is to develop a general framework, which encompasses the LDA (and actually puts it on firmer ground) and provides 
an analytical description of  the edge properties of the Fermi gas in any spatial dimension $d$, 
and at finite temperature $T$.  

This framework, whose main results were recently announced in two short Letters \cite{us_prl,us_epl}, is based on 
Random Matrix Theory (RMT) \cite{mehta,forrester} and, more generally, on the theory of determinantal point processes \cite{johansson,borodin_determinantal,tracy_widom_determinantal}. The simplest example 
is the case of $N$ non-interacting spin-less fermions in a one-dimensional harmonic potential $V(x)=\frac{1}{2}m \omega^2 x^2$ at
zero temperature $T=0$. In this case, by computing explicitly the ground-state wave function, one can show \cite{eisler_prl,mehta,marino_prl} that
there exists a one-to-one mapping between the positions of the fermions $x_1, x_2, \ldots, x_N$ and the (real) eigenvalues $\lambda_1, \lambda_2, \ldots, \lambda_N$ of random $N \times N$ Hermitian 
matrices with independent Gaussian entries, the so called Gaussian Unitary Ensemble (GUE) in RMT. Although this connection has certainly been known for a long time \cite{mehta}, it is only recently that the powerful tools of RMT have been used to compute the statistics of physical observables for fermions \cite{eisler_prl,marino_prl,marino_pre,CDM14,castillo} (see section \ref{section:1d_RMT} for an extended discussion of these applications). In particular, it is well known that, in the large $N$ limit, the (scaled) density of eigenvalues (or equivalently the density of fermions) has a finite support and is
given by the celebrated Wigner semi-circular law 
$f_W(z) = \frac{1}{\pi}\sqrt{2-z^2}$. 

Going beyond the average density,
the fluctuations at the edge of 
the Wigner sea have also generated a lot of interest in RMT. Of particular 
interest is the statistics of the largest
eigenvalue $\lambda_{\max} = \max_{1 \leq i \leq N} \lambda_i$. Indeed, its probability distribution function (PDF), properly shifted and scaled (see section \ref{section:1d_RMT} below for more detail) is given for large $N$ by the celebrated Tracy-Widom (TW) distribution for GUE \cite{TW}, which is now 
a cornerstone of extreme value statistics of strongly correlated variables. Since its discovery in RMT, this TW distribution (and its counterparts for other matrix ensembles) have emerged in a wide variety of systems, a priori unrelated to RMT. These include the longest increasing subsequence of random 
permutations \cite{baik}, directed polymers \cite{johann,poli} and growth 
models \cite{growth} in the Kardar-Parisi-Zhang (KPZ) universality class in 
$(1+1)$ 
dimensions as well as for the continuum $(1+1)$-dimensional KPZ 
equation~\cite{SS10,CLR10,DOT10,ACQ11}, 
sequence alignment problems 
\cite{sequence}, mesoscopic 
fluctuations in quantum dots \cite{dots}, height fluctuations of 
non-intersecting Brownian motions over a fixed time interval
\cite{FMS11,Lie12}, height fluctuations of non-intersecting interfaces
in presence of a long-range interaction induced by a 
substrate~\cite{NM_interface}, and also 
in finance 
\cite{biroli} (see \cite{maj, MS_thirdorder} for reviews). The TW distributions have been recently 
observed in experiments on nematic liquid crystals \cite{takeuchi} and in experiments involving coupled fiber lasers 
\cite{davidson}. From the aforementioned connection between fermions and RMT, it follows that the quantum fluctuations 
of the rightmost fermion $x_{\max}(T=0) = \max_{1 \leq i \leq N} x_i$ are also governed by the TW distribution for GUE. Hence, 
non-interacting spin-less fermions in a one-dimensional harmonic trap at $T=0$ provide one of the simplest physical system where the TW
distribution appears.  

It is natural  to ask what happens at finite temperature $T>0$ and/or in higher dimensions, $d>1$? These are natural and relevant questions both experimentally and theoretically. In this case, the connection with the Gaussian Unitary Ensemble
of RMT is lost. Nevertheless, it is still possible to show that the system is a determinantal process (this is an exact statement at $T=0$ and any $d$ \cite{us_epl} and true only for large $N$ at $T>0$ \cite{us_prl}). This means that all $n$-point correlation functions of the $d$-dimensional Fermi gas, $R_n({\bf x}_1, {\bf x}_2, \ldots, {\bf x}_n)$, can be expressed as the determinant of an $n \times n$ matrix whose entries are given by $K_\mu({\bf x}_i, {\bf x}_j)$ where the function $K_{\mu}({\bf x}, {\bf y})$ is called the kernel and depends on the chemical potential $\mu$. It is thus the central object  
of the theory as it gives access to the computation of all the correlation functions. Therefore, it is important to characterize the behavior of this kernel, in
the limit of a large number of fermions $N \gg 1$, both in the bulk and at the edge of the Fermi gas. In two recent Letters, we announced, giving few details, results for the limiting kernels first
in the case $d=1$ and $T\geq 0$ in Ref. \cite{us_prl} and then in $d>1$, but only for $T = 0$ \cite{us_epl}. In this paper, we present a detailed derivation of these results, which relies in particular on a path-integral representation of the kernel $K_\mu({\bf x}, {\bf y})$. This allows us to study the problem 
in any dimension $d \geq 1$ and for any finite temperature $T \geq 0$, hence generalizing our previous results \cite{us_epl} to finite temperature. We will show that this method allows us to recover, in a fully controlled way, the LDA results in the bulk but also allows one to compute the correlations at the edge. In addition, as we will demonstrate  below, this path-integral method is extremely powerful as it allows us to treat a wide class of trapping potentials of the form $V({\bf x}) \sim |{\bf x}|^p$ at large $|{\bf x}|$, with $p>0$, and demonstrate the {\it universality of our results}, both in the bulk and at the edge.

\subsection{Model}  \label{sec:model} 

We study in this paper a model for $N$ spin-less non-interacting fermions in a 
$d$-dimensional potential $V({\bf x})$. It is described by an $N$- body Hamiltonian $\hat 
{\cal H}_N = \sum_{j=1}^N \hat H_j$, where $\hat H_j= \hat H({\bf p}_j,{\bf x}_j)$ is a 
one-body Hamiltonian of the form $\hat H \equiv \hat H({\bf p}_j,{\bf x}_j)$ with
\bea \label{H} 
\hat H = \frac{\hat {\bf p}^2}{2m}  + V({\bf x})  \quad , 
\quad \hat {\bf p} = \frac{\hbar}{i} \mathbf{\nabla} \;. \label{defhhat}
\eea 
{We will consider here only confining potentials $V({\bf x})$ such that
the one body Hamiltonian $\hat H$ admits {\it an infinite number of bound states.}  
One such class consists of confining potentials of the form $V({\bf x})$ with $V({\bf x}) \sim |{\bf x}|^p$ 
at large $|{\bf x}|$, with $p>0$, and here we will mainly focus here on this class
(and only briefly mention some other cases).} The simplest such confining potential is the isotropic $d$-dimensional 
harmonic oscillator
\bea \label{HO} 
V({\bf x}) \equiv V(r) = \frac{1}{2} m \omega^2 r^2 \quad , \quad r = |{\bf x}| \;.
\eea

\subsection{Outline and summary of the main results}


The paper is organized as follows. In section \ref{sec:zeroT}, we provide the general framework to study the correlations of $N$ non-interacting fermions
in an arbitrary confining potential at zero temperature $T=0$ and in arbitrary dimension $d$. The main result is the determinantal structure of the $n$-point correlations in the ground state, given in Eq. (\ref{eq:Rn_det}), with the associated kernel in Eq.~(\ref{eq:def_kernel}). In section \ref{section:1d_RMT}, we focus on the one-dimensional system ($d=1$), for the special case of a harmonic potential $V(x)= \frac{1}{2} m\omega^2 x^2$, still restricted to $T=0$. This case is particularly interesting because of its connection, valid for any finite number of fermions $N$, with the Gaussian Unitary Ensemble (GUE) of random matrix theory (RMT), see Eqs. (\ref{rm1}) and (\ref{rm_jpdf}). From this connection, many interesting properties, which are summarized in that section, can be obtained for the fermion problem. In particular, this relation with RMT shows that the statistics of the position of the rightmost fermion, at zero temperature, is governed by the celebrated Tracy-Widom distribution for GUE, see Eqs. (\ref{xmax_0T}) and (\ref{fredholm_F2}). In section \ref{sec:1dT} we study the case of $N$ non-interacting fermions in a one-dimensional ($d=1$) arbitrary potential $V(x)$ at finite temperature $T>0$. The main result of this section is to show that the $n$-point correlations  
still have a determinantal structure in the limit of a large number of fermions $N \gg 1$ [see Eqs. (\ref{det_process}), (\ref{kernel_final})]. As explained in detail in that section, this structure is due to a large extent to the equivalence, when $N \gg 1$, between the canonical and the grand-canonical ensemble. In section~\ref{sec:1dTphys} we apply the general analysis performed in section \ref{sec:1dT} to the case of $N$ non-interacting fermions in a one-dimensional harmonic potential at $T>0$. This case is of particular interest as it is exactly solvable and the limit of large $N$ can be studied in detail. The main results of this section concern the local correlations which are described, in the limit $N \to \infty$, by the limiting kernels given by Eq.~(\ref{Kbulk1d}) in the bulk (i.e., near the center of the trap) and by Eq.~(\ref{kff}) at the edge of the Fermi gas. They generalize the well known sine kernel, see Eq. (\ref{sine_kernel.1}), (in the bulk) and the Airy kernel, see Eq. (\ref{airy_kernel.1}), (at the edge), which both play a fundamental role in RMT. The resulting kernel at the edge (\ref{kff}) allows us to compute the fluctuations of the rightmost fermion at finite temperature as a Fredholm determinant (\ref{gap_proba}), which generalizes the Tracy-Widom distribution (\ref{fredholm_F2}) for finite temperatures. Quite remarkably, exactly the same Fredholm determinant has appeared in the context of stochastic growth models in the KPZ equation [see Eqs. (\ref{gen}), (\ref{eq:KPZKernel})]. This establishes an unexpected connection between free fermions at finite temperature and the KPZ equation at finite time $t$ [see Eq.~(\ref{inlaw})]. In section \ref{OHdT=0} we study the case of $N$ non-interacting fermions
in a harmonic potential, in arbitrary dimension $d$ and at zero temperature. Our analysis is based on a path-integral representation of the correlation kernel, given in Eqs.~(\ref{laplace_inverse}) and (\ref{eq:propag_hamonic}), which can then be analyzed in a very elegant way in the large $N$ limit. The most interesting results of that section are certainly the expression of the density profile [see Eqs. (\ref{density_edged}) and (\ref{eq:Fd_laplace})] and the kernel [see Eqs. (\ref{eq:edge_kernel3}) and (\ref{eq:Wz_Airy})] at the edge, the latter being a generalization of the Airy kernel (\ref{airy_kernel.1}) for any finite $d$. In section \ref{sec:Td}, we provide the full analysis of the correlations for a general $d$-dimensional soft potential, of the form $V({\bf x}) \sim |{\bf x}|^p$ (for large $|{\bf x}|$ and $p>1$), at finite temperature $T>0$. Using again a path integral representation of the kernel, we show that the local correlations both in the bulk and at the edge  are universal, i.e., independent of the (smooth) confining potential considered here. The resulting universal correlation kernels in the bulk (\ref{KTd_bulk}) and at the edge (\ref{kernel0Vedge}) generalize respectively the sine kernel (\ref{sine_kernel.1}) and the Airy kernel (\ref{airy_kernel.1}) for any finite dimension $d$ and temperature $T$. The last section \ref{sec:conclusion} contains a discussion of our results, including the connection with the KPZ equation mentioned above, and our conclusions. Some technical details 
have been relegated in Appendix \ref{short_time} and \ref{sec:expansion}.

\section{Correlations for non-interacting fermions at zero temperature} \label{sec:zeroT} 

\subsection{Many body ground state wave function} 

Let us start with $N$-fermions strictly at zero temperature. Consider first the single 
particle eigenfunctions $\psi_{\bf k}({\bf x})$ which satisfy 
the Schr\"odinger equation, 
$\hat H \psi_{\bf k}({\bf x}) = \epsilon_{\bf k} \psi_{\bf k}({\bf x})$, 
where $\hat H = - [\hbar^2/(2m)]\, \nabla^2 + V({\bf x})$ is the Hamiltonian 
and the energy eigenvalues 
$\epsilon_{\bf k}$ are labeled by $d$ quantum numbers denoted by ${\bf k} \in 
\mathbb{Z}^d$. Because of the confining potential these quantum numbers labeled by ${\bf 
k}$ should not be identified with the usual momentum, which we denote by ${\bf p}$.

At zero temperature, the ground state many-body wave function $\Psi_0$ can be expressed as 
an $N \times N$ Slater determinant,
\bea \label{slater} 
\Psi_0({\bf x}_1, \cdots, {\bf x}_N) = \frac{1}{\sqrt{N!}} \, 
\det[ \psi_{{\bf k}_i} ({\bf x}_j)]_{1\leq i,j \leq N} 
\eea 
constructed from the $N$ single 
particle wave functions labeled by a sequence $\{ {\bf k}_i \}$, $i=1, \ldots, N$, with 
non-decreasing energies such that $\epsilon_{{\bf k}_i} \leq \mu$ where $\mu$ is the Fermi 
energy. For a sufficiently confining potential, $\mu$ generically increases with increasing 
$N$~\cite{eisler_prl,Castin}. As an example, we consider the isotropic harmonic oscillator 
(\ref{HO}). In this case the energy levels
\bea \label{spectrumHO} 
\epsilon_{\bf k}=\sum_{a=1}^d \left(k_a + \frac{1}{2}\right) \hbar 
\omega \quad , \quad \psi_{\bf k}({\bf x})= \prod_{a=1}^d 
\phi_{k_a}(x_a) \quad , \quad \phi_k(x) 
= \left[\frac{\alpha}{\sqrt{\pi} 2^k k!}\right]^{1/2} 
e^{-\frac{\alpha^2 x^2}{2}} H_k(\alpha x)
\eea 
where the $k_a$'s are integers which range from $0$ to $\infty$, $H_k(z)$ is the $k$-th 
Hermite polynomial of degree $k$ and 
\be \label{alpha} 
\alpha=\sqrt{m \omega/\hbar}
\ee
is the characteristic 
inverse length scale. Note that, in this example, each, non-groundstate,  single particle energy level is 
degenerate in $d>1$. Hence the $N$-body ground state is degenerate, whenever the last single 
particle level is not fully occupied. This situation will be discussed later in section \ref{sec:1dT}. For 
now, since we are interested in the large $N$ limit where this effect of degeneracy is 
subdominant, we will assume that the last level is fully occupied. In this case, for the 
harmonic oscillator, by filling up completely the levels up to $\mu$ one obtains $N = 
\sum_{{\bf k} \in \mathbf{Z}^d} \theta\left(\mu - \hbar \omega(k_1+..+ k_d)\right)$, where $\theta(x)$ 
is the Heaviside theta function. This leads for large $N$, to 
\begin{equation}
\mu \simeq \hbar \omega 
[\Gamma(d+1) \, N]^{1/d}.\label{musho}
\end{equation}

{For more general potentials, the relation between $\mu$ and $N$ is usually more complicated
(and detailed below), but given our assumption of an infinite number of bound states,
we will always be able to study the limit of large $N \gg 1$, which is the subject of this
paper.}

\subsection{Quantum probability and determinantal structure of correlations}

Consider now the quantum probability, i.e. the squared many-body wave function
\begin{equation}\label{eq:psi_0}
|\Psi_0({\bf x}_1, \cdots, {\bf x}_N)|^2= \frac1{N!} \, \det[ \psi^*_{{\bf k}_i}({\bf x}_j)] 
\det[ \psi_{{\bf k}_i}({\bf x}_j)] \;.
\end{equation} 
Using the fact $\det(A^T) \det(B) = \det(A B)$, it can 
also be written
as a determinant
\bea \label{eq:psi_0.1}
|\Psi_0({\bf x}_1, \cdots, {\bf x}_N)|^2 =\frac{1}{N!}
\det_{1\leq i,j \leq N} K_\mu({\bf x}_i,{\bf x}_j)
\eea
where we have defined the kernel $K_\mu({\bf x},{\bf y})$ as
\begin{equation}\label{eq:def_kernel}
K_\mu({\bf x},{\bf y}) =\sum_{\bf k} \theta(\mu-\epsilon_{\bf k}) 
\psi_{\bf k}^*({\bf x})\psi_{\bf k}( {\bf y}) \;.
\end{equation}

As we will see later, this kernel will play a central role for the calculation of the 
correlations. For instance one usually defines the $n$-point correlation function $R_n({\bf 
x}_1,\cdots, {\bf x}_n)$ as
\begin{eqnarray}\label{eq:def_Rn}
R_n({\bf x}_1,\cdots, {\bf x}_n) = 
\frac{N!}{(N-n)!} \int d{\bf x}_{n+1} \cdots d {\bf x}_N \, 
|\Psi_0({\bf x}_1, \cdots, {\bf x}_N)|^2
\end{eqnarray}
obtained by integrating over $N-n$ coordinates and keeping $n$ 
coordinates $\{{\bf x}_1,\cdots, {\bf x}_n\}$ fixed. 
For $n=1$ this corresponds to the marginal density 
\bea\label{R1.1}
R_1({\bf x}) = N \int d{\bf x}_{2} \cdots d {\bf x}_N \, |\Psi_0({\bf x}, 
{\bf x}_2, \cdots, {\bf x}_N)|^2 \;.
\eea 
Incidentally this is also related to the average local density of fermions 
\bea\label{R1.2}
R_1({\bf x}) = N \rho_N({\bf x}) \quad , \quad 
\rho_N({\bf x}) = \big \langle \hat \rho_N({\bf x}) \big \rangle_0 
\quad , \quad \hat \rho_N({\bf x}) = \frac{1}{N} \sum_{i=1}^N \delta({\bf x}-{\bf x}_i) 
\eea 
where $\langle \cdots \rangle_0$ denotes the average w.r.t. the ground state 
quantum probability in Eq. (\ref{eq:psi_0}). 
The last equality follows from the indistinguishability of the fermions, i.e., the fact 
that the quantum 
probability $|\Psi_0({\bf x}_1, \cdots, {\bf x}_N)|^2$ is invariant
under the exchange of any two coordinates. More generally, the $R_n$'s contain 
information about higher correlations of the local densities, e.g., one has
\bea\label{R2.1}
R_2({\bf x},{\bf y})  = \big \langle \sum_{1 \leq i \neq j \leq N} \delta({\bf x}-{\bf 
x}_i) 
\delta ({\bf y}-{\bf x}_j) \big \rangle_0 
= N^2 \langle \hat \rho_N({\bf x}) \hat \rho_N({\bf y}) \rangle_0 - N \rho_N({\bf x}) \delta({\bf x} - {\bf y})
\eea 
and similarly for higher order correlations. 

Now we note that the kernel $K_\mu({\bf x},{\bf y})$ has the 
reproducing property 
\begin{eqnarray}\label{eq:convolution}
\int K_\mu({\bf x},{\bf z})K_\mu({\bf z},{\bf y})\, d{\bf z} = K_\mu({\bf x},{\bf y}) \;,
\end{eqnarray}
which follows from the ortho-normalization of the single particle wave functions, $\int \psi_{{\bf k}_i}^*({\bf z}) \psi_{{\bf k}_j}({\bf z}) \, d{\bf z} = \delta_{{{\bf k}_i}, {{\bf k}_j}}$. If the kernel satisfies the reproducing property 
in Eq.~(\ref{eq:convolution}), then there is a {\it general theorem}~\cite{mehta} that 
states that $R_n({\bf x}_1,\cdots, {\bf x}_n)$ in Eq. (\ref{eq:def_Rn}) can be expressed as an $n\times n$ determinant 
\begin{eqnarray}\label{eq:Rn_det}
R_n({\bf x}_1,\cdots, {\bf x}_n) = \det_{1\leq i,j \leq n} K_\mu({\bf x}_i,{\bf x}_j) \;.
\end{eqnarray}
{Note that this result (\ref{eq:Rn_det}) has been obtained here within the formalism of first quantization. It
can also be derived within the formalism of second quantization: this is then a consequence of the Wick's theorem applied
to fermionic (i.e., anti-commuting) operators} \cite{Mahan}. Indeed the kernel in Eq. (\ref{eq:def_kernel}) can be expressed, in the second quantization formalism, as 
\begin{eqnarray}
K_\mu(x,y) = \langle \Psi_0 | \Psi^\dagger(x) \Psi(y) | \Psi_0\rangle \;,
\end{eqnarray}
where $\Psi^\dagger(x)$ and $\Psi(y)$ are respectively the creation and the annihilation fermionic operators at positions $x$ and $y$ and $|\Psi_0\rangle$ is the ground state.

Any multi-particle probability distribution, whose $n$-point marginal can 
be expressed as
the determinant of a kernel as in Eq. (\ref{eq:Rn_det}), 
will  
generally 
be referred to as a distribution with a
{\it determinantal structure}. The associated random process corresponding to the random positions of the $N$ fermions is then called a $d$-dimensional
{\it determinantal point process} \cite{johansson,borodin_determinantal}. Let us also point out a very simple consequence of this 
determinantal structure.
Setting $n=1$ in Eq. (\ref{eq:Rn_det}) simply gives 
\begin{equation}
\label{R1.3}
R_1({\bf x})=K_\mu({\bf x},{\bf x})\, .
\end{equation}
This implies, from Eq. (\ref{R1.2}), that the average density is given by
the kernel evaluated at identical points
\begin{equation}
\rho_N({\bf x})= \frac{1}{N}\, K_\mu({\bf x},{\bf x})\, .
\label{density_kernel}
\end{equation}

To summarize, the kernel $K_\mu({\bf x},{\bf y})$ is the key object for any 
determinantal process. Once we know the kernel, we can determine, in principle, any
$n$-point correlation function by computing an $(n\times n)$ determinant [see Eq. (\ref{eq:Rn_det})].

\section{One dimensional harmonic oscillator at zero temperature and RMT}\label{section:1d_RMT}

In this Section, we consider the special case $d=1$ and the harmonic
oscillator potential $V(x)= \frac{1}{2}\, m\,\omega^2\, x^2$. In this 
special case, a host of analytical results for the zero temperature
quantum statistics have been derived over the 
years~\cite{gleisberg,calabrese_prl,vicari_pra,vicari_pra2,vicari_pra3,eisler_prl}. 
It turns out 
there is a close connection between the
ground state quantum probability $|\Psi_0({\bf x}_1, \cdots, {\bf 
x}_N)|^2$ of $N$ fermions in a $1d$ harmonic trap and the joint 
probability distribution
of $N$ real eigenvalues of an $(N\times N)$ Gaussian Hermitian random
matrix, known as the Gaussian Unitary Ensemble (GUE) in RMT. 
Although the connection between free fermions and GUE eigenvalues 
was known implicitly for a long time~\cite{mehta}, 
this connection was first used explicitly, to our knowledge, in the 
context of 
studying the 
statistics of nonintersecting step edges on the vicinal
surface of a crystal~\cite{einstein}. Subsequently, a connection
between nonintersecting lines in presence of a potential $V(x)=x^2/2+ 
c/x^2$ (with $x\ge 0$ and $c>0$) and the eigenvalue statistics of Wishart
ensembles of RMT was established~\cite{NM_interface}.
However, in the precise context of
free fermions in a $1d$ harmonic trap, this 
connection to GUE eigenvalues was noticed and exploited only recently,
first somewhat a posteriori 
in Ref.~\cite{eisler_prl} and then more explicitly in  
Ref.~\cite{marino_prl, marino_pre}, in the context of 
counting the number of fermions in an interval $[-L,L]$ in the ground 
state.

To establish the precise connection to GUE eigenvalues, we
consider the first $N$ single particle levels with
energies $\epsilon_{k}= (k+1/2)\hbar\omega$ where
$k=0,1,2,\ldots, N-1$. The many-body ground state is
constructed by filling up these first $N$ levels with $N$
fermions. Thus the ground state energy is $E_0=\frac{N^2}{2} \hbar\omega$, and the Fermi energy $\mu=(N-1/2)\hbar\omega$
corresponds to the highest occupied single particle energy level
in the many-body ground state.
To construct the many-body ground state wave function, 
we substitute the explicit single particle harmonic oscillator 
wave functions [labeled by $k=0,1,2\ldots, (N-1)$ in  Eq. 
(\ref{spectrumHO})] in the $(N\times N)$ Slater 
determinant (\ref{slater}). This gives
\bea
\Psi_0(x_1,\, x_2\cdots, x_N)\propto e^{-\frac{\alpha^2}{2} 
\sum_{i=1}^N x_i^2}\, \det[H_i(\alpha \, x_j)]\, ,     
\label{HOd1.1}
\eea
where we recall that $H_i(x)$ is the Hermite polynomial of degree $i$
and $\alpha=\sqrt{m \omega/\hbar}$ is an inverse length scale.
By arranging the rows and columns in the determinant, it is easy to see
that it can be reduced to a Vandermonde determinant up to an overall 
constant,  $\det[H_i(\alpha x_j)]\propto \det[x_i^{j-1}] \propto \prod_{i<j}(x_i-x_j)$.
Hence, the ground state quantum probability is given by~\cite{marino_prl}
\begin{equation}
|\Psi_0(x_1, \cdots, x_N)|^2 = 
\frac{1}{z_N} \prod_{i<j} (x_i - x_j)^2 
e^{-\alpha^2\sum_{i=1}^N x_i^2}
\label{rm1},
\end{equation}
where $z_N$ is a normalization constant. 

Consider now a random $(N\times N)$ complex Hermitian matrix $X$ with
independent Gaussian entries, such that the joint distribution
of independent matrix entries is given by, ${\rm Prob.}[X]\,dX\propto
\exp\left[-{\Tr}(X^2)\right]\, dX$. This joint distribution
remains invariant under a unitary transformation, $X\to U^\dagger XU$,
which justifies the name of such an ensemble of random matrices as the 
GUE~\cite{mehta}. Each realization of this matrix can be diagonalized to
give $N$ real eigenvalues $\{\lambda_1,\lambda_2,\cdots,\lambda_N\}$ which
are also random variables. What can be said about the joint distribution
of these $N$ eigenvalues? To obtain this eigenvalue distribution, one
first makes a change of variables from the independent matrix entries to
the eigenvalues and eigenvectors of $X$. Thanks to the rotational 
invariance, the eigenvector degrees of freedom decouple from the
eigenvalues and hence can be integrated out. This provides an explicit
expression for the joint distribution of eigenvalues~\cite{mehta}
\bea
P(\lambda_1,\lambda_2,\cdots,\lambda_N)= \frac{1}{Z_N}\,\prod_{i<j} 
(\lambda_i - 
\lambda_j)^2\, e^{-\sum_{i=1}^N \lambda_i^2}\label{rm_jpdf},
\eea
where $Z_N$ is the normalization constant. The Vandermonde term
$\prod_{i<j}
(\lambda_i -
\lambda_j)^2$ in Eq. (\ref{rm_jpdf}) owes its origin to the Jacobian
of the change of variables from the matrix entries to eigenvalues and 
eigenvectors~\cite{mehta}. 

Comparing Eqs. (\ref{rm1}) and (\ref{rm_jpdf}), it is clear that
the quantum statistics of the Fermion positions 
$x_i$'s in a $1d$ harmonic trap at $T=0$ is identical, up to a trivial rescaling factor $\alpha$, to the classical
statistics of GUE eigenvalues $\lambda_i$'s~\cite{marino_prl}.
The routes of arrival to the identical joint distribution are, however,
quite different in the two problems. Interestingly, in the RMT literature,
to calculate further observables from this joint eigenvalue distribution, 
the
determinantal structure of this joint distribution was noticed and 
exploited
extensively~\cite{mehta,forrester}. In particular, the joint distribution
in Eq. (\ref{rm_jpdf}) was indeed expressed as the determinant
of a kernel,
\bea
P(\lambda_1,\lambda_2,\cdots,\lambda_N)= \frac{1}{N!}   
\det_{1\leq i,j \leq N} K(\lambda_i,\lambda_j),\quad {\rm with}\quad 
K(\lambda,\lambda')= \frac{1}{\sqrt{\pi}}\, 
e^{-\frac{1}{2}(\lambda^2+\lambda'^2)}\, 
\sum_{k=0}^{N-1} \frac{1}{2^k\, k!}\,
 H_{k}(\lambda)\,H_{k}(\lambda')
\label{op.1}
\eea
where $H_{k}(z)$ is the $k$'th Hermite polynomial. We recall that the Hermite polynomials
are orthogonal on the real axis with respect to (w.r.t.) to the Gaussian weight
\bea
\int_{-\infty}^{\infty} e^{-z^2}\, H_n(z)\, H_m(z)\, dz= \sqrt{\pi}\, 
2^n\, n!\, \delta_{n,m}\,.
\label{op.2}
\eea 
In the RMT literature, this orthogonal property of Hermite polynomials
were exploited to derive the determinantal structure of the joint
distribution, hence is known in RMT as 
the orthogonal polynomial method~\cite{mehta}.
But this is precisely equivalent to the many-body quantum mechanics
of $N$ fermions in a trap, once one recognizes that the joint
distribution in Eq. (\ref{rm_jpdf}) is just the square of the Slater 
determinant and that
$e^{-\lambda^2}\,H_k(\lambda)$ is just the $k$-th single particle
eigenfunction of a harmonic oscillator. The kernel $K(\lambda,\lambda')$
in Eq. (\ref{op.1}) is also identical, up to a trivial rescaling factor,
to the one defined for fermions in Eq. (\ref{eq:def_kernel}) in the case 
of a $1d$ harmonic oscillator.

Thus, to summarize, both the RMT and the many-body free fermions 
essentially use the same method to analyze the joint distribution.
Hence, for this analysis, one does not really need to know anything
about random matrices. The starting point is really the joint distribution
in Eq. (\ref{rm1}), which can then be written as the determinant of a
kernel $K_{\mu}(x,y)$. From the general definition of the kernel in Eq. 
(\ref{eq:def_kernel}), upon summing up the first $N$ single particle 
harmonic oscillator eigenfunctions in Eq. (\ref{spectrumHO}), one obtains
\bea
K_{\mu}(x,y)= \frac{\alpha}{\sqrt{\pi}}\, 
e^{-\frac{1}{2}\alpha^2 (x^2+y^2)} \sum_{k=0}^{N-1} 
\frac{1}{2^k\, k!}\,
H_k(\alpha \, x)\,H_k(\alpha \, y)\, .
\label{kernel.1}
\eea
Note that because the Hermite polynomials are orthogonal polynomials (\ref{op.2}), they satisfy the Christoffel-Darboux
identity \cite{mehta,forrester} which allows us to perform the sum over $k$ in (\ref{kernel.1}) explicitly, to yield (for $N\geq 1$):
\begin{eqnarray}\label{CD}
K_{\mu}(x,y) = 
\begin{cases}
&\dfrac{e^{-\frac{\alpha^2}{2}(x^2+y^2)}}{\sqrt{\pi} 2^N (N-1)!}  \dfrac{H_N(\alpha\, x) H_{N-1}(\alpha \, y) - H_{N-1}(\alpha \, x) H_N(\alpha \, y)}{x-y} \;, \; {\rm for} \; x \neq y \\
& \\
& \dfrac{\alpha}{\sqrt{\pi}} \dfrac{e^{-\alpha^2 x^2}}{2^{N-1} (N-1)!}\left(N \left[H_{N-1}(\alpha \, x)\right]^2 - (N-1) H_{N-2}(\alpha \, x) H_N(\alpha \, x) \right) \;, \; {\rm for} \; x = y \;.
\end{cases}
\end{eqnarray}

Consequently, the average density has the exact expression valid for any 
$N$
\bea
\rho_N(x)&=& \frac{1}{N}\, K_{\mu}(x,x)= \frac{\alpha}{N\,\sqrt{\pi}}\,
e^{-\alpha^2\, x^2}\, \sum_{k=0}^{N-1}
\frac{1}{2^k\, k!}\,
H_k^2(\alpha x) \\
&=& \dfrac{\alpha}{\sqrt{\pi}} \dfrac{e^{-\alpha^2 x^2}}{2^{N-1} N!}\left(N \left[H_{N-1}(\alpha x)\right]^2 - (N-1) H_{N-2}(\alpha x) H_N(\alpha x) \right)\, .  
\label{avg.den1}
\eea 
A large number of precise analytical results for the ground state quantum 
statistics of free fermions in a $1d$ harmonic trap have recently been 
predicted~\cite{eisler_prl,marino_prl,castillo,CDM14,marino_pre}, essentially using 
the determinantal structure of the joint distribution and the explicit 
kernel in Eq. (\ref{kernel.1}).

While the results in Eqs. (\ref{kernel.1})-(\ref{avg.den1}) are exact for all $N$, 
it is useful and perhaps more interesting to see how the average density
and the kernel behave asymptotically for large $N$. In the RMT literature,
the large $N$ asymptotics have been studied in great detail, mostly
by using the asymptotic properties of the Hermite 
polynomials in Eqs. (\ref{kernel.1}) and 
(\ref{avg.den1})~\cite{mehta,forrester}.
For the benefit of readers not familiar with the RMT literature, we list
below the principal RMT predictions for large $N$. For details of these
derivations, 
the readers may consult Refs.~\cite{mehta,forrester}.

\subsection{Large $N$ RMT predictions for the average density} \label{rmtsec}

In the large $N$ limit, the average density of fermions (equivalently
that of GUE eigenvalues) is 
given by the celebrated
Wigner semi-circular law \cite{mehta,forrester}:
\begin{eqnarray}\label{wigner}
\rho_N(x) \approx 
\frac{\alpha}{\sqrt{N}} 
f_W\left(\frac{\alpha \, x}{\sqrt{N}} \right) \;, \; 
f_W(z) = \frac{1}{\pi}\sqrt{2-z^2} \;,
\end{eqnarray}
with sharp edges at $\pm \sqrt{2N}/\alpha$ [see Fig. \ref{fig:wigner1}].
Note that the average density is normalized to unity, $\int \rho_N(x)\, 
dx=1$ and $\rho_N(x)$ has the dimension of $\alpha$, i.e, inverse length.
The result in Eq. (\ref{wigner}) indicates that on an average there are 
more particles
near the trap center $x=0$ and less near the two edges. Thus, in the
`bulk' of the Wigner sea, i.e, far away from the two edges, 
the density typically scales as $\rho_N(x)\sim \alpha\, N^{-1/2}$ for 
large $N$. This means that 
the
typical inter-particle separation in the bulk
scales as 
$\ell(x) \sim 1/[N \rho_N(x)]\sim \frac{1}{\alpha}\, N^{-1/2}$ for large 
$N$.

In contrast, near the two edges, the particles are sparse (see Fig.~\ref{fig:wigner1})
and the typical separation between two particles at the same edge scales 
as
$\sim N^{-1/6}$~\cite{mehta,forrester}. 
For finite but large $N$, the 
sharp edges at $\pm \sqrt{2N}/\alpha$ gets smeared over a width
$w_N \sim N^{-1/6}$. This is called the `edge' regime (see Fig.~\ref{fig:wigner1}). The average density near the edge, for finite but large 
$N$, is described 
by a finite size scaling form~\cite{BB91,For93}
\begin{eqnarray}\label{edge_density_1d}
\rho_N(x) \approx \frac{1}{N \, w_N} 
F_1\left[\frac{x - \sqrt{2N}/\alpha}{w_N} \right]
\end{eqnarray}
where we have set the width of the edge regime
\begin{equation}
w_N= \frac{1}{\alpha \sqrt{2}}\, N^{-1/6}\, .
\label{width_def.1}
\end{equation}
The scaling function is given 
exactly by~\cite{BB91,For93} 
\bea\label{edge_density_scaling_function}
F_1(z) = [{\rm Ai}'(z)]^2 - z [{\rm Ai}(z)]^2\,
\eea
where ${\rm Ai}(z)$ is the Airy function
and ${\rm Ai}'(z)$ is its first derivative. The scaling function $F_1(z)$ has 
the
asymptotic behavior
\begin{eqnarray}
F_1(z) \approx
\begin{cases}
&\frac{1}{\pi}\, \sqrt{|z|}\quad\quad\quad {\rm as}\quad 
z\to 
-\infty \label{F1_asymp_left} \\
& \frac{1}{8\pi z}\, e^{-\frac{4}{3}\, z^{3/2}} \quad {\rm 
as}\quad z\to \infty\, .
\label{F1_asymp_right}
\end{cases}
\end{eqnarray}
Far to the left of the right edge, using
$F_1(z)\sim \sqrt{|z|}/\pi$ as $z \to -\infty$ in Eq. 
(\ref{F1_asymp_left}), it is easy to show that  
the scaling form (\ref{edge_density_1d}) smoothly matches with 
the semi-circular density in the bulk (\ref{wigner}). 
Recently, the edge scaling function $F_1(z)$ has been 
shown~\cite{eisler_prl} to be universal, i.e., holds
even for potentials different from the harmonic one, as long as it is 
smooth and confining. 

\subsection{Large $N$ RMT predictions for the kernel}

The kernel in Eq. (\ref{kernel.1}) can be analyzed similarly in the 
large $N$ limit. For 
example, consider first the bulk with two points $x$ and $y$,
both on the scale of the local
inter-particle separation $\ell(x)$, defined
as
\begin{equation}
\ell(x)= \frac{2}{\pi\, N\, \rho_N(x)}\, .
\label{bulk_separation.1}
\end{equation}
Taking the limit $N\to \infty$, and the separation between two points
$|x-y|\to 0$, but keeping the ratio $|x-y|/\ell(x)$ fixed, 
it has been shown that the kernel $K_\mu(x,y)$ satisfies the scaling form
\bea
K_{\mu}(x,y)\approx \frac{1}{\ell(x)} {\cal K}^{\rm 
bulk}\left(\frac{|x-y|}{\ell(x)}\right)\, ,
\label{bulk_kernel.1}
\eea
with the scaling function
\bea
{\cal K}^{\rm bulk}(z)= \frac{\sin(2z)}{\pi z}\, .
\label{sine_kernel.1}
\eea
This is the celebrated sine kernel which also turns out to be
universal, i.e., independent of the precise shape of the trap
potential $V(x)$~\cite{eisler_prl}. Note that when $z\to 0$,
${\cal K}^{\rm bulk}(z)\to 2/\pi$ and consequently, the
kernel $K_{\mu}(x,x)\to N\,\rho_N(x)$, in agreement with
Eq. (\ref{density_kernel}).

Similarly, near the edges (say the right edge at 
$x_{\rm edge}=\sqrt{2N}/\alpha$),
the kernel $K_\mu(x,y)$ in Eq. (\ref{kernel.1}) can be similarly analyzed
in the scaling limit: $N\to \infty$, $x\to x_{\rm edge}$, 
$y\to x_{\rm edge}$ but with the ratios $(x-x_{\rm edge})/w_N$
and $(y-x_{\rm edge})/w_N$ fixed. Here $w_N$ denotes 
the width $w_N$ of the edge regime as defined in Eq.~(\ref{width_def.1}).
In this scaling limit, one finds~\cite{mehta,forrester} 
\bea
K_{\mu}(x,y) \approx \frac{1}{w_N}\, {\cal K}^{\rm 
edge}\left(\frac{x-x_{\rm edge}}{w_N},\frac{y-x_{\rm edge}}{w_N}\right)\, 
\label{edge_kernel.1}
\eea
where the two-variable scaling function is given by the so called
Airy kernel~\cite{mehta,forrester}
\bea
{\cal K}^{\rm edge}(a,b)= K_{\rm Airy}(a,b)= 
\frac{{\rm Ai}(a){\rm Ai}'(b)-{\rm Ai}'(a){\rm Ai}(b)}{a-b}\, = \int_0^{+\infty} du \, {\rm Ai}(a+u) {\rm Ai}(b+u) \, .
\label{airy_kernel.1}
\eea
At coinciding points, it is easy to check that
\bea
K_{\rm Airy}(z,z)= F_1(z)=[{\rm Ai}'(z)]^2 - z [{\rm Ai}(z)]^2\, .
\label{airy_same_point}
\eea
This fact, together with the definition $\rho_N(x)=K_\mu(x,x)/N$
then yields back the edge density result mentioned in Eq.~(\ref{edge_density_1d}).

\subsection{Statistics of the rightmost fermion, of fermion spacings and of number fluctuations}\label{sec:TW_T=0}

From the connection with RMT, one can immediately obtain interesting predictions for various
observables associated to $1d$ free fermions at $T=0$. 
Indeed, in any determinantal point process \cite{johansson,borodin_determinantal}, the full counting statistics can be obtained in terms of
Fredholm determinants (denoted in this paper by $\rm Det$). The Laplace transform of the probability 
$P_J(n)$ that there are exactly $N_J=n$ fermions in a given (arbitrary)
subset $J$ of the real axis is given by
\bea \label{counting} 
\langle e^{- p N_J } \rangle = {\rm Det}[ I - (1- e^{-p}) P_J K_\mu P_J ]
\eea 
where $P_J(x)$ is an indicator function, such that $P_J(x) = 1$ if $x \in J$ and $P_J(x) = 0$ otherwise,
i.e. the projector on the interval $J$. In particular the
{\it hole probability}, i.e., the probability that there is exactly zero fermion
in the subset $J$ is then 
\bea \label{counting2} 
P_J(n=0)  = {\rm Det}[ I -  P_J K_\mu P_J ] \;.
\eea 
There are important applications of this formula both in the bulk and at the edge, which we now discuss.

{\it Fermion spacing distribution.} One example in the bulk concerns the distribution of spacings (or gaps) between fermions, analog of
the famous spacing distribution in RMT.
Denoting by $\{ x^{(i)} \}_{i=1, \ldots, N}$ the ordered set of fermion positions (i.e., $x^{(1)}<x^{(2)}< \ldots< x^{(N)}$), we 
define the spacing $g=|x^{(i+1)}-x^{(i)}|$ between two consecutive fermions. The mean
spacing near the center (which we consider here) is $\bar g=1/(N \rho_N(0))$. The simplest guess for this spacing distribution is the famous {\it Wigner surmise} 
\bea \label{wigner_surmise} 
p_2^W(s) = \frac{32 s^2}{\pi^2} e^{- 4 s^2/\pi}  \quad , \quad s = g/\bar g \;,
\eea 
which is normalized so that $\int_0^{+\infty} ds \, p^W_2(s)=\int_0^{+\infty} ds \, s \; p^W_2(s)=1$, i.e., the mean fermion 
spacing is set to unity.
As is well known this is the exact result for $N=2$. Thus, it is
an exact statement for 2 fermions in a quadratic well (at $T=0$). It also approximates rather well the {\it exact} distribution for a large number $N$ of fermions. The latter, close to the origin, can be obtained as
\bea\label{exact_p2}
p_2(g) = \frac{1}{N \rho_N(0)}\partial^2_g D(g) \;, \; D(g) = {\rm Det}[1 - P_{[0,g]} K_\mu P_{[0,g]}] \;.
\eea 
In the bulk, setting 
$g = s /( N \rho_N(0)) = s \frac{\pi}{2} \ell(0)$ (we made this rescaling by $\pi/2$ to conform to the standard convention used in RMT), one can replace the kernel in Eq. (\ref{exact_p2}) by its limiting form, namely the sine-kernel in Eq. (\ref{sine_kernel.1}). The fermion spacing distribution
is then described by the so called Mehta-Gaudin distribution 
\begin{eqnarray}\label{convergence_gap}
p_2(g) \simeq N \rho_N(0) \tilde D''(s)
\end{eqnarray} 
where $\tilde D(s)$ can be expressed in terms of a particular solution of a Painlev\'e V equation, denoted by $\sigma(x)$, such that
\begin{eqnarray}\label{expr_D_sigma}
\tilde D(s) = \exp\left(\int_0^{\pi s} \frac{\sigma(x)}{x} \, dx \right)
\end{eqnarray}
where $\sigma(x)$ satisfies
\begin{eqnarray}\label{eq:Painleve5}
(x \, \sigma''(x))^2 + 4 (x\sigma'(x) - \sigma)\left(x \sigma'(x) - \sigma(x) + [\sigma'(x)]^2 \right) = 0 \;,
\end{eqnarray}
with the boundary condition $\sigma(x) \sim -x/\pi$ as $x \to 0$. From the Painlev\'e equation (\ref{eq:Painleve5}) the asymptotic behaviors of $\tilde D''(s)$ can be obtained as \cite{mehta,BTW92,Meh92,Grim2004}
\begin{eqnarray}
\tilde D''(s) \sim
\begin{cases}
&\dfrac{\pi^2s^2}{3} - \dfrac{2 \pi^4 s^4}{45} + \dfrac{\pi^6 s^6}{315} - \dfrac{\pi^6 s^7}{4050} - \dfrac{2 \pi^8 s^8}{14175} + \dfrac{11 \pi^8 s^9}{496125} + \dfrac{2 \pi^{10} s^{10}}{467775} + {\cal O}(s^{11}) \;, \; s \to 0 \\
& \\
& \dfrac{\pi^4}{16}\left( \dfrac{2}{\pi s}\right)^{1/4} \left(s^2 - \dfrac{2}{\pi^2} + o(1) \right) \exp{\left(\dfrac{\ln 2}{12} + 3 \zeta'(-1) - \dfrac{\pi^2 s^2}{8}\right)} \:, \; s \to +\infty \;.
\end{cases}
\end{eqnarray}
Note that from Eq. (\ref{convergence_gap}) one obtains that the average gap is given $\bar g = 1/(N \rho_N(0)) = \frac{\pi \ell(0)}{2}$, with $\ell(x)$ given in Eq. (\ref{bulk_separation.1}). \\

{\it Rightmost fermion statistics.} One important 
application of the formula (\ref{counting2}) at the edge is as follows. 
In order to 
probe the statistics at the edge
of the cloud of fermions, it is natural to consider the rightmost fermion, 
$x_{\max}(T=0) = \max_{1\leq i \leq N} x_i$, where the quantum 
fluctuations of the positions of the $N$ fermions are 
described by the quantum probability in Eq. (\ref{rm1}). 
{Now, the cumulative probability distribution of $x_{\rm max}(T=0)$
is precisely related to the hole probability in Eq. (\ref{counting2}):
${\rm Prob.}\left[x_{\rm max}(T=0)\le y\right]$ is precisely the
probability that the interval $J\equiv [y, +\infty)$ is free of 
particles.}
Using the expression (\ref{counting2}) for the hole probability 
associated to the interval $J=[y=x_{\rm edge}+s \,w_N,+\infty)$
(see below), one obtains that the typical quantum fluctuations of 
$x_{\max}(T=0)$, correctly centered and scaled, 
are governed by the celebrated Tracy-Widom (TW) distribution for GUE, ${\cal F}_2(x)$ \cite{TW}. Indeed one has
\begin{eqnarray} \label{xmax_0T}
x_{\max}(T=0) = x_{\rm edge} + w_N \, \chi_2 \;,
\end{eqnarray}
where the cumulative distribution function (CDF) of the random variable $\chi_2$ is ${\cal F}_2(s) = {\Pr}(\chi_2 \leq s)$, which can be written as
a Fredholm determinant \cite{fredholm}
\begin{eqnarray}\label{fredholm_F2}
{\cal F}_2(s) = {\rm Det}(I - P_s K_{\rm Airy} P_s) \;,
\end{eqnarray} 
where $K_{\rm Airy}(a,b)$ is the Airy kernel given in Eq. (\ref{airy_kernel.1}) and $P_s$ is a projector on the interval $[s,+\infty)$. Note that ${\cal F}_2(s)$ can also be written in terms of a special solution $q(x)$ of the following Painlev\'e II equation \cite{TW}
\begin{eqnarray}\label{PII}
q''(x) = x q(x) + 2 q^3(x) \;, \; q(x) \sim {\rm Ai}(x) \;, \; x \to \infty \;. 
\end{eqnarray}
The TW distribution ${\cal F}_2(s)$ can then be expressed as
\begin{eqnarray}
{\cal F}_2(s) = \exp{\left[- \int_s^\infty (x-s) q^2(x) \, dx \right]} \;.
\end{eqnarray}
In particular its asymptotic behaviors are given by \cite{BBD08}
\begin{eqnarray}
{\cal F}_2(s) \sim
\begin{cases}
&\tau_2 \dfrac{e^{-\frac{1}{12}|s|^3}}{|s|^{1/8}}\left(1+ \dfrac{3}{2^6 |s|^3} + {\cal O}(|s|^{-6})\right) \;, \; s \to - \infty \;, \\
& \\
& 1 - \dfrac{e^{-\frac{4}{3} s^{3/2}}}{16\pi s^{3/2}} \left(1-\dfrac{35}{24 s^{3/2}} + {\cal O}(s^{-3}) \right) \;, \; s \to + \infty \;,
\end{cases}
\end{eqnarray}
where $\tau_2 = 2^{1/24}e^{\zeta'(-1)}$ where $\zeta'(x)$ is the derivative of the Riemann zeta function.

\begin{figure}
\begin{center}
\includegraphics[width = 0.5\linewidth]{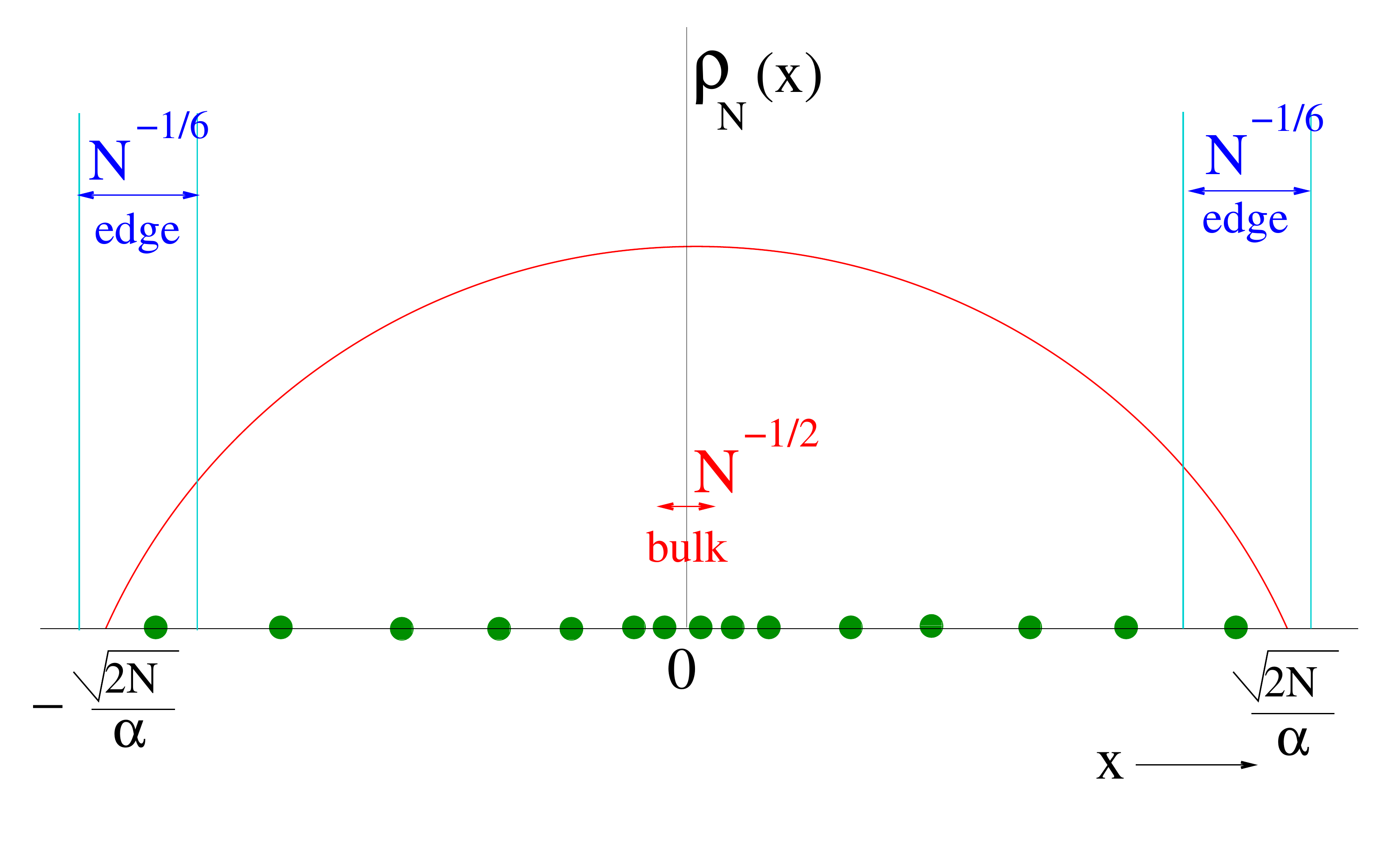}
\caption{The average density, for large $N$, has the
Wigner's semi-circular form: $\rho_N(x)\approx \frac{\alpha^2}{\pi N}\, 
\sqrt{\frac{2N}{\alpha^2}- x^2}$, where $\alpha=\sqrt{m\omega/\hbar}$.
The typical separation between particles in the bulk scales as 
$\sim 
N^{-1/2}$, where as near the edge it is much larger $\sim N^{-1/6}$.}
\label{fig:wigner1}
\end{center}
\end{figure}


\medskip

{\it Number variance.} Another interesting observable is the number of fermions $N_L$ in a symmetric interval $[-L,L]$ (or in any fixed interval), which is also a random variable. Its mean is easily computed from the average density $\rho_N(x)$ in Eq. (\ref{R1.2}) as $\langle N_L \rangle = N \int_{-L}^{+L} \rho_N(x) \, dx$, which for large $N$ can be easily evaluated from the limiting semi-circle law~(\ref{wigner}). What about the higher cumulants of this random variable, for instance the variance ${\rm Var}(N_L)$, and eventually the full distribution of $N_L$? In RMT, the variance was computed a long time ago by Dyson \cite{Dys1962}, but only in the bulk limit when $L = s \, \ell(0)$ where $\ell(0)$ is the inter-particle spacing in the center of the trap given in Eq. (\ref{bulk_separation.1}) for $x=0$, and $s$ is a dimensionless number of order ${\cal O}(1)$. In particular, for large $s$, one has
\begin{eqnarray}\label{variance_dyson}
{\rm Var}(N_{L=s \ell(0)}) \sim \frac{1}{\pi^2} \log s + {\cal O}(1) \;, \; s \to \infty \;,
\end{eqnarray} 
a result that can also be obtained using the LDA \cite{Castin}. In the bulk regime one can show that the full distribution of $N_L$, properly centered and scaled, is a Gaussian \cite{CL1995,FS1995}. It is only recently that the fluctuations of $N_L$, beyond the bulk regime, {\it i.e.} for $L \gg \ell(0)$, were studied. The variance ${\rm Var}(N_L)$ has indeed found a renewed interest \cite{vicari_pra,vicari_pra2,calabrese_prl,vicari_pra2,vicari_pra3,eisler_prl,CDM14,EP2014} in the context of free fermions, thanks to its connection to entanglement entropy (see also below). Its dependence on $L$ was first studied numerically \cite{vicari_pra,eisler_prl} for various trapping potentials, including the harmonic potential, and it was observed that it displays a striking non-monotonic behavior. Recently, a full analytical computation of the variance as well as the full distribution of $N_L$ for fermions in a harmonic potential was performed in \cite{marino_prl,marino_pre}. Using RMT tools, in particular Coulomb gas techniques, the variance was computed for any $L$ and large $N$, thus extending the analysis of Dyson \cite{Dys1962} far beyond the bulk regime.    \\

{\it Entanglement entropy.} The other interesting observable 
at $T=0$ is the R\'enyi 
entanglement entropy $S_q$ of
the interval $[-L,L]$ around the trap center with the
rest of the system. 
Consider the many-body fermionic
system in its ground state, so that the density matrix of the full system in
this pure state is simply, ${\hat \rho}= \left|\Psi_0\rangle\langle
\Psi_0\right|$. Consider now the subsystem $A\equiv [-L,L]$ and let ${\hat
\rho}_A= {\rm Tr}_{\bar A} {\hat \rho}$ denote the reduced density matrix of
the subsystem $A$, obtained by tracing out the complementary subsystem $\bar 
A$ (so that $A$ and $\bar A$ together constitute the full real line). The 
R\'enyi
entanglement entropy of the subsystem $A$, parametrized by $q\ge 1$, is then
defined as, $S_q= \frac{1}{1-q}\, \ln {\rm Tr}\, {\hat \rho}_A^q$.
In the limit $q\to 1$, this reduces to the standard von Neumann entropy, $S_1=
- {\rm Tr}[{\hat \rho_A}\, \ln \hat \rho_A]$. For free fermions in a harmonic  
trap in one dimension at $T=0$, the R\'enyi entropy was studied 
numerically~\cite{vicari_pra,eisler_prl}. Exploiting the connection of the
fermionic system to RMT, the R\'enyi entropy was recently computed 
analytically for large $N$, and for a wide range of $L$~\cite{CDM14}.
For instance, it was shown~\cite{CDM14} that for all $L$ such that 
$\sqrt{2\,N}/\alpha - L \gg w_N$ [where we recall that $w_N = 1/(\alpha \sqrt{2}) N^{-1/6}$, see Eq. (\ref{width_def.1})],
there is an exact relationship between the R\'enyi entropy and the number
variance $V_N(L)$ discussed above
\begin{equation} S_q= \frac{\pi^2}{6}\,
\left(1+\frac{1}{q}\right)\, V_N(L)\, .
\label{renyi2}
\end{equation}
However, around the edge, this relation breaks down and computing 
the scaling behavior of $S_q$ near the edge remains a challenging open
problem~\cite{CDM14}.

Let us close by indicating that, as a property of determinantal point processes which generalizes 
\eqref{counting}, see e.g. \cite{johansson,borodin_determinantal,tracy_widom_determinantal}, there exists a class of 
averages over the positions of the fermions which can be expressed exactly
in terms of a Fredholm determinant (which is valid for any finite $N$)
\bea \label{genf} 
&&  \Big\langle \prod_{i=1}^N f(x_i) \Big \rangle_0 = {\rm Det}( I - L_f) \quad , \quad  L_f(x,y) = (1- f(x)) K_\mu(x,y) \;,
\eea 
where $K_\mu(x,y)$ is the kernel associated with the determinantal process. In Eq. (\ref{genf}), 
the average is the quantum average in the ground state and 
the function $f(x)$ is arbitrary, provided the right hand side exists. 
Such formula can be useful for studying linear statistics of free fermions \cite{Johansson_Lambert}.

\vspace*{1cm}


\section{Correlations for non-interacting fermions at finite temperature $T>0$} \label{sec:1dT} 

We now want to study the effect of non-zero temperature for $N$ non-interacting fermions
in an external confining potential. The discussion below holds for arbitrary dimensions, but
we will focus for simplicity on the $d=1$ case. For the harmonic oscillator at $T=0$ one
could use, as in the previous section, the techniques of RMT. However at finite temperature,
even for the harmonic oscillator potential, these direct RMT connections are lost and
one needs to develop new techniques.

We will focus on the canonical ensemble at temperature $T = 1/\beta$ that corresponds to a fixed number of fermions $N$, 
which is often the situation studied in cold atoms experiments. Before doing that,
we first describe the energy basis of the Hilbert space, and the determinantal properties of
the corresponding wave functions, in a more general setting. This general formalism is then applied to
the harmonic oscillator, in the following section. 

\subsection{$N$-fermion Hilbert space and occupation number basis}

In order to study excited states, we need to consider the full Hilbert space
of the $N$ particles. A natural basis of this Hilbert space is formed by
the eigenstates of the $N$ particle Hamiltonian $\hat {\cal H}_N$. For non-interacting
fermions, these eigenstates and this basis can be constructed from the 
eigenstates of the single particle Hamiltonian $\hat H$. In the case
of the one-dimensional harmonic oscillator, the eigenfunctions $\phi_k(x)$ 
are given in (\ref{spectrumHO}). The associated energy eigenvalues are $\epsilon_k= \hbar\omega(k+1/2)$
where $k$ is an integer which ranges from $0$ to $\infty$. From these single particle eigenstates, one can
construct all many-body eigenfunctions of $\hat {\cal H}_N$ by 
putting $N$ fermions in $N$ different single particle levels indexed by $k_1<k_2<\ldots<k_N$. 
The fermionic nature of the particles allows at most one particle in a given single particle level.
Hence one introduces the set of occupation numbers, denoted by $\{ n_k \}$, $k=0, 1, 2,\ldots$ with 
$n_k =0,1$, to label the
many body states, with $n_{k_1}=n_{k_2}= \ldots = n_{k_N}=1$ for the occupied single particle states and
$n_k=0$ otherwise. They satisfy the constraint $\sum_{k=0}^\infty n_k = N$. 
The corresponding many-body eigenfunction is given by a Slater determinant,
with the corresponding eigenenergy, 
\bea \label{manybodyeigen}
\displaystyle{\Psi_{\{ n_k \}}(\{x_i\})= \frac{1}{\sqrt{N!}}  \det_{1\leq i,j \leq N} \phi_{k_i}(x_j)} \quad , \quad
E \equiv E_{\{ n_k \}} = \sum_{k=0}^\infty n_k \epsilon_k 
\eea 
with, e.g. $E=\hbar\omega\,(k_1+k_2+\dots+k_N+ N/2)$ for the harmonic oscillator. 

An important property, already discussed and used in Section \ref{sec:zeroT} for the ground state, 
but which extends
to any $N$-body eigenstate, i.e. to all the excited states, is that the squared modulus of
the wave function can be written as a determinant 
\bea
|\Psi_{\{ n_k \}}(x_1, \ldots, x_N)|^2 := \frac{1}{N!} \left| \det_{1\leq i,j \leq N} \phi_{k_i}(x_j) \right| ^2 = 
\det_{1\leq i,j \leq N}  K(x_i,x_j; \{n_k\})
\label{kernel.0}
\eea 
where the kernel $K(x,x';\{n_k \})$ is indexed by the set of occupation numbers and is given by
\begin{eqnarray}\label{def_kernel_canonical}
K(x,x';\{n_k \}) = \sum_{j=1}^N \phi^*_{k_j}(x) \phi_{k_j}(x') =
\sum_{k=0}^\infty n_k \, \phi^*_k(x) \phi_k(x') \;.
\end{eqnarray}
Note that there is one kernel associated to {\it each} eigenstate of the energy operator $\hat {\cal H}_N$.
Furthermore, using the ortho-normalization of the single particle eigenfunctions one
easily shows that these kernels obey the following property
\bea \label{convolution2} 
\int dz \ K(x,z;\{n_k \})  K(z,y;\{n'_k \})  = K(x,y;\{n_k n'_k \}) 
\eea 
for any two given sets $\{n_k \}$ and $\{n'_k \}$. Specializing (\ref{convolution2}) to
$\{n_k \}=\{n'_k \}$ and using that $n_k^2=n_k$ for any $k$,
we see that each of these kernels satisfy the reproducing
property (\ref{eq:convolution}). 
An immediate consequence, as in Section \ref{sec:zeroT}, is that
if the system is prepared in any of these states, the density and the correlations
are given by determinants, as in (\ref{density_kernel}) and (\ref{eq:Rn_det}).

\subsection{Canonical measure and observables at finite $T$}

Let us first recall the definition of the canonical partition function $Z_N(\beta)$ for $N$ non-interacting
fermions at temperature $T=1/\beta$
\begin{eqnarray}\label{partition_function}
Z_N(\beta) = {\rm Tr} e^{- \beta \hat {\cal H}_N} = \sum_{k_1<k_2<\ldots< k_N} e^{-\beta\, (\epsilon_{k_1}+\cdots+ \epsilon_{k_N}) } \;,
\end{eqnarray}
where the sum is over all possible $N$ fermion eigenstates of $\hat {\cal H}_N$ labeled
as described above in terms of all possible distinct occupied single particle eigenstates.
It is also convenient to rewrite it using a labeling by occupation numbers~as
\begin{eqnarray}\label{partition_occupation}
Z_N(\beta) =  \sum_{\{n_k\}} \left[ e^{-\beta\, \sum_{k\geq 0} n_k \epsilon_k} 
\delta\left(\sum_{k\geq0} n_k, N\right) \right] \;,
\end{eqnarray}
where $\sum_{\{n_k\}}$ denotes the sum over all the possible occupation numbers 
$n_k = 0,1$ for $k=0,1,2, \ldots$. In Eq. (\ref{partition_occupation}) 
$\delta(i,j) = 1$ if $i=j$ and $\delta(i,j) = 0$ if $i\neq j$: this Kronecker delta function thus 
imposes the total number of particles to be exactly $N$, as we are working in the canonical ensemble.

In the canonical ensemble, the quantum joint probability distribution function 
of the positions $x_i$ of the fermions, $P_{\rm joint}(x_1, \ldots, x_N)$, is defined in terms of the $N$-body density matrix $\hat \rho=e^{- \beta \hat {\cal H}_N}/Z_N(\beta)$ as
\bea \label{jpdf} 
P_{\rm joint}(x_1, \ldots, x_N) = \langle x_1, \ldots ,x_N | \hat \rho | x_1, \ldots ,x_N \rangle 
= \frac{1}{Z_N(\beta)}  \sum_{\{n_k\}} |\Psi_{\{n_k\}} (\{x_i\})|^2 e^{-\beta\, \sum_{k\geq 0} n_k \epsilon_k}
\delta\left(\sum_{k\geq0} n_k, N\right)
\eea 
where the sum is over all many-body eigenstates. Using Eq. (\ref{manybodyeigen}) we can rewrite the
joint PDF of the particle positions in the canonical ensemble  
as the Boltzmann weighted sum of Slater determinants
\begin{eqnarray}\label{p_start}
P_{\rm joint}(x_1, \ldots, x_N) &=& \frac{1}{{N!}Z_N(\beta)}\sum_{k_1< \cdots < k_N} 
\left| \det_{1\leq i,j \leq N} \phi_{k_i}(x_j) \right| ^2 e^{-\beta\, (\epsilon_{k_1}+\cdots+ \epsilon_{k_N})} \;,
\end{eqnarray}
where $Z_N(\beta)$ is the canonical partition function (\ref{partition_function}). 
It is easy to check that $Z_N(\beta)$ is such that the PDF $P_{\rm joint}(x_1, \ldots, x_N)$ 
is normalized to unity.

The first observable we want to compute is the mean density of fermions at finite temperature $T$ defined
as
\bea
\rho_{N}(x) = \frac{1}{N} \sum_{i=1}^N \langle \delta(x-x_i)\rangle
\eea
where from now on $\langle \ldots \rangle$ means an average computed with the
joint PDF (\ref{p_start}). 
This amounts, up to a multiplicative constant, to integrating 
the joint PDF $P_{\rm joint}(x,x_2, \ldots, x_N)$ over the last $N-1$ variables. This 
amounts to the calculation of the following integral
\begin{eqnarray}\label{start_rho}
\rho_N(x) = \int_{-\infty}^\infty dx_1  
\delta(x-x_1) \int_{-\infty}^\infty dx_2 \ldots   
\int_{-\infty}^\infty dx_N P_{\rm joint}(x_1, \cdots x_N) 
\end{eqnarray}
where any of the two forms for $P_{\rm joint}$ in Eqs. (\ref{jpdf}) and (\ref{p_start}) can be inserted. 
More generally, we want to calculate the $n$-point correlation function $R_n(x_1,\cdots,x_n)$ 
at temperature $T$ defined as
\begin{eqnarray}\label{def_correl}
&&R_n(x_1, \cdots, x_n) = \frac{N!}{(N-n)!} \int_{-\infty}^\infty dx_{n+1} \cdots \int_{-\infty}^\infty dx_{N} P_{\rm joint}(x_1, \cdots, x_n, x_{n+1}, \cdots, x_N) \;.
\end{eqnarray}

The question is to handle these multiple integrals in the case of finite $T$. To understand the
difficulty of this calculation one can note that the joint PDF in Eq. (\ref{p_start}) can be written as a determinant,
as it is the case for $T=0$,~\cite{MNS94}
\begin{eqnarray}\label{p_start2}
P_{\rm joint}(x_1, \cdots x_N) &=& \frac{1}{{N!}Z_N(\beta)}
\det_{1\leq i,j \leq N} G(x_i,x_j,\beta \hbar) 
\eea
in terms of the Euclidean propagator associated to the one-body Hamiltonian 
\bea
G(x,y;t) = \langle y | e^{- \frac{t}{\hbar} \hat H} | x \rangle = \sum_k e^{- \frac{t}{\hbar} \epsilon_k} \phi^*_k(x) \phi_k(y)  \;.
\eea
Unfortunately, and at variance with the case of $T=0$, successive integrations over the coordinates 
$x_i$ do not preserve this determinantal structure. This is because the kernel inside the determinant
no longer satisfies the reproducing property since
\bea
\int_{-\infty}^{\infty} dz \, G(x,z; \beta \hbar)  G(z,y; \beta \hbar) = G(x,y; 2 \beta \hbar)
\eea 
which is clearly a different kernel. Hence the evaluation of these integrals for arbitrary $N$ is very difficult.

\subsection{Saddle point calculation and equivalence between canonical and grand canonical ensembles}\label{sec:equivalence}

Fortunately, in the limit of large $N$, it is possible to use a saddle point method to calculate the
density and the $n$-point correlation functions (at fixed $n$). As we will see, this is a manifestation of the
equivalence between the canonical and the grand-canonical ensembles for large $N$.

Consider the density $\rho_N(x)$ in Eq. (\ref{start_rho}). Inserting there the expression (\ref{jpdf})
for $P_{\rm joint}(x,x_2, \ldots, x_N)$, and replacing $|\Psi_{\{n_k\}} (\{x_i\})|^2$ 
by the determinant of the kernel given in (\ref{kernel.0}) we obtain
\bea
&&  \rho_N(x) =  \frac{1}{{Z_N(\beta)}}  \sum_{\{n_k\}} 
\left[ \int_{-\infty}^\infty dx_1  
\delta(x-x_1) \int_{-\infty}^\infty dx_2 \ldots   
\int_{-\infty}^\infty dx_N \det_{1\leq i,j \leq N}  K(x_i,x_j; \{n_k\})  \right] 
e^{-\beta \sum_{k\geq 0} n_k \epsilon_k} \delta\left(\sum_{k\geq 0} n_k,N\right)\;. \nn \\
&&
\eea 
We now use the property of reproducibility of the kernel for each choice of
$\{n_k\}$ noted above in (\ref{convolution2}). From the theorem mentioned
in Section \ref{sec:zeroT} leading to Eqs. (\ref{eq:Rn_det}), (\ref{density_kernel})
we can rewrite this multiple integral as:
\begin{eqnarray}\label{rho_occupation}
&& \rho_N(x) =  \frac{1}{{Z_N(\beta)}} \frac{1}{N} \sum_{\{n_k\}}  K(x,x;\{n_k \}) e^{-\beta \sum_{k\geq 0} n_k \epsilon_k} \delta\left(\sum_{k\geq 0} n_k,N\right) \;
\eea
where
\bea
&& K(x,x;\{n_k \}) =  \sum_{k=0}^\infty n_k |\phi_k(x)|^2 \;.
\end{eqnarray}
Note that in the limit where $T \to 0$, the system is in the ground state characterized by 
$n_k = 1$ if $k=0,1,2, \ldots, N-1$ and $n_k = 0$ if $k \geq N$. 
Hence in this limit, Eq. (\ref{rho_occupation}) reads
\begin{eqnarray}
\rho_N(x) = \frac{1}{N} \sum_{k=0}^{N-1} |\phi_k(x)|^2 \;.
\end{eqnarray}

To calculate the correlation functions given by the integral (\ref{def_correl})
we use the same method, and in particular the determinantal form (\ref{eq:Rn_det})
obtained after the $N-n$ integrations. 
We thus obtain the $n$-point correlations at finite temperature
(\ref{def_correl}) as:
\begin{eqnarray}\label{Rn_occupation}
R_n(x_1, \cdots, x_n) = \frac{1}{Z_N(\beta)} \sum_{\{n_k \}} \left[ \det_{1\leq i,j \leq n} K(x_i,x_j; \{n_k\}) e^{-\beta \sum_{k\geq 0} n_k \epsilon_k} \delta\left(\sum_{k\geq 0} n_k,N\right) \right] \;. \label{R1}
\end{eqnarray}

To compute  the expression in Eq. (\ref{Rn_occupation}) we must evaluate the ratio of two
sums over the occupation numbers ${\{n_k \}}$, each one constrained 
by $\sum_{k \geq 0} n_k = N$. Both in the numerator and the
denominator -- i.e., $Z_N(\beta)$ given by (\ref{partition_function}) -- 
we rewrite the constraint using an integral representation of the Kronecker
delta symbol:
\begin{equation} \label{decoupl} 
\delta\left(\sum_{k\geq 0} n_k,N\right) = \int_{0}^{2 \pi} {d\lambda\over 2\pi}
\exp\left[ i\lambda\left(\sum_{k\geq 0} n_k-N\right) \right] \,.
\end{equation}

Let us first consider the denominator $Z_N(\beta)$. It now reads
\bea
Z_N(\beta) =
\int_{0}^{2 \pi} {d\lambda\over 2\pi}
e^{- i N \lambda} \sum_{\{n_k = 0,1\}} e^{-\beta\, \sum_{k\geq 0} ( n_k \epsilon_k
- i \lambda n_k)} \;.
\eea 
This representation thus makes the $n_k$ variables independent of each other and one can perform
the sum over each $n_k$ separately. Note that each $n_k$ takes values $0$ or $1$. 
Performing the sum over $n_k$'s, for all $k$, we get
\bea
Z_N(\beta) = \int_{0}^{2 \pi} {d\lambda\over 2\pi}
e^{- i N \lambda} \prod_k \left(1 + e^{-\beta(\epsilon_k-i \lambda)} \right) \;.
\eea
Introducing the function 
\bea
 J(\tilde \mu) = -{1\over \beta}\sum_k \ln(1+ e^{-\beta \epsilon_k + \beta\tilde \mu}) \label{Jmu} 
\eea
which is just the free energy in the grand canonical ensemble at chemical potential $\tilde \mu$, 
we can rewrite the partition function simply as
\bea
Z_N(\beta) = \int_{0}^{2 \pi}  {d\lambda\over 2\pi} e^{-\beta J({i\lambda\over \beta})-i\lambda N} \;.
\eea

We now consider the full expression (\ref{Rn_occupation}) and use (\ref{decoupl}) to rewrite it as
\bea
R_n(x_1, \cdots, x_n) &=& \frac{1}{Z_N(\beta)}  \int_{0}^{2 \pi} {d\lambda\over 2\pi}
e^{- i N \lambda}  
\sum_{\{n_k \}}  
\det_{1\leq i,j \leq n} \left[ \sum_{k=0}^\infty n_k \phi^*_k(x_i) \phi_k(x_j)\right]
e^{-\beta \sum_{k\geq 0} (n_k \epsilon_k - i \lambda n_k)} \\
&=& \frac{1}{Z_N(\beta)}  \int_{0}^{2 \pi} {d\lambda\over 2\pi}
e^{- i N \lambda}  
\sum_{\{n_k \}}  
 \sum_{k_1< \ldots < k_N} n_{k_1} \ldots n_{k_N} 
  e^{-\beta \sum_{k\geq 0} (n_k \epsilon_k - i \lambda n_k)}\left|\det_{1\leq i,j \leq n}\phi_{k_i}(x_j)\right|^2 \label{sum1} 
\eea
In the second line we have used the (generalized) Cauchy-Binet-Andreief formula for determinants \cite{Andreief}:
\bea \label{Andreief} 
\det_{1\leq i,j \leq n} \left[ \sum_{k=0}^\infty n_k \phi^*_k(x_i) \phi_k(x_j)\right] 
= \sum_{k_1< \ldots < k_N} n_{k_1} \ldots n_{k_N} \left|\det_{1\leq i,j \leq n} \phi_{k_i}(x_j) \right|^2
\eea 
valid for any set of $\{ n_k \}$. To perform the sum over the occupation
numbers we introduce the notation for the ``expectation value'' of any observable $O[\{n_j \}]$ at fixed $\lambda$
\bea
\big \langle  O[\{n_j \}] \big \rangle_\lambda = e^{\beta J({i\lambda\over \beta})} \sum_{\{n_k \}}  O[\{n_j \}] 
e^{-\beta \sum_{k\geq 0} (n_k \epsilon_k - i \lambda n_k)} \label{avlambda} 
\eea 
which is such that $\langle 1 \rangle_\lambda =1$. This allows to rewrite (\ref{sum1}) as
\bea
R_n(x_1, \ldots, x_n) &=&  \frac{1}{Z_N(\beta)}  \int_{0}^{2 \pi} {d\lambda\over 2\pi} 
e^{- i N \lambda}  e^{- \beta J({i\lambda\over \beta})} 
 \sum_{k_1<..k_N} \langle n_{k_1} \ldots n_{k_N}  \rangle_\lambda 
  \left|\det_{1\leq i,j \leq n}\phi_{k_i}(x_j)\right|^2
\\
&=& 
 \frac{1}{Z_N(\beta)}  \int_{0}^{2 \pi} {d\lambda\over 2\pi} 
e^{- i N \lambda}  e^{- \beta J({i\lambda\over \beta})} 
 \sum_{k_1<..k_N} \langle n_{k_1}  \rangle_\lambda  \ldots \langle n_{k_N}  \rangle_\lambda 
  \left|\det_{1\leq i,j \leq n}\phi_{k_i}(x_j)\right|^2 \\
&=&
 \frac{1}{Z_N(\beta)}  \int_{0}^{2 \pi} {d\lambda\over 2\pi} 
e^{- i N \lambda}  e^{- \beta J({i\lambda\over \beta})}  
\det_{1\leq i,j \leq n} \left[ \sum_{k=0}^\infty \langle n_k \rangle_\lambda \phi^*_k(x_i) \phi_k(x_j)\right] 
\eea 
and $\langle n_k \rangle_\lambda = {1\over 1+e^{\beta \epsilon_k -i\lambda} }$. 
In the second line we have used explicitly the independence of the
variables $n_k$ at fixed $\lambda$, as seen from (\ref{avlambda}). In the
third line we have used again the Cauchy-Binet-Andreief formula (in reverse). Thus we 
finally obtain the $n$- point correlation function in the form
\begin{equation} 
R_n(x_1, \cdots, x_n) = {{ \int_{0}^{2 \pi}  {d\lambda\over 2\pi}
 [ \det_{1\leq i,j \leq n} K(x_i,x_j; \{ \langle n_k \rangle_\lambda \})] \, 
e^{-\beta J({i\lambda\over \beta})-i\lambda N} }\over  
\int_{0}^{2 \pi}  {d\lambda\over 2\pi} e^{-\beta J({i\lambda\over \beta})-i\lambda N} }\label{newR}
 \end{equation}
with
\bea \label{nklambda} 
K(x,x';\{ \langle n_k \rangle_\lambda \}) = \sum_{k=0}^\infty \langle n_k \rangle_\lambda \phi_k(x) \phi_k(x') 
\quad , \quad \langle n_k \rangle_\lambda = {1\over 1+e^{\beta \epsilon_k -i\lambda} } 
\eea 
and $J(\tilde \mu)$ as defined in (\ref{Jmu}). At this stage Eqs. (\ref{newR}), (\ref{nklambda}) and
(\ref{Jmu}) provide an {\it exact} representation
for the correlation function in the canonical ensemble for arbitrary $N$ and $n$, where the integrals over
$\lambda$ still need to be performed. In particular it holds for $R_1(x)=N \rho_N(x)$.

As a remark we note that the crucial property which made this representation possible is 
the following identity 
\bea \label{andre} 
\left\langle \det_{1\leq i,j \leq n} \left[ \sum_{k=0}^\infty n_k \phi^*_k(x_i) \phi_k(x_j)\right]\right
\rangle =
\det_{1\leq i,j \leq n} \left[\sum_{k=0}^\infty \langle n_k \rangle \phi^*_k(x_i) \phi_k(x_j) \right]
\eea 
which holds for arbitrary averaging $\langle \ldots \rangle$ for which the variables $n_k$ are independent. 
It can be proved using  the Andreief formula twice, as done above. 
Here we have further used the fact that the variables $n_k$  at fixed $\lambda$ 
are independent Bernoulli random variables. These identities have been used also in the mathematical literature \cite{hough} in the context of determinantal processes. 

The next step is to evaluate the remaining integral over $\lambda$ using the saddle point method, which becomes exact in the limit of large $N$. Let us first study the denominator in Eq. (\ref{newR}), i.e. the partition sum. 
There the saddle point occurs, in our notation, at $\lambda = \lambda_{\rm sp} =-i\beta \tilde \mu$ 
where $\tilde \mu$ is the chemical potential. It is related to the total number of particles 
$N$ as $N= -\partial J/\partial \tilde \mu$,  which reads
\begin{eqnarray}\label{chemical}
N = \sum_{k = 0}^\infty \frac{1}{e^{\beta(\epsilon_k-\tilde \mu)}+1} \;.
\end{eqnarray}
Hence the chemical potential $\tilde \mu=\tilde \mu(T,N)$ depends on $T$ and $N$. 
In the zero temperature limit, as evident from the above equation,
$\tilde \mu(T=0,N)=\mu$, where $\mu$ is the Fermi level introduced 
in Section \ref{sec:zeroT}.

In the thermodynamic language, this amounts to use the equivalence, in the large $N$ limit, between the canonical ensemble and the grand-canonical ensemble. As is well known,
the values of the average occupation numbers $n_k$ at the saddle point are 
given by the Fermi factor and denoted as
\begin{eqnarray}\label{fermi_factor}
\langle n_k \rangle := \langle n_k \rangle_{\lambda_{\rm sp}} \quad , \quad \langle n_k \rangle = \frac{1}{e^{\beta(\epsilon_k-\tilde \mu)} + 1}   \quad , \quad N = \sum_{k = 0}^\infty \langle n_k \rangle 
\end{eqnarray}
where we recall that for the harmonic oscillator
$\epsilon_k = \hbar \omega(k + \frac{1}{2})$ for $k = 0,1,2,\ldots$. 

The same saddle point analysis can be performed to calculate the
correlation in Eq. (\ref{newR}). It remains valid as long
as the quantity that is averaged does not grow 
too fast with $N$ (e.g., as $\exp(cN^a)$ where $a<1$).
In this case, the value of the chemical potential at the saddle point is not
modified. Here this quantity is 
$\det_{1\leq i,j \leq n} K(x_i,x_j; \{ \langle n_k \rangle_\lambda \})$
and it is reasonable to expect that these conditions are satisfied. 
Therefore, in the large $N$ limit, one obtains from~Eq. (\ref{newR}) our main result
\begin{eqnarray}\label{det_process}
R_n(x_1, \cdots, x_n) \simeq \det_{1 \leq i,j \leq n} K_{\tilde \mu}(x_i,x_j) \;,
\end{eqnarray} 
where the finite temperature kernel is given by 
\bea \label{kernel_final}
K_{\tilde \mu}(x,x') = \sum_{k=0}^\infty \frac{\phi^*_k(x) \phi_k(x')}{e^{\beta(\epsilon_k - {\tilde \mu})} + 1} \;,
\eea
and the chemical potential ${\tilde \mu}$ is fixed by Eq. (\ref{chemical}). 
The case $n=1$ then yields the result for the density 
\begin{eqnarray}\label{rho_N_grand_canonical}
\rho_N(x) \simeq \frac{1}{N} \sum_{k = 0}^\infty \langle n_k \rangle |\phi_k(x)|^2 \;
= \sum_{k=0}^\infty \frac{|\phi_k(x)|^2}{e^{\beta(\epsilon_k - {\tilde \mu})} + 1} \;.\label{dens1}
\end{eqnarray}
Note that we reserve the notations $\mu$ and $K_\mu(x,x')$ for the zero temperature chemical potential and
kernel, respectively, while we denote by $\tilde \mu$ and $K_{\tilde \mu}(x,x')$ their finite temperature versions. When $T \to 0$, we
recall that $\tilde \mu \to \mu$, but at finite $T>0$, not only the chemical potential $\tilde \mu$ differs from $\mu$, but also the full kernel functions are different. Note also that in the limit
$T \to 0$ the Fermi factor becomes $\theta(\mu - \epsilon_k)$ and the 
kernel becomes equal to the one associated
with the ground state, given in Eq. (\ref{eq:def_kernel}) and
the same result holds for the density. 

Hence we find that the correlations at fixed $n$ are asymptotically 
determinantal at large $N$ in the canonical ensemble. Alternatively it is also possible
to define the problem of fermions in an external potential directly in the
grand canonical ensemble. Indeed, the quantities
$R_n(x_1, \cdots, x_n) dx_1\cdots dx_n$ are the probabilities that there is a 
particle in each of the intervals $[x_i,x_i+dx_i]$, $1\leq i\leq n$
(referred to as the correlation density in the mathematics literature).
Clearly such quantities exist and make sense for ensembles where the 
particle number varies. In fact in the grand-canonical ensemble, Eq. (\ref{det_process}) is an exact 
equality. Hence the kernel $K_{\tilde \mu}(x_i,x_j)$ {\it exactly} describes the statistics of a system in the grand canonical 
ensemble at the chemical potential ${\tilde \mu}$ corresponding to $N$ for all values of ${\tilde \mu}$, and not only those 
corresponding to $N$ large. In the physics literature, this determinantal property of the grand canonical ensemble of free fermions 
is usually derived using Wick's theorem \cite{Gaudin,Mahan} and has been known for a long time (see also Ref.~\cite{borodin_determinantal}). Interestingly, in the mathematics literature, this property has been studied rigorously only rather recently, using Cauchy-Binet-Andreief identity in the context of general determinantal point processes \cite{Joh07}. Of course both approaches provide identical results. In this paper, we have preferred  the approach using Cauchy-Binet-Andreief identity to show
that the determinantal property also holds in the canonical ensemble with fixed fermion number $N \gg 1$. We have shown that this is true
provided the saddle-point solution exists. To prove the existence of such a saddle point rigorously is a challenging mathematical problem.   

We end this subsection by commenting on the case where the single particle spectrum is degenerate. In this case,
the many-body ground-state may be degenerate -- an example being the harmonic oscillator in $2d$. Each of these
degenerate many-body ground-states can be written as a Slater determinant and each of them constitutes a separate determinantal point process which can be studied along the lines of section \ref{section:1d_RMT}. However, the zero temperature limit of the finite $T$ measure in Eq.~ (\ref{jpdf}) corresponds to taking a zero temperature density matrix where each of these degenerate many-body 
ground-states appears with equal probability. The resulting mixed state is not determinantal, as in the finite $T$ case.  
However, using the equivalence between the canonical and the grand-canonical ensemble in 
the large $N$ limit, this case can also be treated using Eq. (\ref{kernel_final}) with $N$ given by (\ref{chemical}), in the limit $\beta \to \infty$. Note that, everywhere in the formula given above, the sum over $k$ has to be understood as a sum over all possible single-particle states (including their degeneracies).

We now analyze these formula (\ref{chemical})-(\ref{rho_N_grand_canonical}) first in the bulk 
and then at the edge of the Fermi gas. 


\section{Harmonic oscillator in one dimension at finite temperature $T>0$} \label{sec:1dTphys} 

Before applying the general formula for the finite temperature kernel and correlations
to the harmonic oscillator case, it is useful to discuss the relevant scales in the problem.
We consider the harmonic oscillator potential $V(x)=\frac{1}{2} m \omega^2 x^2$ in $d=1$.
We have seen that at $T=0$ there is a natural length scale associated to quantum fluctuations in the confining
potential $1/\alpha= \sqrt{ \hbar/m \omega}$. At $T=0$ there are two length scales 
$\ell(x)=\pi/ N \rho_N(x)$ and $w_N=N^{-1/6}/(\alpha \sqrt{2})$ 
denoting respectively the inter-particle distance in the bulk and the edge (see Fig. \ref{fig:wigner1}). A finite temperature introduces a length scale characterizing the width of
a wave packet associated with a quantum particle, the thermal de Broglie wavelength 
$\lambda_T=\hbar \sqrt{2 \pi/(m T)}$ obtained by the equating kinetic energy and
$T$. Therefore the thermal effects dominate over the quantum effects only if $\lambda_T$ is smaller than
the typical inter-particle distance, in which case the system behaves classically. In the bulk
of the Fermi gas, comparing $\lambda_T$ and $l$ the quantum to classical crossover
occurs at a temperature scale
\bea
T \sim N \hbar \omega \label{cross1} \;.
\eea 
Similarly at the edge, comparing $\lambda_T$ and $w_N$ we find that the 
corresponding crossover occurs at a much lower temperature
\bea
T \sim N^{1/3} \hbar \omega \label{cross2} \;.
\eea 
We will thus focus on these two regimes (\ref{cross1}) and (\ref{cross2}) in the following. 

In addition, we know that the average density follows the Wigner semi-circle
law at zero temperature, a clear signature of the quantum effects. In the other limit 
of large temperature $T \gg N \hbar \omega$ the system behaves classically
and we expect the standard Gibbs-Boltzmann distribution of independent 
particles
\bea \label{boltzmann} 
\rho_N(x) \xrightarrow {T \gg N \hbar \omega} \sqrt{\frac{\beta m \omega^2}{2 \pi}}e^{- \frac{\beta}{2} m \omega^2 x^2} 
\eea 
Our goal below is to study the quantum to classical crossover in the density as well as in the kernel.

\subsection{High temperature scaling and results in the bulk}

As anticipated by Eq. (\ref{cross1}) the bulk scaling regime corresponds to the limit $T \to \infty$, $N \to \infty$, 
while keeping fixed the following dimensionless variable
\bea\label{def_yz}
y = N \hbar \omega/T = \beta \tilde \mu \;.
\eea 
Similarly there is a length scale $\ell_T = \sqrt{2 T/m \omega^2}$ associated 
with the high temperature thermal fluctuations from Eq. (\ref{boltzmann}), hence we can
consider the dimensionless length scale, also kept fixed 
\bea\label{def_yz2}
z= x/\ell_T = x \sqrt{m \omega^2/2 T} \;.
\eea 
In this scaling limit, Eq. (\ref{chemical}) fixing the chemical potential ${\tilde \mu}$, reads
\bea
&& N = \sum_{k=0}^\infty \frac{1}{e^{\beta (\hbar \omega (k+ \frac{1}{2})} e^{- \beta {\tilde \mu}} +1} 
\simeq \int dk \frac{1}{e^{k y/N} e^{- \beta {\tilde \mu}} +1} = \frac{N}{y} \ln(1+ e^{\beta {\tilde \mu}}) \;,
\eea 
where we could replace the sum by an integral since $\beta \hbar \omega\ll 1$.
This yields the relation
\bea\label{mu_bulk}
e^{\beta {\tilde \mu}} = e^y - 1 \;.
\eea
Hence in that scaling regime $\beta \tilde \mu$ is also of order ${\cal O}(1)$.

\subsubsection{Density of fermions in the bulk} 

We now analyze the density $\rho_N(x)$ given in Eqs. (\ref{rho_N_grand_canonical}) and (\ref{fermi_factor}) which we evaluate for $x = z/\sqrt{\beta m \omega^2/2} = \sqrt{2N}/(\alpha \sqrt{y}) \gg 1$. After performing the change of variable 
$k = N p$ in the sum over $k$ in (\ref{rho_N_grand_canonical}), one obtains:
\bea\label{rho_bulk_inter}
&& \rho_N(x) \simeq \frac{1}{N} \sum_{k=0}^\infty \frac{\phi_k(x)^2}{e^{\beta (\hbar \omega (k+ \frac{1}{2})} e^{- \beta {\tilde \mu}} +1}  
\simeq \int_0^\infty dp \frac{\left[\phi_{N p}\left(x={z}\sqrt{2N}/(\alpha\, \sqrt{y}) \right)\right]^2}{e^{y p} (e^y -1)^{-1} +1}
\eea 
where we have replaced ${\tilde \mu}$ by its value given in Eq. (\ref{mu_bulk}) and where $\phi_k(x)$ is given in 
Eq. (\ref{spectrumHO}). We now need an asymptotic expansion of $\phi_k(x)$ for large $k$ (and large $x$). This expansion is provided by the Plancherel-Rotach formula [as given for instance in Eqs. (3.10) (3.11) of Ref. \cite{FFG06}]. For $-1<X<1$, one has
\bea
&& e^{- M X^2} H_M(\sqrt{2 M} X) = \left(\frac{2}{\pi}\right)^{1/4} \frac{2^{M/2}}{(1-X^2)^{1/4}} M^{-1/4} (M!)^{1/2}  g_M(X) \left( 1 + {\cal O}\left(\frac{1}{M}\right) \right) \label{Plancherel_1}\\
&& {\rm with} \;\; g_M(X) = \cos\left( M X \sqrt{1-X^2} + (M + 1/2) \sin^{-1} X - M \pi/2\right)  \;. \label{Plancherel_2}
\eea 
Using these formulas (\ref{Plancherel_1}) and (\ref{Plancherel_2}) for $M=N p$ and $X=z/\sqrt{p y}$ -- see Eq. (\ref{rho_bulk_inter}) -- (taking into account that $X<1$, i.e., $p>z^2/y$) one finds
\bea
\rho_N\left(x = z/\sqrt{\beta m \omega^2/2} \right) &=& \frac{\alpha \sqrt{2}}{\pi} \int_{z^2/y}^\infty dp \frac{1}{e^{y p} (e^y -1)^{-1} +1} 
(N p)^{-1/2} \frac{1}{\sqrt{1- \frac{z^2}{p y}}} \left[g_{N p}(z/\sqrt{p y})\right]^2  \\
&=& \frac{\alpha \sqrt{2}}{\pi \sqrt{y} \sqrt{N}} \int_{z^2}^\infty dq \frac{1}{e^{q} (e^y -1)^{-1} +1} 
 \frac{1}{\sqrt{q - z^2}} \left[g_{N p}(z/\sqrt{q})\right]^2 \;, \label{rho_bulk_inter2}
\eea
where, in the second line, we have simply performed the change of variable $p = q/y$. To obtain the large $N$ limit of Eq. (\ref{rho_bulk_inter2}) we notice that, thanks to the identity $\cos^2 x = 1/2 + \cos{(2 x)}/2$, one can replace $\left[g_{N p}(z/\sqrt{q})\right]^2$, given in Eq.~(\ref{Plancherel_2}), in the integral over $q$ in (\ref{rho_bulk_inter2}) by $1/2$ (the remaining cosine being highly oscillating for large $N$ and thus subleading). If one finally performs the change of variable $q \to q + z^2$ in (\ref{rho_bulk_inter2}), one obtains finally
\bea\label{rho_bulk_Final}
\rho_N\left(x = z/\sqrt{\beta m \omega^2/2} \right) = \frac{\alpha}{\pi \sqrt{y} \sqrt{2 N}}  
\int_0^\infty dq \frac{1}{e^{q+z^2} (e^y -1)^{-1} +1} 
 \frac{1}{\sqrt{q}} = - \frac{\alpha}{\sqrt{2 N \pi y }} {\rm Li}_{1/2}(- (e^y-1) e^{-z^2}) \;,
\eea 
where ${\rm Li}_n(x) = \sum_{k=1}^\infty x^k/k^n$ is the polylogarithm function. Hence from Eq. (\ref{rho_bulk_Final}) we obtain the main result of this section: the fermion density in the bulk takes the
scaling form
\bea \label{densityT}
\rho_N(x) \sim \frac{\alpha}{\sqrt{N}} R\left(y = \beta N \hbar \omega, z= x \sqrt{\beta m \omega^2/2} \right) \;,
\eea
with the bulk scaling function
\bea\label{scaling_function}
R(y,z) = - \frac{1}{\sqrt{2 \pi y}} {\rm Li}_{1/2}(- (e^y-1) e^{-z^2}) \;,
\eea
which is plotted in Fig. \ref{fig_density_bulk}.

\begin{figure}[hh]
\includegraphics[width=0.5\linewidth]{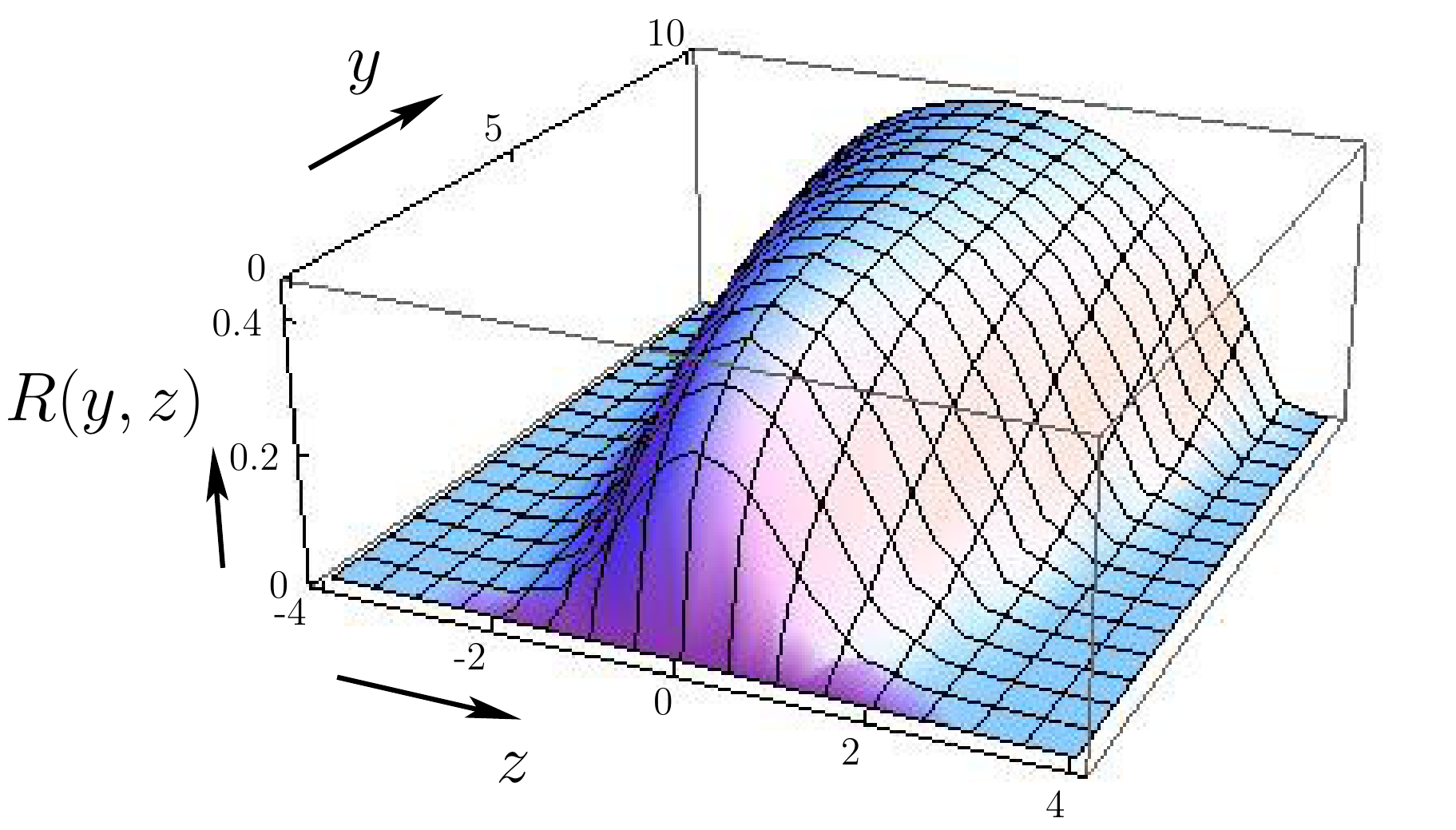}
\caption{(Color online) Plot of the scaling function $R(y,z)$ associated with the density $\rho_N(x)$ (\ref{densityT}), given in Eq.~(\ref{scaling_function}).}\label{fig_density_bulk}
\end{figure}

We now show, from an asymptotic analysis of $R(y,z)$, that Eq. (\ref{densityT}) interpolates between the Wigner semi-circle~(\ref{wigner}) in the limit $T \to 0$
and the classical Gibbs-Boltzmann distribution for $T \to \infty$:
\begin{equation}
\rho_N(x) \xrightarrow {T \to + \infty} \sqrt{\frac{\beta\,m\, \omega^2}{2\pi}}\, 
\exp\left[-\frac{\beta}{2}\, m\,\omega^2\, x^2\right]\, \;,
\label{Tinf}
\end{equation}
which holds also in the scaling limit $\beta\to 0$, $x\to \infty$ but 
keeping $x\,\sqrt{\beta}$ fixed (with the limit $N\to \infty$ already 
taken). Note that the physical mechanism behind this interpolation is very
different from those found earlier in other matrix models~\cite{JP1,JP2}.

To analyze the $T \to \infty$ and the $T \to 0$ limits of $\rho_N(x)$ in Eqs. (\ref{rho_bulk_Final}) and (\ref{scaling_function}), we need the following asymptotic behaviors of the polylogarithm function:
\begin{eqnarray}\label{asympt_small}
{\rm Li}_{1/2}(X) \sim X \:, \; X \to 0 \;,
\end{eqnarray}
and
\bea\label{asympt_large}
{\rm Li}_{1/2}(-e^X) \sim - \frac{2}{\sqrt{\pi}} X^{1/2} \;, \; X \to \infty \;.
\eea
From these behaviors in (\ref{asympt_small}) and (\ref{asympt_large}), one finds the asymptotic behaviors of the scaling function $R(y,z)$ in (\ref{scaling_function}):
\begin{eqnarray}\label{asympt_scaling}
R(y,z) \sim
\begin{cases}
&\sqrt{\dfrac{y}{2 \pi}} e^{-z^2} \;, \; y \to 0 \\
& \\
& \dfrac{\sqrt{2}}{\pi} \sqrt{1 - \dfrac{z^2}{y}} \;, \; y \to \infty \;, \; z \to \infty \;{\rm with} \;\; z^2/y \;\; {\rm fixed} \;.
\end{cases}
\end{eqnarray}
From the first line of Eq. (\ref{asympt_scaling}), one recovers the $T \to \infty$ limit where the density converges to the Gibbs-Boltzmann (Gaussian) form, given in Eq. (\ref{Tinf}). On the other hand, from the second line of Eq. (\ref{asympt_scaling}), one obtains the $T \to 0$ limit of the density, which is given by the Wigner semi-circle law Eq.(\ref{wigner}).

\subsubsection{Kernel and correlations in the bulk}

We analyze the kernel $K_{\tilde \mu}(x,x')$ in the bulk where both $x = u N^{-1/2}/\alpha$ and $x' = u' N^{-1/2}/\alpha$ are close to the center of the trap (and $x-x'$ is of the order of the typical inter-particle distance). Hence we analyze the formulas (\ref{chemical}) and (\ref{kernel_final}) in the limit $N \to \infty$, $\beta \to 0$ keeping $y = \beta N \hbar \omega$ in (\ref{def_yz}) fixed. In this limit the chemical potential ${\tilde \mu}$ is given by Eq. (\ref{mu_bulk}) and the large $N$ analysis of $K_\mu(x,x')$ (\ref{kernel_final}) can be performed along the same lines as done before for the density $\rho_N(x)$ yielding eventually Eq. (\ref{rho_bulk_inter2}). Indeed, using the Plancherel-Rotach asymptotic expansions (\ref{Plancherel_1}) and (\ref{Plancherel_2}) we obtain:
\bea\label{kernel_bulk_inter}
K_{\tilde \mu}(x = u N^{-1/2}/\alpha, x' = u' N^{-1/2}/\alpha) \simeq \frac{\alpha \sqrt{2 N}}{\pi} \int_{0}^\infty \frac{dp}{\sqrt{p}} \frac{1}{e^{y\,p} (e^y -1)^{-1} +1} 
  g_{N p}\left(\frac{u}{N \sqrt{p}}\right) g_{N p}\left(\frac{u'}{N \sqrt{p}}\right) \;.
\eea
From the explicit expression of $g_N(X)$ in Eq. (\ref{Plancherel_2}), one obtains straightforwardly
\bea\label{simplif_gNp}
g_{N p}\left(\frac{u}{N \sqrt{p}}\right) \simeq \cos{\left(2 u \sqrt{p} - N p \frac{\pi}{2} \right)} \;.
\eea
And therefore the product $g_{N p}\left(\frac{u}{N \sqrt{p}}\right) g_{N p}\left(\frac{u'}{N \sqrt{p}}\right)$ in Eq. (\ref{kernel_bulk_inter})
 reads, for large $N$:
\bea\label{prod_g}
g_{N p}\left(\frac{u}{N \sqrt{p}}\right) g_{N p}\left(\frac{u'}{N \sqrt{p}}\right) \simeq \frac{1}{2} \cos{\left(2 \sqrt{p}(u-u') \right)} + \frac{1}{2} \cos{\left(2 \sqrt{p}(u+u') - N p \pi \right)}  \;.
\eea 
The second term in Eq. (\ref{prod_g}) is highly oscillating in the large $N$ limit and hence the leading contribution, once inserted in the integral over $p$ in Eq. (\ref{kernel_bulk_inter}), comes from the first term of Eq. (\ref{prod_g}), which is independent of $N$. Therefore we finally obtain
\bea
K_{\tilde \mu}(x = u N^{-1/2}/\alpha, x' = u' N^{-1/2}/\alpha) \sim \alpha \sqrt{N} \frac{1}{\pi\sqrt{2}} \int_0^\infty \frac{dp}{\sqrt{p}} \frac{\cos(2 \sqrt{p}(u-u'))}{e^{yp}(e^y-1)^{-1}+1} \;.
\eea
Finally, performing the change of variable $p \to p/y$, one obtains the final form of
the finite temperature kernel in the bulk 
\bea
K_{\tilde \mu}(x,x') = \alpha N^{1/2} {\cal K}^{\rm bulk}_{y}\left(\alpha \sqrt{N} (x-x')\right)\; , 
\eea 
where 
\bea\label{Kbulk1d}
{\cal K}^{\rm bulk}_{y}(v) = \frac{1}{\pi \sqrt{2 y}} \int_0^{+\infty} dp \frac{\cos(\sqrt{\frac{2 p}{y}} v)}{(1+ e^p/(e^y-1)) \sqrt{p}} 
\eea 
(see also Refs. \cite{Verba,Joh07} for alternative derivations of this kernel). In the inset of Fig. \ref{fig_kernels} we show a plot of the 2-point correlation function in the bulk $g^{\rm bulk}_y(v) = {\cal K}_y^{\rm bulk}(0)^2 - [{\cal K}_y^{\rm bulk}(v)]^2$ for different scaled temperature parameter $y$. 

\subsection{Low temperature scaling: density and kernel at the edge} \label{shotedge}

\subsubsection{Density at the edge} 

We now focus on the density near the zero temperature edge at $x_{\rm edge}=\frac{\sqrt{2N}}{\alpha}$. 
We recall that at zero temperature the density strictly vanishes at the edge, and the edge region
has a width $w_N =  \frac{1}{\alpha \sqrt{2}} N^{-\frac{1}{6}}$ which corresponds to the 
typical separation between particles near the edge. To analyze how this density profile
gets modified near the edge we set
\bea
x = \frac{\sqrt{2N}}{\alpha} + s \,w_N  \;.
\eea 
We have also seen from Eq. (\ref{cross2}) that in the edge region the
crossover from quantum to classical regime occurs at temperature 
$T \sim N^{1/3} \hbar \omega$. Hence we define the dimensionless
(inverse temperature) parameter $b$ 
\bea
b = \frac{ \hbar \omega}{T} N^{1/3} \label{def_b_1}
\eea 
which will be kept fixed in the large $N$ and large $T$ limit in this edge regime. 
From (\ref{def_yz}) we see
that the variable $y = b N^{2/3} \gg 1$ in this regime. Hence, from
(\ref{mu_bulk}) we can set $\beta \tilde \mu \simeq y = b N^{2/3}$.
We insert this value of $\beta \tilde \mu$ in Eq. (\ref{dens1}), and use $\epsilon_k = \hbar \omega ( k + 1/2)$.
Making further a shift $k-N = m$ (neglecting the $1/2$ factor compared to $N$) we 
obtain the following expression for the density
%
\begin{eqnarray}\label{rho_N_inter}
\rho_N(x) \simeq \frac{1}{N} \sum_{m=-N}^\infty \frac{[\phi_{N+m}(x)]^2}{\exp{(b m/N^{1/3})}+1} \;.
\end{eqnarray} 
Using the Plancherel-Rotach formula for Hermite polynomials at the edge (see for instance Ref. \cite{FFG06}) yields:
\begin{equation}\label{Plancherel_Airy}
\phi_{N+m}\left(\frac{\sqrt{2N}}{\alpha} + \frac{s}{\sqrt{2}\alpha} N^{-\frac{1}{6}} \right) \sim {\sqrt{\alpha}} \frac{2^{\frac{1}{4}}}{N^{\frac{1}{12}}} {\rm Ai}\left(s - \frac{m}{N^{\frac{1}{3}}} \right)  \;,
\end{equation} 
up to terms of order ${\cal O}(N^{-2/3})$. Hence, by 
inserting this asymptotic formula (\ref{Plancherel_Airy}) into Eq. (\ref{rho_N_inter}) and replacing the discrete sum over $m \sim N^{1/3}$ by an integral, we obtain the scaling form of the density near the edge
%
\bea
\rho_N(x) \simeq \frac{1}{N w_N} F_{1,b}\left( \frac{x - x_{\rm edge}}{w_N}\right) 
\eea 
where the finite temperature scaling function $F_{1,b}(s)$ is
obtained as
\bea \label{F1bs} 
F_{1,b}(s) = \int_{-\infty}^{+\infty} du \frac{{\rm Ai}(s+u)^2}{1 + e^{-b u}} \;.
\eea 


In the zero-temperature limit $b \to \infty$, the Fermi factor becomes
a Heaviside step function, and we recover  $F_{1,b=+\infty}(s)=F_1(s)$ 
given in (\ref{edge_density_scaling_function}). 
%
%
%
In Fig. \ref{fig_density_edge}, we show how $F_{1,b}(s)$ behaves for different values of the reduced inverse temperature $b$. Note that the oscillations are more and more attenuated as temperature increases. 
\begin{figure}
\includegraphics[width=0.5\linewidth]{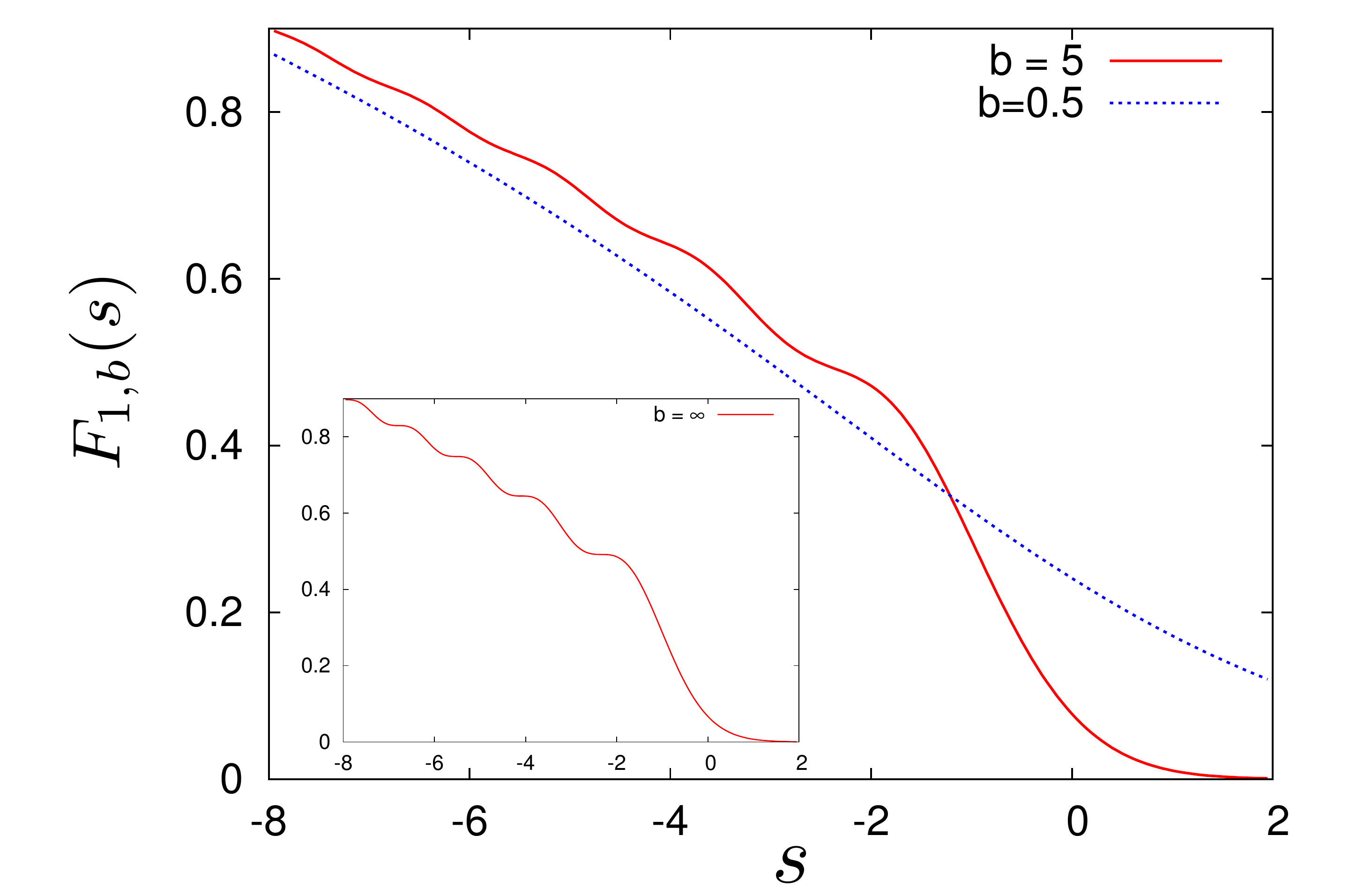}
\caption{(Color online) Plot of the scaling function $F_{1,b}(s)$ for the density, given in Eq. (\ref{F1bs}) corresponding to two different (scaled) temperatures $b=0.5$ (dotted line) and $b=5$ (solid line). {\bf Inset:} plot of $F_{1,b}(s)$ corresponding to $b \to \infty$ (zero temperature) shown here for comparison with the main plot.}\label{fig_density_edge}
\end{figure}

{\it Density at the edge: temperature dependence}. It is interesting to discuss the value of the density exactly at the edge $x=x_{\rm edge}$. One has
\bea
N \rho_N(x_{\rm edge}) \simeq N^{1/6} \alpha \sqrt{2} f_e(T/N^{1/3} \hbar \omega) \quad , \quad 
f_e(t) = t \int_{-\infty}^{+\infty} du \frac{{\rm Ai}(u t)^2}{1 + e^{-u}} \;.
\eea 
In the low temperature scaling regime $T \sim N^{1/3} \hbar \omega$ the two limiting behaviors are
\bea\label{rho_edge_lowT}
\rho_N(x_{\rm edge}) \simeq \begin{cases} 
& \dfrac{\alpha \sqrt{2}}{3^{2/3} \Gamma \left(\frac{1}{3}\right)^2 } N^{-5/6}
   \quad , \quad T \ll N^{1/3} \hbar \omega \\
   \\
& \quad \dfrac{\left(1-\sqrt{2}\right) \zeta \left(\frac{1}{2}\right) \alpha}{\sqrt{2 \pi  }} 
\dfrac{1}{N} \sqrt{\dfrac{T}{\hbar \omega}} \quad , \quad 
T \gg N^{1/3} \hbar \omega \;.
\end{cases} 
\eea
{Note that we display the complete low temperature series, as an expansion
in power of $T^2$, in Appendix \ref{sec:expansion}.}

We recall that in the bulk high temperature regime $T \sim N \hbar \omega$, one has
from (\ref{densityT}) and \eqref{scaling_function}
\bea
\rho_N(x_{\rm edge}) \simeq  
- \frac{\alpha}{\sqrt{2 \pi N y}} {\rm Li}_{1/2}(e^{-y}-1)  \quad , \quad y= N \hbar \omega/T
\eea 
which gives the two limiting behaviors
\bea\label{rho_edge_highT}
\rho_N(x_{\rm edge}) \simeq \begin{cases} 
& \dfrac{\left(1-\sqrt{2}\right) \zeta \left(\frac{1}{2}\right) \alpha}{\sqrt{2 \pi  }} 
\dfrac{1}{N} \sqrt{\dfrac{T}{\hbar \omega}} 
   \quad , \quad T \ll N \hbar \omega \\
   \\
& \quad \sqrt{\dfrac{m\, \omega^2}{2\pi T}} \quad , \quad 
T \gg N \hbar \omega \;,
\end{cases} 
\eea
which shows a perfect matching of the high temperature end of the
low $T$ (edge) scaling regime [second line of Eq.~(\ref{rho_edge_lowT})], with the low temperature end of the 
high $T$ (bulk) regime [first line of Eq. (\ref{rho_edge_highT})]. We recall that in the high $T$ regime, the 
Fermi gas extends well beyond the $T=0$ edge. Note that for
fixed (large) $N$ this density first increases as a function of 
$T$ in the low $T$ regime and then exhibits a maximum
for $T \sim N$ in the high $T$ regime before decreasing again. \\

{\it Right tail of the density.} Let us consider the behavior of the density scaling function $F_{1,b}(s)$ to the right of
the edge, for large positive $s$. The analysis of the integral in Eq.~(\ref{F1bs}) in that 
limit was performed in \cite{LargeDev_KPZ}. It is found that there are two regimes
depending on whether the parameter $\tilde s= s/b^2$ is smaller or larger 
than the critical value $\tilde s_c= 1/4$
\bea\label{F1_right_tail}
F_{1,b}(s) \simeq \begin{cases}
& \dfrac{1}{4 b^2 \sqrt{\tilde s} \sin(2 \pi \sqrt{\tilde s}) } \exp \left(- \frac{4}{3} s^{3/2} \right)  
\quad , \quad 1 \ll s <  \frac{b^2}{4} \\
\\
& \dfrac{1}{\sqrt{4 \pi b} } \exp \left(- b s + \frac{b^3}{12} \right) \quad , \quad s > \frac{b^2}{4} \;.
\end{cases} 
\eea 
Hence we obtain a transition between a stretched exponential tail in the density, as in the
zero temperature case, to a pure exponential decay in the far tail for $s > b^2/4$. Thus, 
for a fixed value of the reduced temperature $b$ (not necessarily large), 
the decay is always exponential. Note that the pre-exponential factor
exhibits a crossover from the two limiting cases indicated above, in the
vicinity of $\tilde s=1/4$ \cite{LargeDev_KPZ}.

{\it Left tail of the density.} For $s$ large and negative, the integral over $u$ in Eq. (\ref{F1bs}) is dominated by the region $u \in (-\infty, |s|]$. Within
this interval one can thus replace, for large negative $s$, the Airy function by its asymptotic form for large negative argument, ${{\rm Ai}}(z) \simeq \frac{1}{\sqrt{\pi}} |z|^{-1/4} \sin{(\frac{2}{3} |z|^{3/2} + \pi/4)}$, for $z \to -\infty$. By substituting the Airy function by its asymptotic behavior in the integral over $u$ in (\ref{F1bs}), one finds straightforwardly that $F_{1,b}(s)$ behaves asymptotically as
\begin{eqnarray}
F_{1,b}(s) \simeq \frac{1}{\pi} \sqrt{|s|}\;, \; {\rm for} \; s \to - \infty \;,
\end{eqnarray}
which matches, as it should, with the Wigner semi-circle expanded close to $x_{\rm edge} = \sqrt{2N}/\alpha$ (\ref{wigner}).

\begin{figure}[h]
\includegraphics[width = 0.6\linewidth]{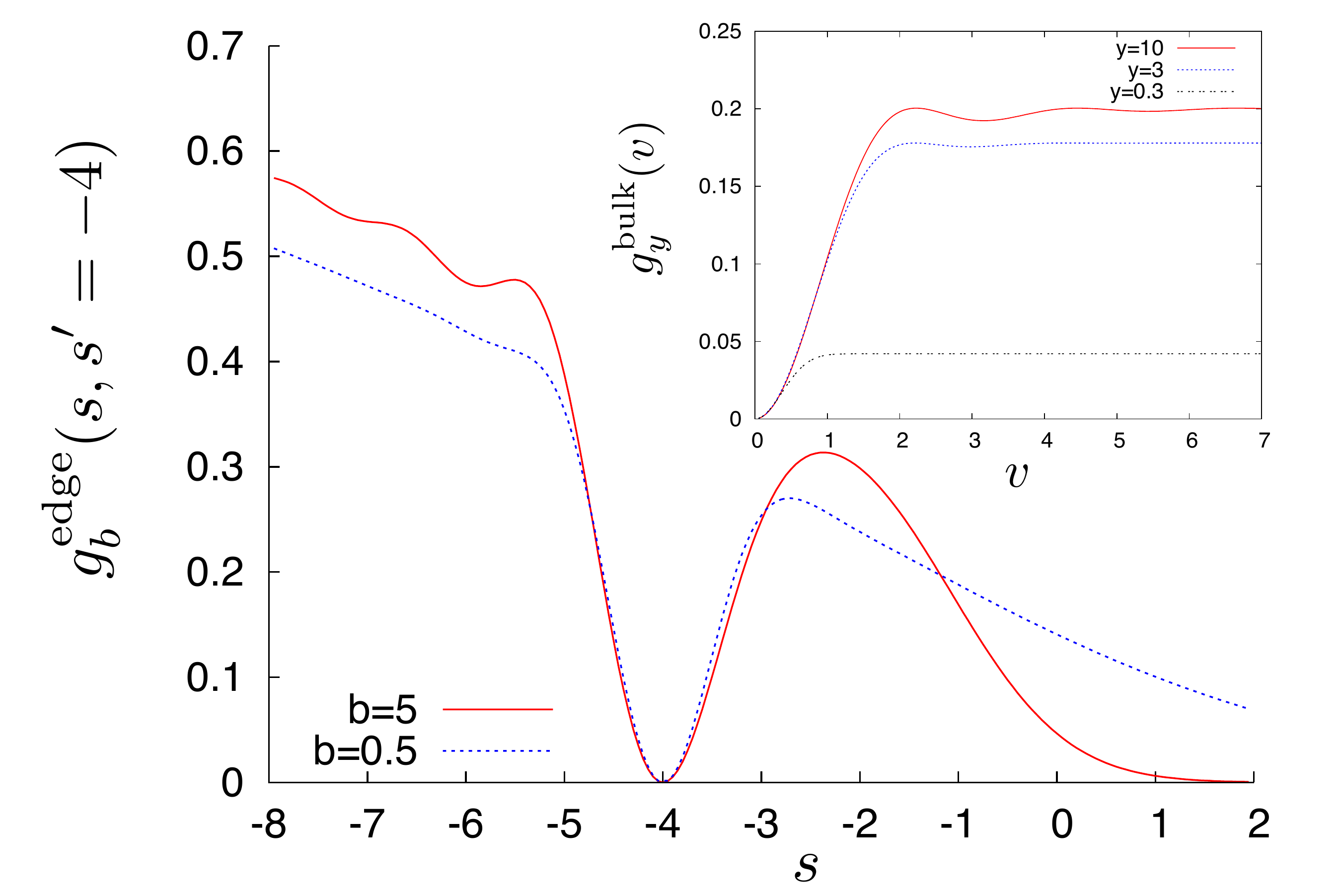}
\caption{(Color online) Plot of the 2-point correlation function at the edge $g^{\rm edge}_b(s,s'=-4)$ as a function of $s$ and for  scaled inverse temperatures $b=0.5$ and $5$. {\bf Inset:} Plot of the 2-point correlation function in the bulk $g^{\rm bulk}_y(v)$ versus $v$ for scaled inverse temperatures $y = 0.3, 3$ and $10$.} \label{fig_kernels}
\end{figure}

\subsubsection{Kernel at the edge} 

We now consider the finite temperature kernel with both $x$ and $x'$ close to the edge $x_{\rm edge}=\sqrt{2 N}/\alpha$.
We set $x = x_{\rm edge} + w_N s$ and $x' = x_{\rm edge} + w_N s'$. We insert these coordinates into
Eq. (\ref{kernel_final}) and follow the same analysis as in the case of the density (see above). This finally
gives in the scaling limit
\bea \label{kff0}
K_\mu(x,x') \simeq \frac{1}{w_N} {\cal K}^{\rm edge}_b \left( \frac{x-x_{\rm edge}}{w_N} , \frac{x'-x_{\rm edge}}{w_N} \right)
\eea 
where the scaled finite temperature edge kernel is given by 
\begin{eqnarray}\label{kff}
{\cal K}^{\rm edge}_{b}(s,s') = \int_{-\infty}^\infty \frac{{\rm Ai}(s+u){\rm Ai}(s'+u)}{e^{-b\, u} +1} du \;.
\end{eqnarray}
For $s=s'$ it reduces to the expression (\ref{F1bs}) for the density. Note that in the limit of zero temperature, when $b \to \infty$, the non-zero contribution to the integral over $u$ on the right hand side of Eq. (\ref{kff}) comes from $u \in [0, +\infty)$ and one gets, using Eq. (\ref{airy_kernel.1})
\begin{eqnarray}\label{limit_airy}
\lim_{b \to \infty} {\cal K}^{\rm edge}_{b}(s,s') = \int_0^\infty \, {\rm Ai}(s+u){\rm Ai}(s'+u) \, du = K_{\rm Airy}(s,s') \;.
\eea
The kernel in Eq. (\ref{kff}) is thus the finite temperature
generalization of the Airy kernel \eqref{airy_kernel.1}. Finally, in Fig. \ref{fig_kernels} we show a plot of the 2-point correlation function at the edge $g^{\rm edge}_b(s,s') = {\cal K}_b^{\rm edge}(s,s) {\cal K}_b^{\rm edge}(s',s') - [{\cal K}_b^{\rm edge}(s,s')]^2$ for different scaled inverse temperatures $b$. 

\subsection{Extremal statistics near the edge at finite temperature} \label{statedge} 

\subsubsection{Statistics of the rightmost fermion: exact distribution and its tails} 

We are now in position to study the fluctuations of the 
position $x_{\max}(T)$ of the rightmost fermion at finite temperature $T$. In principle it can be derived from the joint PDF of 
the fermion positions in Eq.~(\ref{jpdf}) as 
\bea\label{eq:wall}
\Pr(x_{\max}(T) \leq w) = \int_{-\infty}^w dx_1 \, \ldots  \int_{-\infty}^w dx_N \, P_{\rm joint}(x_1, \cdots, x_N) \;.
\eea
At $T=0$, this distribution has a limiting scaling form given by the Tracy-Widom distribution [see Eqs. (\ref{xmax_0T}), (\ref{fredholm_F2})], as discussed in section \ref{sec:TW_T=0}. As we have shown in section \ref{sec:equivalence}, in the limit of large $N$, the positions of the fermions at finite $T$ form a determinantal point process, with the kernel parametrized by temperature as given in Eq.~\eqref{kff0}. As a result (following the discussion in section \ref{sec:TW_T=0}), the multiple integral in Eq. (\ref{eq:wall}) can be written as a Fredholm determinant \cite{johansson}. Indeed, at finite temperature $T \sim {\cal O}(N^{1/3})$ (with $b = \hbar \omega N^{1/3}/T$ fixed) and in the $N \to \infty$ limit, the 
scaled cumulative distribution function (CDF), denoted by $Q_b(s)$, of $x_{\max}(T)$ can be expressed as the following Fredholm determinant \cite{fredholm}
\begin{equation}\label{gap_proba}
\Pr\left(x_{\max}(T) \leq \frac{\sqrt{2N}}{\alpha} + \frac{N^{-\frac{1}{6}}}{\alpha \sqrt{2}}s\right) 
\underset{N \to \infty}{\to} Q_b(s) := {\rm Det}(I - P_s {\cal K}^{\rm edge}_{b} P_s) \;,
\end{equation}
where $P_s$ is the projector on the interval $[s,+\infty)$, the kernel ${\cal K}^{\rm edge}_{b}$ is given in \eqref{kff} and $\alpha = \sqrt{m \omega/\hbar}$.

The first property to note is that in the limit $T \to 0$, i.e. $b \to +\infty$, since
${\cal K}^{\rm edge}_{b} \to K_{\rm Airy}$ this distribution (\ref{gap_proba}) converges to 
the TW distribution for GUE given in  \eqref{fredholm_F2}. Hence $Q_b(s)$ is a generalization of the TW distribution to finite temperature. The calculation of the Fredholm determinant (FD) in \eqref{gap_proba} is quite involved. 
As we have pointed out in \cite{us_prl} the
same FD occurs in the exact solution of the KPZ equation with droplet
initial conditions. This correspondence is recalled in section \ref{sec:KPZ} 
and here we will borrow some of the results obtained in that context.
This FD can be expressed in terms of the solution of a non-local generalization of the
Painlev\'e II equation \cite{ACQ11}, namely one has
\bea
\partial_s^2 \ln Q_b(s) = - \int_{-\infty}^{+\infty} dv\, \sigma_b'(v) [q_b(s,v)]^2 \;,
\eea 
where
\bea
\sigma_b(v) = \frac{1}{1+ e^{- b v}} \;,
\eea 
and $\sigma'_b(v) = \partial_v \sigma_b(v)$. The function $q_b(s,v)$ satisfies 
a non-linear integro-differential equation in the variable $s$ 
\bea\label{PII_nonlocal}
\partial_s^2 q_b(s,v) = \left(s+v+2 \int_{-\infty}^{+\infty} dw \, \sigma_b'(w) [q_b(s,w)]^2\right)
q_b(s,v) \;,
\eea 
with the boundary condition $q_b(s,v)\simeq_{s \to +\infty} {\rm Ai}(s+v)$. 
Note that in the zero temperature limit $b \to \infty$, $\sigma'_b(v)
\to \delta(v)$, hence one recovers that $q_b(s,0)$ satisfies the standard Painlev\'e II equation (\ref{PII})
which is related to the TW distribution for GUE. The analysis of this equation (\ref{PII_nonlocal}) is rather non trivial. Alternatively, a numerical evaluation
of the FD is possible, along the lines of Ref.~\cite{SpohnProlhac} using the method developed by Borneman~\cite{borneman}.
This is left for future studies. 

{\it Tails of $Q_b(s)$.} While we have a formal expression for the full scaled distribution function $Q_b(s)$ in terms of a FD in Eq. (\ref{gap_proba}), can we determine its tails for large $|s|$ explicitly for fixed $b$? Indeed this is possible for arbitrary $b$ for the right tail $s \to + \infty$. However, for the left tail, we can only provide results in the scaling limit $s \to -\infty$, $b \to \infty$, but keeping the ratio $s/b^2$ fixed.    

We start with the right tail. In the limit of large positive $s$ the FD in (\ref{gap_proba}) can be approximated
by the first term in a trace expansion. Indeed in this limit, since $K^{\rm edge}_b(s,s')$ is ``small'', one can expand the FD in Eq. (\ref{gap_proba}) in the following way:
\begin{eqnarray}\label{FD_expansion}
\ln Q_b(s) := \ln {\rm Det}(I - P_s {\cal K}^{\rm edge}_{b} P_s) = -\sum_{p=1}^\infty \frac{1}{p} {\rm Tr} \, \left[P_s\,K^{\rm edge}_b\,P_s\right]^p \;.
\eea
Keeping only the leading $p=1$ term gives, for large $s$
\bea
\ln Q_b(s) \approx - {\rm Tr} \, \left[P_s\,K^{\rm edge}_b\,P_s\right] = - \int_s^\infty K^{\rm edge}_b(s',s') \, ds' \;.
\eea
Taking derivative w.r.t. $s$, and using $\lim_{s \to \infty} Q_b(s) = 1$, yields
\bea
Q_b'(s) \simeq K^{\rm edge}_b(s,s) = F_{1,b}(s) = \int_{-\infty}^{+\infty} du \frac{{\rm Ai}(s+u)^2}{1 + e^{-b u}} \;,
\eea 
where $F_{1,b}(s)$ is the density near the edge in Eq. (\ref{F1bs}). Therefore the right tail of $Q_b(s)$ coincides to leading order with the edge density. 
It turns out that, just the leading term already provides a numerically accurate estimation of the right tail of $Q_b(s)$. Indeed, this is also the case at $T=0$, where the edge density provides a numerically accurate approximation of the right tail of the TW distribution. From the analysis of $F_{1,b}(s)$ in Eq. (\ref{F1_right_tail}), we see that for fixed $b$, as $s \to \infty$, 
\begin{eqnarray}\label{Qb_right}
Q_b(s) \sim \dfrac{1}{\sqrt{4 \pi b} } \exp \left(- b s + \frac{b^3}{12} \right) \;.
\end{eqnarray}
However, if one scales $b$ and $s$ such that $\tilde s = s/b^2$ is fixed, then the right tail of $Q_b(s)$ undergoes the same crossover as $F_{1,b}(s)$ at $\tilde s_c=1/4$ [as in Eq. (\ref{F1_right_tail})].

We now turn to the left tail of $Q_b(s)$ as $s \to - \infty$. Here, analyzing the FD for fixed $b$ with $s \to - \infty$ turns out to be difficult. 
However, one can make progress in the scaling limit when $b \to \infty$, $s \to -\infty$, keeping the ratio $\tilde s = s/b^2$ fixed. In fact, in the context of the height distribution of the KPZ equation at late times (and will be discussed later in section \ref{sec:KPZ}), this
scaling limit was already investigated in \cite{LargeDev_KPZ} by analyzing the solution of the non-local Painlev\'e equation (\ref{PII_nonlocal}).
There it was argued that in this scaling limit the CDF behaves as
\bea
Q_b(s) \sim e^{- b^6 \Phi_-(s/b^2) }    \quad , \;{\rm where} \;\quad   \Phi_-(z)= \frac{1}{12} |z|^3 \;.
\eea 

\subsubsection{Statistics of the rightmost fermion: {{finite}} 
temperature behavior 
of the distribution of the position of the rightmost fermion}  \label{sec:crossover} 


In the previous subsection we discussed the limiting distribution of the position 
of the rightmost fermion
in the limit where {{$T \sim {\cal O}(N^{1/3})$}} and $N$ is large. In 
that 
analysis, we kept the scaling parameter $b = \hbar \omega N^{1/3}/T$ fixed
and investigated the CDF $Q_b(s)$ as a function of $s$ for fixed $b$. We were able to obtain explicit results for the tails of $Q_b(s)$ for $b$ large, i.e., $T \ll {\cal O}(N^{1/3})$. Thus, in some sense, the system was still in the vicinity of the $T=0$ limit. In this subsection, we consider the opposite high temperature limit  where $T \gg {\cal O}(N^{1/3})$, i.e., the $b \to 0$ limit.

To proceed, we use some recent results from the connection between the KPZ equation in droplet geometry
and the fermion problem \cite{us_prl} (for details, see section \ref{sec:KPZ}). The height distribution in the KPZ problem was recently analyzed exactly
in the short time limit \cite{Short_time_PRL}, which corresponds to high temperature in the fermion problem. This allows  one to obtain the high temperature expansion,
in powers of the small parameter $b$, of the cumulants of the position
of the rightmost fermion $x_{\max}(T)$, and to provide approximate 
interpolation formula which should be useful for comparison with cold atom experiments.

{\it Cumulants}. We start by discussing the mean position and the variance, as obtained
from the analysis of Section~\ref{sec:KPZ}. Let us define
the rescaled variable
\bea\label{def_xi}
\xi = \frac{x_{\max}(T) - x_{\rm edge}}{w_N} 
\eea 
where $x_{\rm edge}= \sqrt{2 N}/\alpha$ is the edge at $T=0$. 
Note that this definition differs from the one of the variable
$\xi$ defined in~\cite{Short_time_PRL}. The first 
few terms in the series expansion of the mean position read
\bea
&& \langle \xi \rangle = - \frac{1}{2 b} \ln (4 \pi b^3) +  \frac{\gamma_E}{b}
- \sqrt{\frac{\pi}{2}} \frac{b^{1/2}}{2}  
+ \left( \frac{32 \pi}{9 \sqrt{3}} -2 - \frac{3 \pi}{2}\right) \frac{b^2}{4} + {\cal O}(b^{7/2}) 
\eea
and the variance behaves as
\bea  \label{var} 
&& \langle \xi^2 \rangle^c  = \langle \xi^2 \rangle - \langle \xi \rangle^2 = \frac{\pi^2}{6 b^2} + 
\sqrt{2 \pi} \frac{1}{2 b^{1/2}}   + \left( 4 + 5 \pi - \frac{32 \pi}{3 \sqrt{3}}\right) \frac{b}{4}  + {\cal O}(b^{5/2}) \;.
\eea

{These formula give the behavior for small $b$, and we know that they should
crossover at large $b$ to the zero temperature limit given by the cumulants of
the GUE Tracy-Widom distribution. In particular, the mean $\langle \xi \rangle$ converges to $m_{\rm TW}=-1.771086807411\ldots$, 
and the variance $\langle \xi^2 \rangle^c $ converges to 
$\sigma^2_{\rm TW}  = 0.8131947928329 \ldots$. In fact we know a bit more: as shown
in the Appendix \ref{sec:expansion} the large $b$ expansion takes the form
\bea
&& \langle \xi \rangle = m_{\rm TW} + {\cal O}(b^{-4})  \\
&& \langle \xi^2 \rangle^c = \sigma^2_{\rm TW}  + \frac{\pi^2}{3 b^2} + {\cal O}(b^{-4}) \;.
\eea 
Note that the form of the correction to the variance as $a_2/b^2$ is common to the
KPZ universality class \cite{Ferrari_Frings}, although the prefactor $a_2$ may be
non-universal. For the continuum KPZ equation its value is fixed and is computed in Appendix~\ref{sec:expansion}.

Since the variance is the easiest cumulant to measure in a cold atom experiment we 
give here an approximation to the crossover
from low to high values of $b$, by constructing a Pad\'e approximant $\langle \xi^2 \rangle^c|_{\rm Pad{e}} $
which (i) reproduces
all known terms in the small $b$ expansion given in (\ref{var}) (ii) converges at large 
$b$ to $\sigma^2_{\rm TW} $ as $\langle \xi^2 \rangle^c|_{\rm Pad{e}} \simeq \sigma^2_{\rm TW} + \frac{a_2}{b^2}$. It reads, with the value $a_2=\pi^2/3$,}
\bea
 \langle \xi^2 \rangle^c|_{\rm Pade} =
\frac{1.64493 +1.13494 \, b^{3/2} + a_2 \, b^4+0.813195 \, b^6}{b^2 \left(1 -0.0719632
   \, b^{3/2}+b^4\right)} \label{pade} 
   \eea
and is plotted in Fig. \ref{FigPade}. 
\begin{figure}
\includegraphics[width=0.5\linewidth]{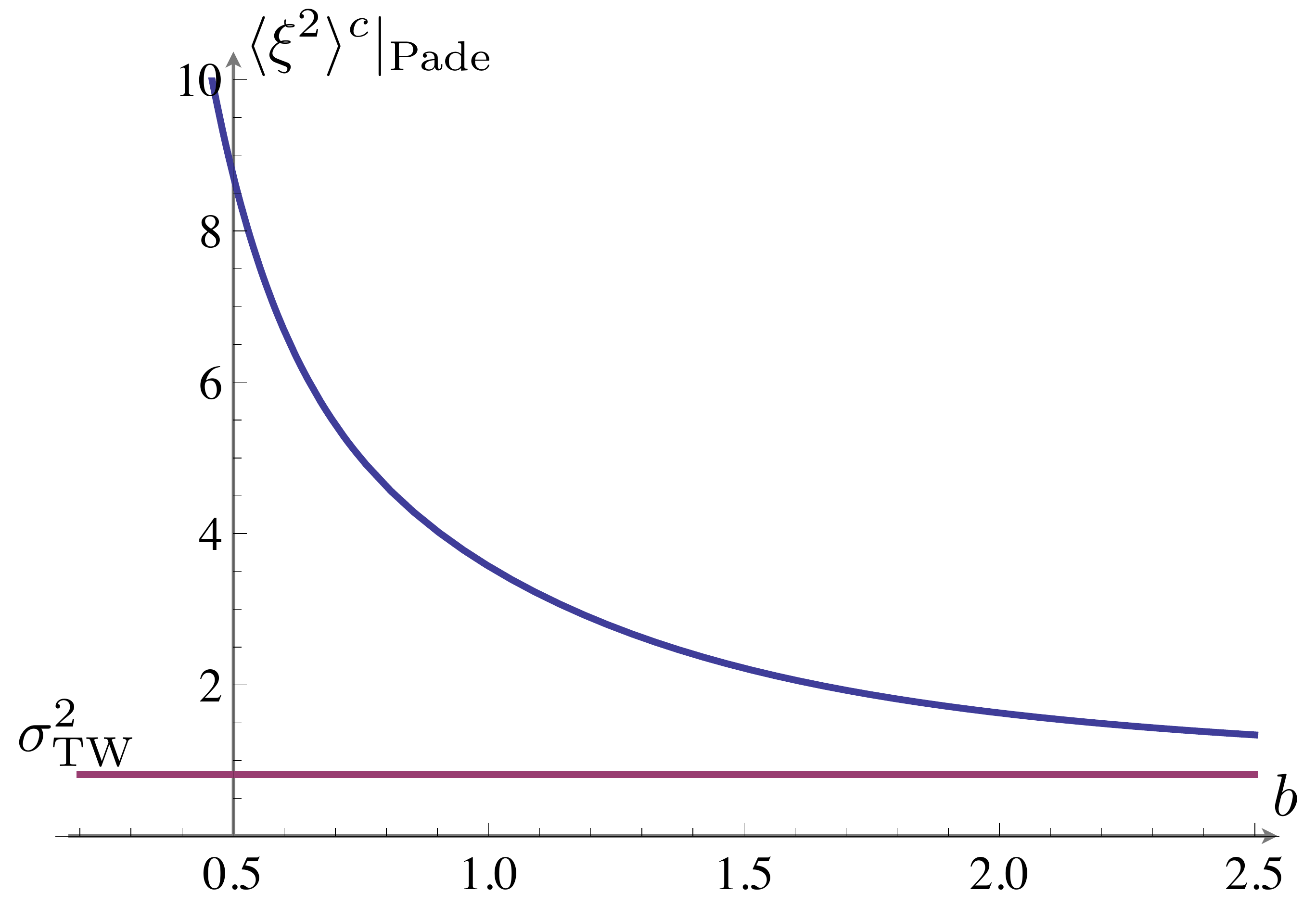}
\caption{Plot of the Pad\'e approximate $\langle \xi^2 \rangle^c|_{\rm Pade}$ given in Eq. (\ref{pade}) with $a_2=\pi^2/3$. The horizontal line is corresponds to the variance of the TW distribution $\sigma^2_{\rm TW}$, which is the exact result for $b \to \infty$, i.e., in the $T \to 0$ limit.}\label{FigPade}
\end{figure}
Although there is some degree of arbitrariness, this curve should be
useful to calibrate the experiments, given that the range of values of $b$ presently available
is $b \sim 0.5-2$. Another way to present the result for the variance of the
fluctuations of the rightmost fermion, is to divide it by the (half) size of the 
fermi cloud (at $T=0$) $x_{\rm edge}$, and write
\bea
\Bigg \langle \left(\frac{x_{\max}(T) - x_{\rm edge}}{x_{\rm edge}} \right)^2 \Bigg \rangle^c= \frac{T^2}{T_F^2}  ~ {\cal V} \left( b = \frac{T_F}{N^{2/3} T} \right)
\eea 
where $T_F= N \hbar \omega$, and the scaling function ${\cal V}(z)$ is such that ${\cal V}(0)=\frac{\pi^2}{24}$ [see Eq. (\ref{var})] and ${\cal V}(z) \simeq_{z \to +\infty}  \sigma^2_{\rm TW} z^2/4$ to yield back the zero temperature limit (i.e., as $b \to \infty$). A Pad\'e approximation of the function ${\cal V}(z)$ is easily obtained from (\ref{pade}). 

{Finally, we also give the third cumulant of the scaled variable $\xi$ in Eq.~(\ref{def_xi}). For small
$b$ it reads}
\bea
&& \langle \xi^3 \rangle^c  = \frac{2 \zeta(3)}{b^3} +  \left( \frac{32}{3 \sqrt{3}} -6 \right) \frac{\pi}{4}  + {\cal O}(b^{3/2}) 
\eea
which allows to calculate the skewness of the position of the right-most fermion (note that the skewness
is independent of any rescaling)
\bea
&& {\rm Sk} := \frac{  \langle \xi^3 \rangle^c }{  [ \langle \xi^2 \rangle^c ]^{3/2}} 
= 1.13955-1.30237 \, b^{3/2}+1.20563 \, b^3+{\cal O}(b^{7/2}) \;.
\eea 
It decreases from the skewness of the Gumbel distribution (see below)
${\rm Sk}_{\rm Gumbel}=1.13955$ at high temperature (small $b$) to the skewness
of the TW distribution for GUE ${\rm Sk}_{\rm TW}=0.224084203610 \ldots$ at
zero temperature (large $b$). {Note however that since $\langle \xi^3 \rangle^c = \kappa_3 + {\cal O}(b^{-4})$ (see Appendix 
\ref{sec:expansion}),  
we find that the approach to the zero-temperature limit is from below
\bea
{\rm Sk}  = {\rm Sk}_{\rm TW} \left( 1 - \frac{\pi^2}{2  \sigma^2_{\rm TW} b^2} + {\cal O}(b^{-4})\right) \;.
\eea 
Hence the skewness is (weakly) non-monotonic as a function of the temperature. 
}

\bigskip

\begin{figure}
\includegraphics[width=0.7\linewidth]{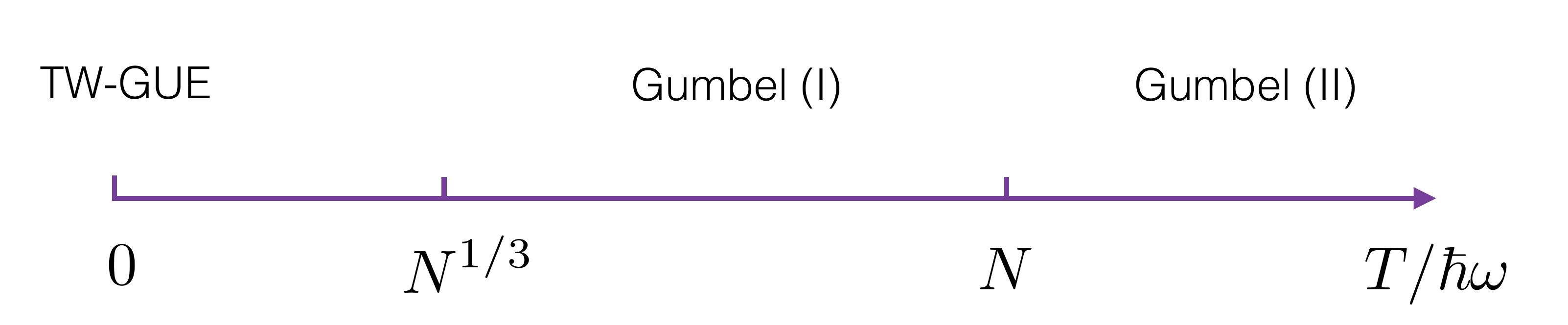}
\caption{Sketch of the behavior of the distribution of the rightmost fermion as a function of $T/\hbar \omega$. When $T/\hbar \omega \ll N^{1/3}$, the distribution is given by the TW distribution for GUE [see Eq. (\ref{summary})] while for $T/\hbar \omega \gg N^{1/3}$ -- and $T/\hbar \omega \ll N$ -- it crosses over to a Gumbel distribution (\ref{summary}), denoted as Gumbel (I) in the figure. The full crossover, when $T/\hbar\omega \sim {\cal O}(N^{1/3})$ is described the finite temperature generalization of the TW distribution (\ref{gap_proba}), which also appears in the KPZ equation in a droplet geometry (see also Eq. (\ref{pade}) for the second cumulant). Finally, for $T/\hbar \omega \gg N$, the distribution is described by yet another Gumbel distribution (denoted as Gumbel (II) in the figure), as discussed in the text [see Eq. (\ref{gumbel1})].}\label{Fig_rightmost}
\end{figure}

{\it High temperature limit and the Gumbel distribution.} In the high temperature limit of the edge regime, i.e. $T \gg N^{1/3} \hbar \omega$ (equivalently $b \ll 1$), the full PDF of $\xi$ becomes a Gumbel
distribution, up to a temperature dependent constant shift~\cite{Joh07,Short_time_PRL}. More precisely
one finds
\bea \label{gumbel2} 
\frac{x_{\max}(T) - x_{\rm edge}}{x_{\rm edge}} \simeq \frac{1}{4y} \ln \left(\frac{N^2}{4 \pi y^3}\right) + \frac{1}{2y}\, \gamma \quad , \quad y= \frac{\hbar \omega N}{T} = \frac{T_F}{T} \;,
\eea 
where $\gamma$ is a Gumbel random variable with PDF $p(\gamma)=e^{-\gamma - e^{-\gamma}}$. Note that we have used the variable $y$ which was introduced to study the high temperature bulk regime (which is
defined by $y={\cal O}(1)$). Although this formula is derived here from studying the high temperature limit of the low temperature edge regime, we expect that it should match in some way with the low temperature limit of the 
high temperature bulk regime, although as we will see this matching is not trivial. This result for high temperature in Eq. (\ref{gumbel2}) together with the
$T=0$ result [see Eqs. (\ref{xmax_0T}), (\ref{fredholm_F2})] show that the fluctuations of the position of the rightmost fermion, in the edge regime, interpolates between a TW-random variable $\chi_2$ when $T\ll N^{1/3} \hbar \omega$ and Gumbel random variable $\gamma$ when $T \gg N^{1/3} \hbar \omega$ \cite{Joh07,Short_time_PRL}, i.e. (in terms of the original parameters $T$ and $N$),
\begin{eqnarray}\label{summary}
\frac{x_{\max}(T)-x_{\rm edge}}{x_{\rm edge}} \sim
\begin{cases}
&\dfrac{1}{2} N^{-2/3} \, \chi_2 \;, \; T \ll N^{1/3} \hbar \omega  \\
& \\
& \dfrac{T}{4 N \hbar \omega} \ln \left(\dfrac{T^3}{4 \pi N (\hbar \omega)^3}\right) + \dfrac{T}{2N \hbar \omega} \, \gamma \;, \; T \gg N^{1/3} \hbar \omega  \;.
\end{cases}
\end{eqnarray}

It is instructive to compare this Gumbel distribution to the one that we can infer, from extreme value statistics arguments (see e.g., Ref. \cite{Gumbel}), in the very high
temperature regime $y \ll 1$. In that regime, we know that the positions of the fermions are independent and identical
random variables drawn from the Boltzmann distribution given in Eq.~(\ref{boltzmann}),  and therefore $x_{\max}(T)$ is also distributed according to a Gumbel law. This implies
\bea \label{gumbel1} 
\frac{x_{\max}(T) - x_{\rm edge}}{x_{\rm edge}} \simeq \sqrt{\frac{\ln N}{y}} - 1 - \frac{\ln(4 \pi \ln N)}{4  \sqrt{y \ln N}}  
+ \frac{1}{2  \,\sqrt{y \ln N}} \tilde \gamma \;,
\eea 
where $\tilde \gamma$ is also a Gumbel random variable (we used the different notation $\tilde \gamma$ in Eq. (\ref{gumbel1}) to emphasize that this random variable is different from the Gumbel variable in Eq. (\ref{gumbel2}), while both of them have the same statistics). The behavior (\ref{gumbel1}) is 
expected to hold for $N \gg 1$, and fixed $y \ll 1$. Note that $x_{\rm edge}$ in this
formula is the $T=0$ edge, while at these temperatures the density does not
display a true edge, and extends well beyond $x_{\rm edge}$. 
We note that, although we obtain Gumbel laws in the two
regimes (\ref{gumbel2}) and (\ref{gumbel1}) the detailed dependence in $y$ and
$N$ is quite different. From its derivation, the regime (\ref{gumbel2}) should hold for large $y$. If one
compares the deterministic terms in (\ref{gumbel2}) and (\ref{gumbel1}), keeping in mind that the fluctuations
increase with temperature, one obtains 
the stronger condition for (\ref{gumbel2}) to hold
\bea
y \gg {\ln N} \quad , \quad {\rm i.e.} \quad N^{1/3} \ll \frac{T}{\hbar \omega} \ll \frac{N}{\ln N} \;.
\eea 
The interpolation between these two regimes remains an open problem. It requires
to study the Fredholm determinant associated with the full finite temperature kernel (\ref{kernel_final})
in the region $y = {\cal O}(1)$. Note that, although the fluctuations of $x_{\max}(T)$ 
are universal (this is shown in section \ref{sec:6d}) in the low temperature scaling
regime $b={\cal O}(1)$ ([i.e., in Eq.~(\ref{summary})], the extreme value statistics arguments leading to (\ref{gumbel1}) also show that 
the (logarithmic) dependence in $N$ depends on the power law index $p$ of
the confining potential $V(x) \sim |x|^p$. These different behaviors of the rightmost fermion, as described in Eqs. (\ref{summary}) and (\ref{gumbel1}) above, are sketched in Fig. \ref{Fig_rightmost}.

Note that besides the regime of typical fluctuations of $x_{\max}(T)$ one can also study the
large deviations. In the limit of high temperature of the low temperature edge regime
this was done in detail in \cite{Short_time_PRL}. In addition, one can also study the counting statistics of the number of
fermions in a given interval $J$, or the PDF of the spacing
between nearest neighbor fermions near the edge at finite temperature. As a consequence of the determinantal structure of all correlations in the large $N$ limit, these observables 
are given in terms of Fredholm determinants by exactly the same formula
as (\ref{counting}) and (\ref{exact_p2}) replacing the kernel $K_J$ there by the finite $T$ edge kernel~(\ref{kff}). 
The detailed analysis of such FD is left for future investigations. 

{Here we add a few interesting remarks. First one can extend the
property shown in  Eq.~\eqref{genf} to finite temperature, assuming that again the
equivalence canonical-grand canonical goes through at large $N$ (the property
is exact in the grand-canonical ensemble). It can be written everywhere
(bulk and edge) but let us display it here near the edge. Defining the rescaled positions of the fermions near 
the edge $a_i = \frac{x_i-x_{\rm edge}}{w_N}$ and taking the $N \to \infty$ limit, keeping $a_i$'s fixed
\bea \label{genff} 
&&  \left \langle \prod_{i=1}^\infty f(a_i) \right \rangle_{T} = {\rm Det}( I - L_{f,b}) \quad , \quad  L_{f,b}(r,r') = (1- f(r)) 
{\cal K}^{\rm edge}_{b}(r,r') \;,
\eea 
where ${\cal K}^{\rm edge}_{b}(r,r')$ is given in Eq. (\ref{kff}). In Eq. (\ref{genff}), 
the average is performed at finite temperature $T$ and 
the function $f(x)$ is arbitrary, provided the right hand side exists. Next,
since one can rewrite the CDF of the position of the rightmost 
fermion in \eqref{gap_proba} more explicitly in terms of the Airy kernel as
\bea \label{gap_proba2}
&& Q_b(s) := {\rm Det}(I - P_s {\cal K}^{\rm edge}_{b} P_s) =  {\rm Det}(I - \bar K_{b,s}) \\
&& \bar K_{b,s}(u,u') = K_{\rm Ai}(u,u') \sigma_{s,b}(u) \quad , \quad  \sigma_{s,b}(u) = 1/(1 + e^{- b (u-s)}) 
\eea
we immediately obtain (with $P_J(x)$ denoting the indicator function of the interval $J$)
\bea
\left \langle \prod_{i=0}^\infty P_{(-\infty,s]}(a_i) \right \rangle_{T} = Q_b(s)  
= \left \langle \prod_{i=0}^\infty \left(1- \sigma_{s,b}(a_i)\right) \right \rangle_{T=0} 
= \left \langle \prod_{i=0}^\infty \frac{1}{1 + e^{b (a_i-s)}}  \right \rangle_{T=0} 
\eea 
A similar product appeared in a recent paper by Borodin and Gorin \cite{BoGo}. 
}

\section{The $d$-dimensional isotropic harmonic oscillator at zero temperature}\label{OHdT=0}

We now consider the model for non-interacting fermions in a harmonic potential in arbitrary dimension $d$,
as defined in Section \ref{sec:model}. Here we focus on $T=0$, the finite $T$ case is
studied in the next Section. To study this more general problem it is convenient to use
a method based on the one-body Euclidean propagator that we describe in Section \ref{sec:5a}. This will lead to
results both in the bulk (Section \ref{sec:5b}), as well as at the edge (Section \ref{sec:5c})
of the $d$-dimensional Fermi gas. In particular, this provides a useful
alternative (also in $d=1$) to the method relying on the 
Plancherel-Rotach asymptotic formulas for Hermite polynomials.

\subsection{Representation of the $T=0$ kernel using the one-body Euclidean
propagator in arbitrary $d$} \label{sec:5a} 

\subsubsection{General framework: kernel and propagator} 

We start by considering the zero temperature kernel for $N$ non-interacting spin-less fermions with an arbitrary one-body Hamiltonian $\hat H$, as defined in (\ref{H}). 
As discussed in Section \ref{sec:zeroT}, the kernel corresponding to a system with Fermi energy $\mu$ is then defined by
\begin{equation} \label{K0} 
K_\mu({\bf x},{\bf y}) = \sum_{{\bf k}}\theta(\mu-\epsilon_{\bf k})\psi^*_{\bf k}({\bf x}) \psi_{\bf k}({\bf y}) ,
\end{equation}
where $\theta(z)$ is the Heaviside step function and the energy eigenvalues 
$\epsilon_{\bf k}$ are labeled by $d$ quantum numbers denoted by ${\bf k} \in 
\mathbb{Z}^d$.
We now compute
\begin{eqnarray}
 G({\bf x},{\bf y};t)  &=& {t\over \hbar}\int_0^\infty d\mu \exp\left(-{t\mu \over \hbar}\right)K_\mu({\bf x},{\bf y}) \nonumber \\
 &=& \sum_{\bf k}\psi^*_{\bf k}({\bf x}) \psi_{\bf k}({\bf y})\exp\left(-{\epsilon_{\bf k} t\over \hbar}\right)
 \end{eqnarray}
 and immediately see that $G({\bf x},{\bf y};t)$ is in fact  the one body Euclidean propagator associated to the one body  Hamiltonian $\hat H$. By definition, it obeys the imaginary time 
 Schr\"odinger equation
 \begin{equation}
 -\hbar{\partial G({\bf x},{\bf y};t)\over \partial t}=  \hat H G({\bf x},{\bf y};t),\label{itshro}
 \end{equation}
 where $\hat H= \hat H({\bf y}, \frac{\hbar}{i} \nabla_{\bf y})$ is the quantum Hamiltonian defined in Eq. (\ref{defhhat}) acting on the variable ${\bf y}$ (in our convention). From the completeness of the basis of the eigenfunctions, it satisfies the initial condition
 \begin{equation}
 G({\bf x},{\bf y};0) = \delta^d({\bf x}-{\bf y}).
 \end{equation}
If the propagator $G$ is known as a function of $t$, then the kernel can be obtained via the Bromwich inversion formula for Laplace transforms as
\begin{equation}
K_\mu({\bf x},{\bf y}) = \int_{\Gamma} {dt\over 2\pi i t} \exp\left({\mu t\over \hbar}\right)G({\bf x},{\bf y};t) \;,\label{laplace_inverse}
\end{equation}
where $\Gamma$ indicates the Bromwich integration contour in the complex plane.

For the isotropic $d$-dimensional harmonic oscillator, which we focus on here, the Euclidean propagator  is known exactly at all times, and is given by \cite{feynman_hibbs}
\begin{eqnarray}\label{eq:propag_hamonic}
G({\bf x},{\bf y};t) = \left(\frac{\alpha^2}{2 \pi \sinh{(\omega \, t)}}\right)^{d/2} \exp{\left[-\frac{\alpha^2}{2 \sinh{(\omega\,t)}}  \left(({\bf x}^2 + {\bf y}^2) \cosh{(\omega \, t) - 2 \, {\bf x}\cdot {\bf y}}\right) \right]}
\end{eqnarray}  
where $\alpha = \sqrt{m \omega/\hbar}$. 
The Bromwich integral in Eq. (\ref{laplace_inverse}) is in general difficult to compute explicitly. However, for a system with a large number of fermions, the Fermi energy $\mu$ will be large and so we can assume that the integral is dominated by small values of $t$. The validity of the approach will be made more precise below.

\subsubsection{Short-time expansion of the propagator and of the kernel} 

As we will make clear later, to obtain all of our results, we will need the short time expansion of the propagator
up to ${\mathcal O}(t^3)$. For this purpose it is useful to write the propagator $G({\bf x},{\bf y};t)$ given in Eq. (\ref{eq:propag_hamonic}) in the following form
\begin{equation}\label{propag_bulk}
G({\bf x},{\bf y};t) = \left(\frac{\alpha^2}{2 \pi \sinh{(\omega \, t)}}\right)^{d/2} \exp{\left[-\frac{\alpha^2}{2 \sinh{(\omega\,t)}}  \left( ({\bf x}- {\bf y})^2 + ({\bf x}^2 + {\bf y}^2)(\cosh(\omega\,t)-1) \right) \right]} \;,
\end{equation}
which gives at coinciding points with $r= |{\bf x}|$
\begin{equation}\label{eq:propag_hamonic2}
G({\bf x},{\bf x};t) = \left(\frac{\alpha^2}{2 \pi \sinh{(\omega \, t)}}\right)^{d\over 2}\exp{\left[-\tanh{\left(\frac{\omega \, t}{2} \right)} \alpha^2 r^2 \right]} \;.
\end{equation}

Expanding the formula (\ref{propag_bulk}) to ${\mathcal O}(t^3)$ gives
\begin{eqnarray}
G({\bf x},{\bf y};t) \simeq \left(\frac{m}{2 \pi \hbar t }\right)^{d/2} \exp\left(-\frac{m}{2\pi\hbar t}({\bf x}-{\bf y})^2 -\frac{m\omega^2 t}{12\hbar}\left[3({\bf x}^2+{\bf y}^2)-({\bf x}-{\bf y})^2\right]\right.\nonumber \\
\left.-\frac{d\omega^2t^2}{12} + \frac{m\omega^4 t^3}{720\hbar}\left[15({\bf x}^2+{\bf y}^2)
-7({\bf x}-{\bf y})^2\right]\right) \;.\label{expsho}
\end{eqnarray}
In particular at coinciding points ${\bf y}={\bf x}$ with $r= |{\bf x}|$
and using $\alpha=\sqrt{m \omega/\hbar}$, one obtains to ${\mathcal O}(t^3)$
\begin{eqnarray}
G({\bf x},{\bf x};t) &\simeq& 
\left[ {\alpha^2\over 2\pi\omega t}\right]^{d\over 2} \exp\left(-{\alpha^2 r^2 \omega t\over 2} -{d\,\omega^2t^2\over 12} +{\alpha^2 r^2\omega^3 t^3\over 24}\right) .\label{shot3}
\end{eqnarray} 
Substituting the expression (\ref{shot3}) into Eq. (\ref{laplace_inverse}) then yields the
short time expansion to order ${\mathcal O}(t^3)$ of the kernel
\begin{eqnarray}
K_\mu({\bf x},{\bf y}) \simeq 
\left[{\alpha^2\over \omega}\right]^{d\over 2}\int_\Gamma {dt\over (2\pi t)^{1+{d\over 2}} i}
\exp\left(-\frac{m}{2\pi\hbar t}({\bf x}-{\bf y})^2 + \frac{t}{\hbar} (\mu  -\frac{m\omega^2 }{12}\left[3({\bf x}^2+{\bf y}^2)-({\bf x}-{\bf y})^2\right]) \right.\nonumber \\
\left.-\frac{d\omega^2t^2}{12} + \frac{m\omega^4 t^3}{720\hbar}\left[15({\bf x}^2+{\bf y}^2)
-7({\bf x}-{\bf y})^2\right]\right) \;. \label{expsho2}
\end{eqnarray}
At coinciding points the kernel expanded to ${\mathcal O}(t^3)$ becomes
\begin{equation}
K_\mu({\bf x}, {\bf x}) \approx \left[{\alpha^2\over \omega}\right]^{d\over 2}\int_\Gamma {dt\over (2\pi t)^{1+{d\over 2}} i}\exp\left({t\over \hbar}[\mu -{m\omega^2 r^2 \over 2}] -{d\omega^2t^2\over 12} +{\alpha^2 r^2\omega^3 t^3\over 24}\right) \;. \label{lap1}
\end{equation}
The latter two formula will be used extensively below.

\subsection{Results in the bulk} \label{sec:5b} 

\subsubsection{The bulk density profile}\label{shobd}

We start by analyzing the kernel at coinciding points, {\em i.e.} ${\bf x}={\bf y}$, to obtain the density $\rho_N({\bf x}) = (1/N) K_\mu({\bf x},{\bf x})$ for large $N$ (equivalently for large $\mu$ as we assume now and verify a posteriori). 
Our starting point is Eq. (\ref{lap1}) where we see that the effective energy scale entering the Laplace transform is not simply $\mu$ but 
\begin{equation}
\epsilon(r)= \mu -m\omega^2 r^2/ 2 = \mu \left( 1 - \frac{r^2}{r^2_{\rm edge}} \right) \;, \label{defeps}
\end{equation}
which is the difference between the Fermi energy and the classical potential energy at the radial coordinate $r$. 
We have defined
\be \label{redged} 
r_{\rm edge} = \sqrt{2\mu/m\omega^2} \;,
\ee
the radius at which $\epsilon(r)$ vanishes. We will see below that $r_{\rm edge}$ is also the
edge where the bulk density vanishes. When $\epsilon(r)$ is large, which is the case
for large $\mu$ and away from the edge, the integral in Eq. (\ref{lap1}) can be approximated by the term of  ${\cal O}(t)$ in the exponential, and is dominated by times of order $t \sim \hbar/\epsilon(r)$. 
The corrections to this approximation coming from the ${\cal O}(t^2)$ and 
${\cal O}(t^3)$, are respectively 
\bea
\frac{d}{12} \left(\frac{\hbar \omega}{\epsilon(r)} \right)^2 
+ \frac{\alpha^2 r^2}{24}  \left(\frac{\hbar \omega}{\epsilon(r)} \right)^3 \;.
\eea 
Hence the first one is unimportant when
\begin{equation}
\epsilon(r) \gg \hbar\omega,\label{ineqa}
\quad  \Longleftrightarrow \quad
|r-r_{\rm edge}| \gg (r_{\rm edge} \alpha)^{-1}/\alpha .
\end{equation}
Similarly the cubic terms can be seen to be unimportant when 
\begin{equation}
\frac{\alpha^2 r^2}{24}  \left(\frac{\hbar \omega}{\epsilon(r)}\right)^3 \ll 1 \quad  \Longleftrightarrow \quad
|r-r_{\rm edge}| \gg (r_{\rm edge} \alpha)^{-{1\over 3}}/\alpha \;.
\label{ineqb}
\end{equation}
We will see below that $r_{\rm edge}$ is large for large $N$.
In that case, sufficiently away from the edge, Eq. (\ref{ineqb}) is
satisfied and this automatically ensures that Eq. (\ref{ineqa})
is also satisfied. Hence, keeping only this leading term to describe the bulk density, we can use the standard 
identity
\begin{equation}\label{eq:besseltheta}
\int_\Gamma \frac{dt}{2\pi i} \frac{1}{t^{d/2+1}} \exp(zt) = \frac{z^{d/2}}{\Gamma(1+ \frac{d}{2})} 
\theta(z) 
\end{equation} 
to write 
\begin{eqnarray}
K_\mu({\bf x}, {\bf x})&\approx&{1\over \Gamma(1+{d\over 2})}\left[{m\over 2\pi\hbar^2}\right]^{d\over 2}\left(\mu - {m\omega^2 r^2\over 2}\right)^{d\over 2} \theta \left(\mu - {m\omega^2 r^2\over 2}\right)
\\ &=& 
{1\over 2^d \pi^{d\over 2} \Gamma(1+{d\over 2})}\alpha^{2d}\left(r^2_{\rm edge} -r^2\right)^{d\over 2} \theta \left(r_{\rm edge} -r\right) \;, \label{kxxsho}
\end{eqnarray}
where $\theta$ is the Heaviside step function. Consequently the bulk density, as a function of $r$ = $|{\bf x}|$ is given by
\begin{equation}
\rho_N(r) = {1\over 2^d N \pi^{d\over 2} \Gamma(1+{d\over 2})}\alpha^{2d}\left(r^2_{\rm edge} -r^2\right)^{d\over 2} \theta \left(r_{\rm edge} -r\right) \;.\label{rhosho}
\end{equation}
Hence, as anticipated above, the bulk density vanishes at the radius $r=r_{\rm edge}$ given in
Eq. (\ref{redged}). It should also be noticed that the expression given here for the density is the one obtained by the local density approximation (LDA) \cite{Castin}. 

The value of the chemical potential $\mu$ corresponding to a fixed value of $N$, the number of fermions, can be evaluated from the normalization condition $\int d^{d}{
\bf x}\ \rho_N({\bf x}) =1$ which yields 
\begin{equation}
{S_d \, r_{\rm edge}^d \over \Gamma(1+{d\over 2})}\mu^{d\over 2} \left({m\over 2\pi \hbar^2}\right)^{d\over 2} \int_0^1 du\ u^{d-1}(1-u^2)^{d\over 2} = N,  \label{norm1} 
\end{equation}
where $S_d= 2 \pi ^{d\over 2}/\Gamma(d/2)$ is the surface unit area of a unit sphere in $d$ dimensions.
Using 
\begin{equation}
\int_0^1 du\ u^{d-1}(1-u^2)^{d\over 2}= {1\over 2}{\rm B}\left({d\over 2}, {d\over 2}+1\right),
\end{equation}
where ${\rm B}(p,q)$ denotes the beta function, as well as Eqs. (\ref{redged}) and (\ref{norm1}),
we obtain the Fermi level and the value of the edge radius as a function of $N$
\begin{equation}
\mu = \hbar\omega [N\Gamma(d+1)]^{1\over d} \quad , \quad r_{\rm edge} 
= \frac{2^{1\over 2} [\Gamma(d+1)]^{1\over 2d}}{\alpha}N^{1\over 2d} \;, \label{muredge} 
\end{equation}
as anticipated in Eq. (\ref{musho}). Finally, this leads to the final result for the normalized density as a function
of $N$
\begin{equation}
\rho_N({\bf x}) = {1\over 2^d N^{1\over 2} \pi^{d\over 2}\Gamma(1+{d\over 2})}
\left({m\omega\over \hbar}\right)^{d\over 2}\left(2 \Gamma(d+1)^{1\over d} - {m\omega\over \hbar N^{1\over d}}r^2\right)^{d\over 2} \;.\label{bulkdensityshoeq}
\end{equation} 
Using the explicit dependence of $\mu$ and thus $r_{\rm edge}$ on $N$ 
from (\ref{muredge}) in (\ref{ineqb}), we see that this formula for the bulk density is valid
in the regime $r_{\rm edge}-r \gg N^{-\frac{1}{6d}}/\alpha$. 
Below we will see how to obtain both the density and the kernel in the edge regime
where $r_{\rm edge}-r \sim N^{-\frac{1}{6d}}/\alpha$.

As an example we consider the simple harmonic oscillator in two dimensions. We have numerically computed the density $\rho_N({\bf x})$ for $N=28$ (corresponding to a full Fermi level) free fermions in a two dimensional harmonic trap for  $\alpha =\sqrt{m\omega/\hbar}=1$, by directly summing the modulus squared of the first $28$ wave functions. Shown in Fig.~\ref{bulkdensityshofig} is the  comparison of the numerical
summation with the bulk asymptotic formula Eq. (\ref{bulkdensityshoeq}).
\begin{figure}[hh]
\includegraphics[width=0.5\linewidth]{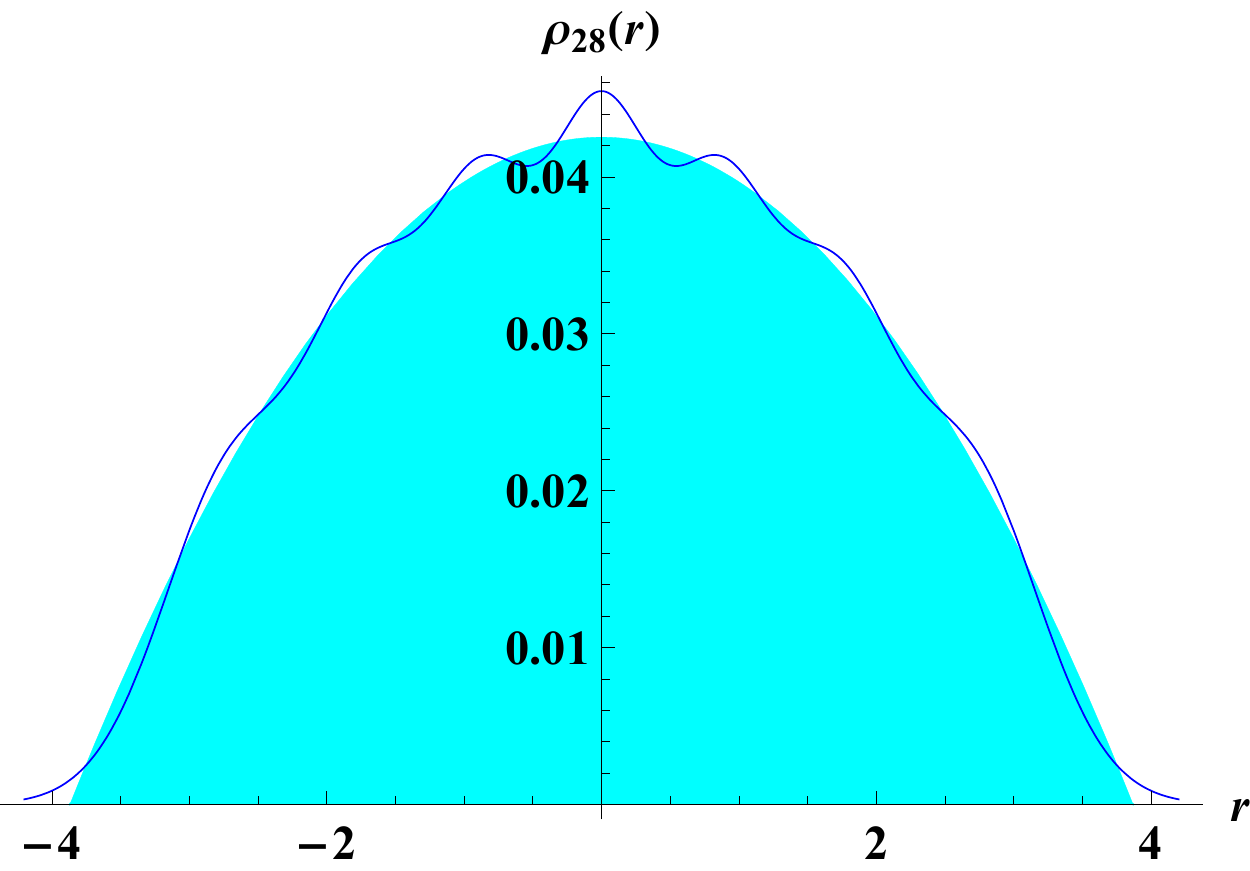}
\caption{(Color online) Plot of the bulk density (solid line)  $\rho_{28}(r)$ for $28$ fermions in an isotropic  harmonic potential in two dimensions compared with the asymptotic formula for the bulk density Eq. (\ref{bulkdensityshoeq}), (the region below the asymptotic result is shown via the  shaded region).}\label{bulkdensityshofig}
\end{figure}

\subsubsection{The kernel in the bulk}\label{shobk}

In order to compute the bulk kernel for the isotropic simple harmonic oscillator, 
we extend our analysis for the bulk density, and thus we keep 
only terms up to order $\mathcal{O}(t)$ in Eq.~(\ref{expsho2})
which gives
\begin{equation}\label{eq:kernel_bulk1}
K_{\mu}({\bf x},{\bf y}) \approx \left(\frac{m}{2 \pi \hbar}\right)^{\frac{d}{2}} \int_{\Gamma} \frac{dt}{2\pi i} \, \frac{1}{t^{\frac{d}{2}+1}} \, \exp\left(\frac{\mu t}{\hbar} - \frac{m ({\bf x} - {\bf y})^2}{2 \hbar t}  - \frac{m\omega^2({\bf x}^2 +{\bf y}^2 +{\bf x}\cdot{\bf y})t}{6\hbar}\right).
\end{equation}  
To proceed one can use the following integral representation of the standard Bessel function of the first kind of index $\nu$, denoted by $J_\nu(x)$ \cite{Grad}, 
\begin{equation}\label{eq:bessel}
\int_\Gamma \frac{dt}{2\pi i} \frac{1}{t^{d/2+1}} \exp(zt - \frac{a}{t}) = \left(\frac{z}{a}\right)^{d/4} \, J_{\frac{d}{2}}\left(2\sqrt{a\,z}\right) \;.
\end{equation} 
 We use this identity to evaluate the representation of the kernel in Eq. (\ref{eq:kernel_bulk1}) to obtain
 
 \begin{equation}\label{eq:density}
 K_{\mu}({\bf x},{\bf y})\approx \left(\frac{\alpha^2}{2\pi}\right)^{\frac{d}{2}}
 \left(\frac{3 r_{\rm edge}^2- {\bf x}^2-{\bf y}^2 -{\bf x}\cdot{\bf y}}{3 ({\bf x}-{\bf y})^2}\right)^{\frac{d}{4}}J_{\frac{d}{2}}\left( \frac{\alpha^2}{\sqrt{3}}|{\bf x}-{\bf y}|\sqrt{3r_{\rm edge}^2- {\bf x}^2-{\bf y}^2-{\bf x}\cdot{\bf y}}\right) \;,
\end{equation}
where we recall that $\mu=\frac{1}{2} m \omega^2 r_{\rm edge}^2$. 
Now if we consider two points ${\bf x}'$ and ${\bf y}'$ both close to a point ${\bf x}$ in the bulk we find that the kernel has the scaling form
\begin{eqnarray}\label{scaling_bulk}
K_{\mu}({\bf x}+ {\bf x}',{\bf x}+{\bf y}') \approx \ell({\bf x})^{-d} {\cal K}^{\rm bulk}_d(|{\bf x'}-{\bf y'}|/\ell({\bf x}))
\end{eqnarray}
where 
\be \label{lx} 
\ell({\bf x}) = [N \rho_N({\bf x}) \gamma_d]^{-1/d}  \quad , \quad \gamma_d =  S_d/d=\pi^{d/2} [\Gamma(d/2+1)]
\ee
is the local typical separation between fermions in the bulk. 
The explicit formula for the scaling function in Eq. (\ref{scaling_bulk}) is given by
\begin{eqnarray} \label{eq:kernel_bulk2}
{\cal K}^{\rm bulk}_d(x) = \frac{J_{d/2}(2 x)}{(\pi x)^{d/2}},\label{bksho}
\end{eqnarray}
which has a well defined limit at the origin with ${\cal K}^{\rm bulk}_d(0)= 1/\gamma_d$. In $d=1$, using
$J_{1/2}(z) = \sqrt{2/(\pi z)} \sin{z}$, we recover the standard sine-kernel ${\cal K}^{\rm bulk}(x) = \frac{\sin(2 x)}{\pi x}$ of RMT. 
The domain of validity of the bulk analysis carried out above can be determined by examining the corrections coming from the quadratic and cubic corrections in $t$, as was done above for the study of the density.
The result for the bulk kernel in Eq. (\ref{eq:kernel_bulk1}) is found to be valid when
{\em both} ${\bf x}$ and ${\bf y}$ are in the bulk, away from the edge $r_{\rm edge}$.

%

The result for the kernel (\ref{eq:kernel_bulk2}) for relative distances of order the typical particle separation 
can also be obtained from the LDA \cite{Castin,scardicchio}. Indeed we remark here that  the Fourier transform of this kernel is
Fermi-step function, this can be seen using the formula:
\begin{eqnarray}
\int_{|{\bf k}|<k_F} \frac{d^d k}{(2 \pi)^d} e^{i \bf k \cdot \bf x} =  \left(\frac{k_F}{2 \pi |{\bf x}|}\right)^{d/2} J_{d/2}(k_F |{\bf x}|).\label{ftker}
\end{eqnarray}
In the expression in Eq. (\ref{ftker})
the local Fermi momentum must be chosen as  $k_F = k_F({\bf x})= 2/\ell({\bf x}) =2 (N \rho_N({\bf x}) \gamma_d)^{1/d}$.
This value is exactly consistent with the one obtained by the counting of states for a uniform system of density
$\rho_N({\bf x})$, by setting $N \, \rho_N({\bf x}) = \int_{|{\bf k}|<k_F}  \frac{d^d k}{(2 \pi)^d}  = k_F^d/(4 \pi)^{d/2} \Gamma(1+d/2)$. Thus, to describe correlations on scale $\ell({\bf x})$ in the bulk, one can approximate locally the system by free
fermions without any external potential, but at a fixed density $\rho_N({\bf x})$,
assumed to be slowly varying on that scale. Our method thus gives  a rigorous derivation of the results, within the bulk, obtained from the heuristic LDA. 

\subsection{Results at the edge} \label{sec:5c} 

\subsubsection{The density near the edge}\label{shodensity}

We have seen how the density and kernel in the bulk region can be obtained via a leading order short-time expansion of the propagator for the simple harmonic oscillator in Section (\ref{sec:5b}). However this  expansion becomes insufficient when the inequalities in Eqs. (\ref{ineqa}) and (\ref{ineqb})
are violated: this occurs near the edge where the density vanishes. As demonstrated below, the description of the edge regime requires one to expand the propagator to higher orders in $t$ as in Eq. (\ref{shot3}).

We start by investigating the density near $r=r_{\rm edge}= \sqrt{2\mu/m\omega^2}$, for finite but large $N$
by setting
\begin{equation}
r = |{\bf x}| = r_{\rm edge} + z \, w_N \quad , \quad w_N = b_d \, N^{-\phi}, \label{choicer}
\end{equation}
where the distance from the edge is parametrized by the scaled dimensionless distance $z$, while 
the edge exponent $\phi$ is to be determined a posteriori. For convenience we
introduced the factor $b_d=\left[\Gamma(1+d)\right]^{-\frac{1}{6d}}/(\alpha \sqrt{2})$
which has the dimension of a length.

To calculate the kernel in this edge region, we need to use the expansion of
the kernel $K_\mu({\bf x},{\bf x})$ up to ${\cal O}(t^3)$, as given 
in Eq.~(\ref{lap1}). Substituting (\ref{choicer}) into 
(\ref{lap1}), we see that the two terms of ${\cal O}(t)$ cancel each other inside the 
argument of the exponential, leaving only three terms
\begin{equation}
\label{eq:propag_hamonic3}
K_\mu({\bf x},{\bf x}) \approx \left(\frac{\alpha^2}{2\pi \omega}\right)^{d/2} \int_\Gamma \frac{d\,t}{2\pi i} \frac{1}{t^{d/2+1}} \exp{\left[ -\sqrt{\frac{2\mu}{m}} \alpha^2 \, z\,b_d N^{-\phi} t - \frac{d}{12}\omega^2 \,t^2+ \frac{\mu \, \omega^2}{12 \hbar} t^3\right]} \;.
\end{equation}
We now determine the exponent $\phi$ by comparing the magnitude of the three terms inside the
exponential and using $\mu \sim N^{1/d}$. We obtain, in the order that they appear
\bea
T_1 \sim z \sqrt{\mu} N^{- \phi} t \sim z N^{\frac{1}{2 d}- \phi } t \quad , \quad T_2 \sim t^2
\quad , \quad T_3 \sim \mu t^3 \sim N^{\frac{1}{d}} t^3 \;.
\eea 
Since the first term must be of order ${\cal O}(1)$, this implies that $t \sim N^{\phi-\frac{1}{2d}}$.
We then have only two possibilities a priori. The first one is to choose $\phi = 1/(2d)$ that keeps $T_2 \sim {\cal O}(1)$, but then $T_3 \sim N^{1/d} $ and diverges as $N \to \infty$, which is inconsistent. Hence, in order to make the term $T_3 = {\cal O}(1)$,
we must choose
\bea 
\phi= \frac{1}{6 d} \quad \Longrightarrow \quad t \sim N^{- \frac{1}{3 d}} \;, \label{scale} 
\eea 
which is consistent with the assumption of an expansion in small $t$. 
It is easy to check in Eq. (\ref{eq:propag_hamonic2}) that with this scaling exponent the 
terms of order higher than ${\cal O}(t^3)$ vanish as $N \to \infty$. Note that this 
result can be understood qualitatively by arguing that there should be of order one particle
in the typical fluctuation region, i.e. in a box of linear size $w_N$ around the edge, which leads to
\bea
\left(\frac{w_N}{r_{\rm edge}}\right)^{d-1} \, \int_{r_{\rm edge}- w_N}^{r_{\rm edge}} \rho_N(r) r^{d-1} dr  \sim \frac{1}{N} 
\eea 
which in turn implies $w_N \sim N^{-1/(6 d)}$, using the formula (\ref{bulkdensityshoeq}) for the density $\rho_N(r) = \rho_N(|{\bf x}|)$. Hence, rescaling $t = N^{-{1}/{(3d)}}/(\omega [\Gamma(d+1)]^{{1}/{(3d)}}) \, \tau$, we obtain our main result for the density 
\begin{equation}\label{density_edged}
N\rho_N({\bf x}) = K_\mu({\bf x},{\bf x}) = \frac{1}{w_N^d} F_d\left(\frac{r-r_{\rm edge}}{w_N} \right)
\end{equation}
where we have defined the width of the edge regime in $d$-dimension
\bea \label{wN} 
w_N = b_d \, N^{-\frac{1}{6d}} = \frac{1}{\alpha \sqrt{2}} \left[\Gamma(1+d) N\right]^{-\frac{1}{6d}}
\eea
and the scaling function $F_d(z)$ is given by 
\begin{eqnarray}\label{eq:Fd_laplace}
F_d(z) = (4\pi)^{-d/2} \int_{\Gamma} \frac{d \tau}{2\pi i} \, \frac{1}{\tau^{d/2+1}} \, e^{- \tau\,z + {\tau^3}/{12}} \;.
\end{eqnarray}
The expression  Eq. (\ref{eq:Fd_laplace}) can be rewritten
in the following way. First we make use of the identity 
\begin{equation}\label{eq:identity}
\frac{1}{\tau^{d/2+1}} = \frac{1}{\Gamma(d/2+1)} \int_0^\infty e^{-\tau x} \, x^{d/2} \, dx
\end{equation}
in Eq. (\ref{eq:Fd_laplace}) to obtain
\begin{equation}\label{eq:identity2}
F_d(z) = \frac{1}{\Gamma(d/2+1) \, (4 \pi)^{d/2}} \int_{0}^\infty dx \, x^{d/2} \int_\Gamma \frac{d\tau}{2\pi i} e^{-\tau(x+z)+ {\tau^3}/{12}} \;.
\end{equation}
Rescaling $\tau \to 2^{2/3} \tau$ and using the integral representation of the Airy function 
\begin{equation}\label{eq:airy_integral}
{\rm Ai}(z) = \int_{\Gamma} \frac{d\tau}{2\pi i}\, e^{-\tau z + \tau^3/3}
\end{equation}
then gives, after a further rescaling {$x \to 2^{-2/3} x$}, in the integral expression for $F_d(z)$ in Eq. (\ref{eq:identity2}) then yields
\begin{equation}\label{Fd_explicit}
F_d(z)= {1\over \Gamma({d\over 2}+1)2^{\frac{4d}{3}} \pi^{d\over 2}}\int_0^\infty du\  u^{d\over 2}{\rm Ai}(u+2^{2/3}\,z) \;.
\end{equation}

{\it General properties of the density scaling function in $d=1,2,3$}. This scaling function satisfies some general properties in any $d$. For instance 
differentiating Eq. (\ref{Fd_explicit}) w.r.t. $z$ once and using integration by part one
finds the recursion formula 
\begin{equation}
\frac{dF_d(z)}{dz}=- \frac{1}{4\pi}\, F_{d-2}(z)\,,
\label{scaling_d.1}
\end{equation}
which allows to obtain $F_d(z) = \frac{1}{4\pi}\, \int_z^{+\infty} F_{d-2}(z)\, $ from
the knowledge of $F_{d-2}(z)$. In addition it is easy to see that $F_d(z)$
satisfies a third-order differential equation
\begin{equation}
\frac{d^3 F_d(z)}{dz^3} - 4\, z\, \frac{dF_d(z)}{dz} + 2\, d\, F_d(z)=0
\label{diff_eqn.1}
\end{equation}
which must be complemented by appropriate boundary conditions (see below). 

{\it Explicit forms of the density scaling function in $d=1,2,3$}. In $d=1$, the above integral can be performed exactly. We start with the identity \cite{vallee}
\begin{eqnarray}\label{eq:def_Iz}
\int_0^\infty {\rm Ai}(z+u) \frac{du}{\sqrt{u}} = 2^{2/3} \pi {\rm Ai}^2\left(\frac{z}{2^{2/3}}\right) \equiv I(z) \;
\end{eqnarray}
and differentiate it twice with respect to $z$. Using the Airy differential equation ${\rm Ai}''(z) = z {\rm Ai}(z)$, one obtains 
\begin{eqnarray}\label{final_identity}
\int_0^\infty du \, \sqrt{u} {\rm Ai}(z+u)  = I''(z) - z I(z) = \pi 2^{1/3} \left(\left[{\rm Ai}'\left(\frac{z}{2^{2/3}} \right)\right]^2  - \frac{z}{2^{2/3} }{\rm Ai}^2\left(\frac{z}{2^{2/3}} \right) \right).
\end{eqnarray} 
It then follows from Eq. (\ref{Fd_explicit}), upon setting $d=1$, that
\begin{equation}
F_1(z) = {\rm Ai}'^2(z) - z {\rm Ai}^2(z) \;,
\end{equation}
thus recovering the result obtained in Section \ref{rmtsec} [see Eq. (\ref{edge_density_scaling_function})], 
which coincides with the well known RMT result \cite{BB91,For93}. {One obtains similar quadratic
forms in ${\rm Ai}(z)$ and ${\rm Ai}'(z)$ with polynomial coefficients in $z$ in
any odd space dimension by repeated application of the Airy operator $(\partial_z^2 - z)$ on $I(z)$.
For instance in $d=3$:
\begin{eqnarray}
F_3(z) = \frac{1}{12 \pi} ( 2 z^2 {\rm Ai}(z)^2 - {\rm Ai}(z) {\rm Ai}'(z) - 2 z {\rm Ai}'(z)^2 ) \;.
\end{eqnarray}
In $d=2$ one can use the Airy equation and find 
\begin{eqnarray}
F_2(z) = \frac{1}{2^{\frac{8}{3}} \pi} ( - {\rm Ai}'(2^{\frac{2}{3}} z) - 2^{\frac{2}{3}} z {\rm Ai}_1(2^{\frac{2}{3}} z)) 
\end{eqnarray}
where ${\rm Ai}_1(z)=\int_z^\infty dx \, {\rm Ai}(x)$.


{\it Asymptotic behavior of $F_d(z)$.} Here we give the asymptotic behavior of the scaling functions $F_d(z)$, the full form of which are plotted in Fig.~\ref{edgedensityfd} for $d=1,2,3$. 
\begin{figure}[t]
\includegraphics[width=0.5\linewidth]{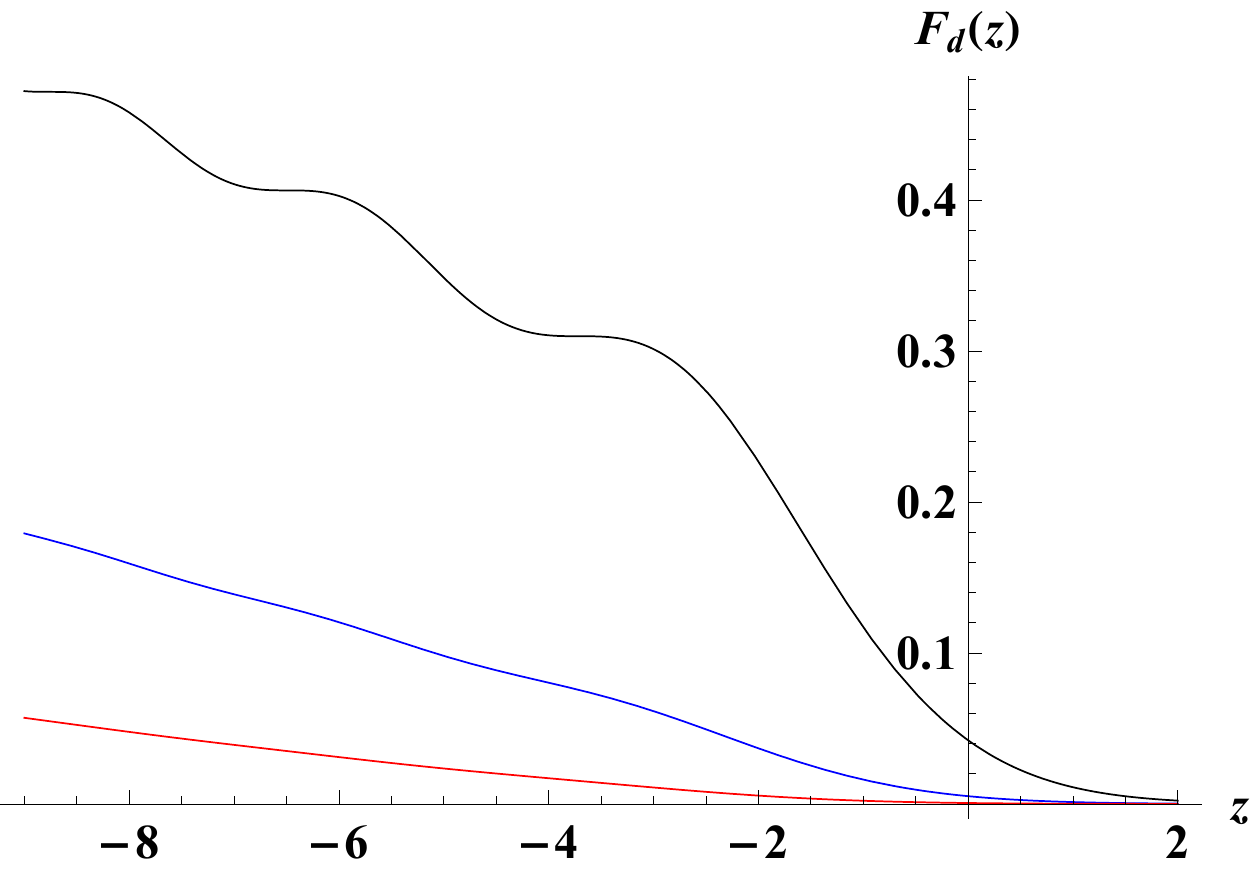}
\caption{(Color online) Plot of the scaling functions $F_d(z)$ in Eq. (\ref{Fd_explicit}) for $d=1,2,3$ (top to bottom) for the density near the edge. The oscillatory structure of the scaling function becomes less pronounced as the dimension $d$ increases.}
\label{edgedensityfd}
\end{figure}
We first consider the $z\to +\infty$ limit. In this limit the Airy function has the leading asymptotic behavior \cite{vallee}
\begin{equation}
\label{airy_largez}
{\rm Ai}(z) \sim \frac{1}{2\sqrt{\pi}} \, z^{-1/4} \, \exp{\left(-\frac{2}{3} z^{3/2}\right)} \;.
\end{equation}
Substituting this asymptotic behavior in Eq. (\ref{Fd_explicit}), expanding for large $z$, one gets to leading order  
\begin{equation}
F_d(z) \approx  (8\pi)^{-\frac{d+1}{2}}\, z^{-\frac{d+3}{4}}\, 
\exp{\left(-\frac{4}{3}\, 
z^{3/2}\right)}\;{\rm as}\; z\to \infty \label{asymp_plus} \;.
\end{equation}
For the other side $z \to -\infty$, it is more convenient to use the representation in Eq. (\ref{eq:Fd_laplace}). We set $z = -|z|$ and scale $\tau\,|z| = t$. This makes the order $\tau^3$ term to be $|z|^{-3} \, t^3/12$ which can then be dropped for large $|z|$. The resulting Bromwich contour integral  can be easily evaluated to give the leading asymptotic behavior
\begin{equation}
F_d(z)\approx \frac{(4\pi)^{-\frac{d}{2}}}{\Gamma(d/2+1)}\, |z|^{\frac{d}{2}}
\quad {\rm as}\quad z\to -\infty \;.
\label{asymp_minus}
\end{equation}
One can show that when $z \to -\infty$, i.e., when $r \ll r_{\rm edge}$, the asymptotic behavior in Eq.~(\ref{asymp_minus}) matches smoothly with the bulk density given in Eq. (\ref{kxxsho}). This can be seen by noting that one can write $r_{\rm edge} = 1/(2 w_N^3\alpha^4)$ and then writing $r=r_{\rm edge}- |z| w_N$ in Eq. (\ref{kxxsho}), which then becomes
\begin{equation}
w_N^d\rho(r_{\rm edge}-|z|w_N) \approx \frac{|z|^{\frac{d}{2}}}{(4\pi)^{\frac{d}{2}}\Gamma(1+\frac{d}{2})} \;,
\end{equation}
which coincides with the behavior in Eq. (\ref{asymp_minus}). 

\subsubsection{Kernel near the edge}\label{edgekernelsho}
\label{edge_kernel}

In order to analyze the full kernel $K_\mu({\bf x},{\bf y})$ near the edge, 
we introduce the following scaled dimensionless coordinates near 
a point ${\bf r}_{\rm edge}$ on the circle of radius $r_{\rm edge}$ as
\bea \label{ab} 
{\bf x} = {\bf r}_{\rm edge} + w_N\, {\bf a} \quad , \quad {\bf y} = {\bf r}_{\rm edge} + w_N\, {\bf b}
\eea 
where $w_N$ is given in (\ref{wN}) and 
\bea
r_{\rm edge}= \sqrt{2\mu/(m \omega^2)} 
\simeq  [\Gamma(d+1)]^{1/(2d)}\, N^{1/(2d)} \sqrt{2}/\alpha \;. \label{redge2} 
\eea

Following the analysis for the edge density in the previous section,
we insert (\ref{ab}) into the expansion of $K_\mu({\bf x},{\bf y})$
up to order ${\cal O}(t^3)$ given in (\ref{expsho2}). We note that if one takes $|{\bf a}|, |{\bf b}| = {\cal O}(1)$, the diffusion
part in (\ref{expsho2}) scales as 
\bea
\frac{m}{2\pi\hbar t}({\bf x}-{\bf y})^2 \sim \frac{w_N^2}{t} \sim \frac{N^{-1/(3 d)}}{t}= {\cal O}(1) \;,
\eea 
where we have used Eq. (\ref{scale}). And the analysis for the other terms is similar to the one performed for the density,
since one has
\bea
&& {\bf x}^2+{\bf y}^2 = 2 \, r_{\rm edge}^2 (1 + w_N (a_n + b_n)) + {\cal O}(w_N^2) \\
&& ({\bf x}-{\bf y})^2 = w_N^2 ({\bf a} - {\bf b})^2  \ll {\bf x}^2+{\bf y}^2
\eea 
where $a_n = {\bf a} \cdot {\bf r}_{\rm edge}/r_{\rm edge}$ and $b_n = {\bf b} \cdot {\bf r}_{\rm edge}/{r}_{\rm edge}$ are the projections of ${\bf a}$ and ${\bf b}$ in the radial direction.
Putting all together gives 
\begin{equation}\label{eq:edge_kernel}
K_{\mu}({\bf x},{\bf y})\approx \frac{1}{C_d w_N^d} \int_{\Gamma} \frac{d\tau}{2\pi i} \frac{1}{\tau^{\frac{d}{2}+1}} \, e^{-\frac{({\bf a} - {\bf b})^2}{2^{8/3}\tau} - \frac{(a_n + b_n)\tau}{2^{1/3}} + \frac{\tau^3}{3}} \;,
\end{equation}   
with $C_d = (2^{\frac{4}{3}}\sqrt{\pi})^{d}$ and where we have rescaled
$t = N^{-{1}/{(3d)}}/(\omega [\Gamma(d+1)]^{{1}/{(3d)}}) \, \tau$, followed by $\tau \to  2^{2/3} \tau$ as 
in the previous section.

One can make a further simplification of Eq.~(\ref{eq:edge_kernel}) by using the integral representation of the diffusive propagator
\begin{eqnarray}\label{eq:diffusion}
\frac{e^{-\frac{({\bf a} - {\bf b})^2}{4 \,D \,\tau}}}{(4\pi D\,\tau)^{\frac{d}{2}}} = \int \frac{d^d q}{(2 \pi)^d} \, e^{- D\,q^2 \tau - i {\bf q}\cdot({\bf a} -{\bf b} ) } \;.
\end{eqnarray}  
We choose $D = 2^{2/3}$ and use this in Eq.~(\ref{eq:edge_kernel}). This gives 
our final result for the scaling behavior of the edge kernel, 
\begin{eqnarray}\label{eq:def_edge_kernel}
K_{\mu}({\bf x},{\bf y}) \approx \frac{1}{w_N^d} {\cal K}^{\rm edge}_d\left(\frac{{\bf x} - {\bf r}_{\rm edge}}{w_N},\frac{{\bf y} - {\bf r}_{\rm edge}}{w_N}\right) \;,
\end{eqnarray}
where the scaling function is given explicitly by
\begin{equation}\label{eq:edge_kernel3}
{\cal K}^{\rm edge}_d({\bf a},{\bf b}) = \int \frac{d^d q}{(2 \pi)^d}  e^{-i {\bf q} \cdot ({\bf a} - {\bf b})  } {\rm Ai}_1\left(2^{\frac{2}{3}} q^2 + \frac{a_n+b_n}{2^{1/3}}\right)\;,
\end{equation}   
while the function ${\rm Ai}_1(z)$ is given by
\beq\label{eq:Wz_Airy}
{\rm Ai}_1(z) =  \int_\Gamma \frac{d \tau}{2\pi i} \frac{1}{\tau} e^{-z \tau + \tau^3/3} = 
\int_z^\infty  {\rm Ai}(u) \, du \;.
\eeq

This edge kernel is a novel result and generalizes the standard Airy kernel in $d=1$ to
higher dimensions. Indeed, putting $d=1$ in Eq. (\ref{eq:edge_kernel3}) we get
\beq\label{eq:airy_kernel1}
{\cal K}^{\rm edge}(a,b) =  \int_{-\infty}^\infty \frac{dq}{2\pi} e^{i q (a-b)} \int_{2^{2/3}q^2+2^{-1/3}(a+b)}^\infty {\rm Ai}(z) \, dz \;. 
\eeq
Making a shift $z = 2^{2/3}q^2 + 2^{-1/3}(a+b) +u$ gives
\beq\label{eq:airy_kernel2}
{\cal K}^{\rm edge}(a,b) =  \int_{-\infty}^\infty \frac{dq}{2\pi} e^{i q (a-b)} \int_{0}^\infty {\rm Ai}(u+2^{2/3}q^2 + 2^{-1/3}(a+b)) \, du \;. 
\eeq
We next use a non-trivial identity involving Airy functions \cite{vallee}
\beq\label{eq:identity_airy}
\int_{-\infty}^\infty \frac{dq}{2 \pi}\, e^{-iq\,(v-v')} \, {\rm Ai}(2^{2/3}q^2 + 2^{-1/3}(v+v')) = 2^{-\frac{2}{3}} {\rm Ai}(v){\rm Ai}(v') \;.
\eeq 
Choosing $v = a + 2^{-2/3} u$ and $v' = b + 2^{-2/3} u$, substituting this identity in Eq. (\ref{eq:airy_kernel2}) and rescaling $u \to 2^{-2/3}u$ gives
\beq\label{eq:airy_kernel3}
{\cal K}^{\rm edge}(a,b) =  \int_0^\infty du \, {\rm Ai}(a+u) \, {\rm Ai}(b+u) \;.
\eeq
Since ${\rm Ai}(z)$ satisfies the differential equation ${\rm Ai}''(z)- z {\rm Ai}(z)=0$ we replace ${\rm Ai}(z)$ by ${\rm Ai}''(z)/z$ in Eq. (\ref{eq:airy_kernel3}). Next we use the identity 
\beq
\frac{1}{(u+a)(u+b)} = \frac{1}{b-a} \left[\frac{1}{u+a} - \frac{1}{u+b} \right] 
\eeq
and integrate by parts. This then reduces Eq. (\ref{eq:airy_kernel3}) to the standard Airy kernel form 
\beq
{\cal K}^{\rm edge}(a,b) = K_{\rm Airy}(a,b) = ({\rm Ai}(a)\,{\rm Ai}'(b) - {\rm Ai}'(a)\,{\rm Ai}(b))/(a-b)  \;.
\eeq

%
%

\section{General $d$-dimensional soft potentials at finite $T$ and universality}\label{sec:Td} 

In this section we finally consider the most general case of arbitrary $d$ and finite $T$.
In addition we show that the results obtained for the harmonic oscillator in the
previous sections can be extended to a very broad class of smooth confining potentials.
We start by generalizing the method based on the Euclidean propagator, introduced in
the previous section, to finite temperature (Section \ref{sec:6a}). To study the 
large $N$ limit, we only need the small time expansion of the propagator, which is
obtained for a general potential in Section \ref{sec:6b}. Using this expansion, we 
obtain results in the bulk in Section \ref{sec:6c} and at the edge of the Fermi gas in 
Section \ref{sec:6d}. Most of our formulas will be valid for arbitrary confining smooth potentials
$V({\bf x})$. For the spherically symmetric potentials, in particular those of the type
\bea
V({\bf x}) = V(r) = V_0 \, \left(\frac{r}{r_0}\right)^p \label{V0} 
\eea 
we will obtain additional explicit results. Note that this parametrization contains
some arbitrariness but final results only depend on $V_0/r_0^p$. 
For convenience we choose 
\be
V_0= \frac{\hbar^2}{2 m r_0^2} \label{V00}
\ee
so that $V_0, r_0$ are respectively of the order of the single particle ground state energy 
and its radius. The harmonic oscillator is recovered for $p=2$, 
$r_0=1/\alpha$ and $V_0= \frac{1}{2} \hbar \omega$. 

\subsection{Representation of the finite $T$ kernel using the Euclidean propagator} \label{sec:6a}

To deal with finite $T$ we need to generalize to arbitrary $d$ the method explained in detail for $d=1$
in Section \ref{sec:1dT}. There we showed that the canonical and grand canonical ensembles lead to the same
results for the $n$-point correlations for large $N$. This allows us to work here in the grand-canonical ensemble, where the
method of the one-body Euclidean propagator can also be applied.

To study the problem at finite $T$, one considers arbitrary excited eigenstates of the $N$ body Hamiltonian, 
$\hat {\cal H}_N$, with arbitrary potential $V({\bf x})$ in Eq. (\ref{H}). These states 
are labeled by a set of occupation numbers $n_{\bf k}=0,1$. As discussed
in section \ref{sec:1dT}, to each excited state one associates a kernel given by
\begin{equation}
K({\bf x},{\bf y}; \{ {n}_{\bf k} \}) = \sum_{\bf k} n_{\bf k} \psi^*_{\bf k}({\bf x}) \psi_{\bf k}({\bf y}) ,
\end{equation}
which generalizes the $d=1$ formula (\ref{def_kernel_canonical}). We recall that the 
$\epsilon_{\bf k}$ and $\psi_{\bf k}$ are the eigenenergies and eigenfunctions of
the one-body Hamiltonian. 
Following the steps presented in Section \ref{sec:1dT} we obtain the generalization of the
set of formula (\ref{chemical})-(\ref{rho_N_grand_canonical}).
In particular the $n$-point correlation functions $R_n$ are
given, in any $d$, by determinants constructed from the
grand-canonical kernel, as in formula (\ref{det_process}). 
This kernel is obtained by performing the average over the $n_{\bf k}$ 
in the grand-canonical ensemble,
leading to
\begin{equation}
 K_{\tilde \mu}({\bf x},{\bf y}):=\langle K({\bf x},{\bf y};  \{ n_{\bf k} \})\rangle = \sum_{\bf k} 
 {1\over 1+ \exp\left(\beta(\epsilon_{\bf k}-{\tilde \mu})\right)}\psi^*_{\bf k}({\bf x}) \psi_{\bf k}({\bf y}) \;,
 \end{equation}
where the chemical potential ${\tilde \mu}$ is related to $N$ and the temperature $T$ via
the relation
\bea \label{Nmutilde} 
N = \sum_{\bf k}  {1\over 1+ \exp\left(\beta(\epsilon_{\bf k}-{\tilde \mu})\right)} \;.
\eea 
We recall that, as in $d=1$, we reserve the notations $\mu$ and $K_{\mu}({\bf x},{\bf y})$ 
for the zero temperature chemical potential and kernel, respectively, while
we denote 
$\tilde \mu$ and $K_{\tilde \mu}({\bf x},{\bf y})$ their finite temperature counterparts.
We recall that, as the temperature goes to zero, $\tilde \mu \to \mu$, but at finite 
temperature, not only the chemical potential $\tilde \mu$ differs from $\mu$,
but also the full kernel functions are different.

A useful representation of the finite temperature kernel can now be derived in terms of the Euclidean propagator
associated to the one-body Hamiltonian (\ref{H}) with arbitrary $V({\bf x})$, defined
in (\ref{itshro}). Using the formula (\ref{K0}) for the zero temperature kernel 
$K_\mu({\bf x},{\bf y})$, one can rewrite
 \begin{equation}
K_{\tilde \mu}({\bf x},{\bf y}) = \int_0^\infty d\mu {\partial K_{\mu}({\bf x},{\bf y})\over \partial \mu}
{1\over 1+ \exp\left(\beta(\mu-{\tilde \mu})\right)}.
\end{equation} 
 Now using Eq. (\ref{laplace_inverse}) we obtain the following representation of the 
 finite temperature kernel
 \begin{equation}
 K_{\tilde \mu}({\bf x},{\bf y}) = 
  \int_0^\infty d\mu' {1\over 1+ \exp\left(\beta(\mu'-{\tilde \mu})\right)}\int_{\Gamma} {dt\over 2\pi \hbar i}
  \exp\left({\mu' t\over \hbar}\right)G({\bf x},{\bf y};t) \;, \label{tkgen}
 \end{equation}
where we have used $\mu'$ as a running integration variable, rather than $\mu$, which in
the remaining will always denote the chemical potential at zero temperature. 
 

\subsection{Small time expansion of the propagator for generic potential} \label{sec:6b}

We have seen in the previous section, for the $d$-dimensional simple harmonic oscillator
at $T=0$, 
that in the limit of large $\mu$ the kernel and density 
can be extracted from the short time behavior of the Euclidean propagator.
As shown below, this remains true even at finite temperature, in the
relevant regimes studied here. This extends the short time expansion analysis 
initiated in \cite{MM88}.
However, for the sake of completeness, we now display the result for a general soft
potential $V({\bf x})$, and provide an independent probabilistic derivation of this expansion
in the Appendix \ref{short_time}.

The expansion of $G({\bf x},{\bf y};t) $ for a soft potential is, to ${\mathcal O}(t^3)$, given by
\begin{equation}
G({\bf x},{\bf y};t) \sim \left(m\over 2\pi \hbar t\right)^{d\over 2}\exp\left[-{m\over 2\hbar t}({
\bf x}-{\bf y})^2\right]\exp\left[-{t\over \hbar}S_1({\bf x},{\bf y})  -{t^2\over 2m}S_2({\bf x},{\bf y}) + {t^3\over 2m\hbar}S_3({\bf x},{\bf y})\right],\label{expt3}
\end{equation}
where
\begin{eqnarray}
&&S_1({\bf x},{\bf y}) = \int_0^1du\   V({\bf x}+ u({\bf y}-{\bf x})),\label{S1}\\
&&S_2({\bf x},{\bf y}) =\int_0^1du\  u(1-u)  \nabla^2V({\bf x}+ u({\bf y}-{\bf x}))\label{S2},\\
&&S_3({\bf x},{\bf y}) \nonumber = \int_0^1du\int_0^1dw\ \left[
{\rm min}(u,w)-uw\right] 
\nabla V({\bf x}+ u({\bf y}-{\bf x}))\cdot\nabla V({\bf x}+ w({\bf y}-{\bf x}))\\ &&-{\hbar^2\over 4m} \int_0^1du \ u^2(1-u)^2 \nabla^2\nabla^2 V({\bf x}+ u({\bf y}-{\bf x})).\label{S3} 
\end{eqnarray}
With some work it can be checked that the above result agrees with the short time expansion given in Eq. (\ref{expsho}) when applied to the simple harmonic oscillator. We see that the occurrence of derivatives of first and higher order in $V$,  resulting from the expansion of the term $V( {\bf x}{(1-u)} + {\bf y}{u} + \sqrt{D_0 t}{\B}_u)$
in Eq. (\ref{repbb}) -- where ${\B}_u$ denotes the $d$-dimensional Brownian bridge -- requires that the potential is sufficiently smooth within the neighborhood of the direct path between ${\bf x}$ and ${\bf y}$.
At coinciding points ${\bf x}={\bf y}$ (the case where one computes the density), one finds
\begin{eqnarray}
&&S_1({\bf x},{\bf x}) = V({\bf x}),\label{s1xx}\\
&&S_2({\bf x},{\bf x}) ={1\over 6}   \nabla^2V({\bf x}),\\
&&S_3({\bf x},{\bf x}) = {1\over 12}\left[\nabla V({\bf x})\right]^2-{\hbar^2\over 120 m}  \nabla^2\nabla^2 V({\bf x}).
\end{eqnarray}

Using these results we now analyze successively the bulk and the edge regimes.

\subsection{Results in the bulk} \label{sec:6c}

\subsubsection{Zero-temperature}

The calculation of the bulk density and kernel follows exactly those of  the simple harmonic 
oscillator in subsections \ref{shobd} and \ref{shobk}. Here only the term to ${\cal O}(t)$ is retained 
and one obtains
\begin{eqnarray}
K_\mu({\bf x},{\bf y})&=&\theta\left[\mu - S_1({\bf x},{\bf y})\right]\left({[\mu -S_1({\bf x},{\bf y})]m
\over 2 \pi^2 \hbar^2 ({\bf x}-{\bf y})^2}\right)^{d\over 4}  J_{d\over 2}\left(\sqrt{2m({\bf x}-{\bf y})^2[\mu -S_1({\bf x},{\bf y})]\over \hbar^2}\right) \;.\label{eq:kernel_gen}
\end{eqnarray}

Let us first discuss the normalized particle density is given by $\rho_N({\bf x}) = K_\mu({\bf x},{\bf x})/N$.
Again we see that its support $S_1({\bf x},{\bf x}) = V({\bf x}) < \mu$ is finite and,
using Eq. (\ref{s1xx}) is obtained as
\begin{equation}
\rho_N({\bf x}) = \left(m\over 2\hbar^2 \pi\right)^{d\over 2}{\theta\left[\mu - V({\bf x})\right]\over N \Gamma(1+{d\over 2})}\left[\mu -V({\bf x})\right]^{d\over 2}.\label{eq:global}
\end{equation}
From this formula, by integrating over ${\bf x}$, one can calculate the Fermi energy (or temperature) $T_F=\mu$, from the condition $\int d{\bf x} \rho_N({\bf x})=1$.

For explicit calculations, let us focus on the case of spherically symmetric potential of the form 
(\ref{V0}) with $p>0$. The density vanishes at the edge at $r=r_{\rm edge}$
such that
\bea \label{redge2} 
V(r_{\rm edge})= \mu  \Longleftrightarrow  r_{\rm edge}= r_0 \left(\frac{\mu}{V_0}\right)^{1/p} \;.
\eea 
For the harmonic oscillator in $d=1$ one recovers $r_{\rm edge} = \sqrt{2 N}/\alpha$
using that in that case $\mu=N \hbar \omega$. In the case of general $p$,
we obtain
\bea\label{TF}
T_F = \mu = a_{p,d} V_0 N^{\frac{2 p}{d(p+2)}} 
\eea 
with $a_{p,d}= (4 \pi)^{\frac{p}{2+p}} [p \, \Gamma(1+d/2)/(S_d {\rm B}(1+d/2,d/p))]^{2p/(d(p+2))}$
where $S_d=2 \pi^{d/2}/\Gamma(d/2)$ and ${\rm B}(p,q) = \int_0^1 u^{p-1}(1-u)^{q-1} du$ and where we used (\ref{V00}). Consequently one obtains that
\be 
r_{\rm edge} \sim N^{2/(d(p+2))} \label{redgep}
\ee
for large $N$. 

Consider now the kernel given in Eq. (\ref{eq:kernel_gen}). As in the case of the simple harmonic oscillator, we consider two generic points ${\bf x}$ and ${\bf y}$ in the bulk (far from the edges) close to each other, with a separation of order $|{\bf x}- {\bf y}| \sim [N \rho_N({\bf x})]^{-1/d}$.
Eq. (\ref{eq:kernel_gen}) then simplifies to the scaling form 
\begin{eqnarray}\label{scaling_bulk_gen}
K_{\mu}({\bf x},{\bf y}) \approx \ell^{-d}({\bf x}) {\cal K}_d^{\rm bulk}(|{\bf x}-{\bf y}|/\ell({\bf x}))
\end{eqnarray}
where $\ell({\bf x}) = [N \rho_N({\bf x})\gamma_d]^{-1/d}$ is the typical separation between fermions in the bulk, {$\gamma_d =  \pi^{d/2} [\Gamma(d/2+1)]$} and ${\cal K}_d^{\rm bulk}$ is the same scaling function, 
as given in Eq. ({\ref{bksho}), respectively, for the harmonic oscillator. The dependence on the potential $V({\bf x})$ thus enters only through the local density $\rho_N({\bf x})$ via  the scale factor $\ell({\bf x})$. However, the scaling function associated with the bulk kernel in Eq. (\ref{eq:kernel_bulk2}) is completely universal for all $V({\bf x})$.  

The above result is the general form of the  LDA
\cite{GPS08}, which is normally obtained from semi-classical or physical arguments. The range of validity of this approximation can, as was the case for the simple harmonic oscillator, be established by examining the corrections due to the quadratic and cubic terms 
in $t$ in the short time expansion of the propagator. Here we define the two point energy function
\begin{equation}
\epsilon_2({\bf x},{\bf y}) = \mu - S_1({\bf x},{\bf y}),
\end{equation}
the range of validity of the LDA approximation is thus given by
\begin{equation}
\epsilon_2({\bf x},{\bf y}) \gg  \left[\hbar^2 \frac{S_2({\bf x},{\bf y})}{m}\right]^{\frac{1}{2}}
\end{equation}
and 
\begin{equation}
\epsilon_2({\bf x},{\bf y}) \gg  \left[\hbar^2 \frac{ S_3({\bf x},{\bf y})}{m}\right]^{\frac{1}{3}}.
\end{equation}
If we consider a trapped system with a potential which grows in a polynomial fashion we see, from power counting, that  for large values of ${\bf x}$ and ${\bf y}$, the second of the 
above inequalities determines the validity of the LDA.

\subsubsection{Bulk statistics at finite temperature and $T \sim T_F$}

At zero temperature there is a unique length scale associated 
to the quantum fluctuation in the confining potential, denoted by $r_0$ for
the class of potentials (\ref{V0}) above. By contrast, at finite temperature, for
a generic confining potential, there are two natural length
scales in the problem. The first one is the thermal de Broglie wave-length 
\bea
\lambda_T = \hbar \sqrt{\frac{2\pi}{m T}} \label{dBlength} 
\eea
of the free fermions. The second length is set by the temperature and the
confining potential and is purely classical. In the case of a power law spherically
symmetric potential (\ref{V0}) it is given by
\bea
\beta V(r_T) = 1 \Leftrightarrow r_T = r_0 \left(\frac{T}{V_0}\right)^{1/p}  \;. \label{rT} 
\eea 

Now let us discuss the energy scales involved. At zero temperature the natural
energy scale is the Fermi energy $\mu =T_F$ defined and discussed
in the previous section, where we showed that $\mu$ is large for large $N$,
i.e. $\mu \sim N^{2p/(d(p+2))}$ for the class of potentials defined in (\ref{V0}). 
Significant changes from the $T=0$ properties are expected in the bulk only 
when $T \sim T_F$. In what follows we will focus on this bulk regime 
and consider $T \sim T_F=\mu$. Consequently the fugacity which we denote as
\be
\zeta = \exp(\beta \tilde \mu) \label{zeta} 
\ee
is also of $\mathcal{O}(1)$ when $N$ is large. 
Introducing the variable $u=\beta \mu'$ in Eq. (\ref{tkgen}) and making the change of variable 
$t = \tau \beta \hbar$ we find 
 \begin{equation}
 K_{\tilde \mu}({\bf x},{\bf y})
 =  \int_0^\infty du{1\over 1+ \zeta^{-1}\exp\left(u\right)}\int_{\Gamma} {d\tau\over 2\pi i} \exp(u\tau)G({\bf x},{\bf y};\beta \hbar\tau) \;,\label{tkgen2}
 \end{equation}
which at this stage is still an exact equation.

We now analyze this equation in the large $N$ limit and the regime $T \sim T_F$. This means that
$\beta$ is small, hence the time variable $t$ in the above propagator 
$G({\bf x},{\bf y}; t=\beta \hbar\tau)$ is also small. Hence we can use the
short time expansion of the Euclidean propagator, as was done for the
harmonic oscillator case. As discussed, there to describe the bulk, one
only needs to keep the first term of order ${\cal O}(t)$ and one obtains 
\begin{equation}
K_{\tilde \mu}({\bf x},{\bf y})
 =  \int_0^\infty du{1\over 1+ \zeta^{-1}\exp\left(u\right)}\int_{\Gamma} {d\tau\over 2\pi i}  
 \left(m\over 2\pi\tau \beta \hbar^2\right)^{d\over 2}
\exp\left[- \frac{\pi}{\tau} \frac{({\bf x}-{\bf y})^2}{\lambda_T^2}  
 + \tau u -\tau \beta S_1({\bf x}, {\bf y})) \right] \label{205}
 \end{equation}
 where $\lambda_T$ is given in (\ref{dBlength}), $\zeta$ by (\ref{zeta}) where
 $\tilde \mu$ is fixed as a function of $N$ by Eq. (\ref{Nmutilde}). 

To obtain the density, we start by considering the kernel at coinciding points
\begin{equation}
K_{\tilde \mu}({\bf x},{\bf x})
 =  \int_0^\infty du{1\over 1+ \zeta^{-1}\exp\left(u\right)}\int_{\Gamma} {d\tau\over 2\pi i}  \left(m\over 2\pi\tau \beta \hbar^2\right)^{d\over 2}
 \exp(\tau u -\tau \beta V({\bf x})).
 \end{equation}
 The appropriate bulk scaling is thus to choose ${\bf x}$ such that $\beta V({\bf x}) \sim 
 \mathcal{O}(1)$. Performing the explicit integral over the Bromwich contour using
(\ref{eq:besseltheta}) we obtain
 \begin{equation}
 K_{\tilde \mu}({\bf x},{\bf x}) =
\left(m\over 2\pi \beta \hbar^2\right)^{d\over 2} \int_0^\infty du{1\over 1+ \zeta^{-1}\exp\left(u\right)}{\theta\left(u-\beta V({\bf x})\right)\left(u-\beta V({\bf x}))\right)^{{d\over 2}-1}\over \Gamma({d\over 2})}.
 \end{equation}
This can then be rewritten, shifting the integral over $u$, and leads to our main result for
the density
\bea
\rho_N({\bf x}) = \frac{1}{N} K_{\tilde \mu}({\bf x},{\bf x}) =
{1\over N \Gamma({d\over 2})\lambda_T^d} \int_0^\infty dq{q^{{d\over 2}-1}\over 1
+ \zeta^{-1}\exp\left(q+ \beta V({\bf x}) \right)}
 =
 - \frac{1}{N \lambda_T^d} \; {\rm Li}_{d/2}(- \zeta e^{- \beta V({\bf x})})  \label{gdenst} 
\eea
where $\lambda_T$ is the thermal wavelength of the free fermions
given in (\ref{dBlength}) and ${\rm Li}_n(x) = \sum_{k=1}^\infty x^k/k^n$ is the polylogarithm function. 

Let us provide an explicit example in the case of the power-law potentials (\ref{V0}). 
We define the scaling variables, which generalize the $1d$ result given in Eqs. (\ref{def_yz}) and (\ref{def_yz2})
\bea
y = \left( \frac{T_F}{T}\right)^d \quad , \quad z = r/r_T \label{Tf_d} \;,
\eea 
where $T_F$ is given in Eq. (\ref{TF}) and $r_T$ is the classical thermal confining length introduced in Eq.~(\ref{rT}).
The explicit calculation predicts that the density is given by eliminating
the fugacity $\zeta$ in the following pair of equations
\bea
&& \rho_N({\bf x}) = - \frac{B_2}{\sqrt{y} N^{\frac{2}{p+2}}} {\rm Li}_{\frac{d}{2}}(- \zeta e^{-z^p}) \label{power1}  \\
&& 1 = - \frac{B_1}{y^{\frac{p+2}{2 p}}}   {\rm Li}_{\frac{d}{2} + \frac{d}{p}}(- \zeta) \label{power2} 
\eea 
where $B_2 = (4\pi)^{-d/2} r_0^{-d} a_{p,d}^{d/2}$, $a_{p,d}$ is given below Eq. (\ref{TF}) and $B_1 = \Gamma(1+d/2+d/p)$. 
Using the identity 
\bea
\int_0^{+\infty} dr \, r^a {\rm Li}_b(- \zeta e^{-r^p}) = \frac{1}{p} \Gamma\left(\frac{a+1}{p}\right) {\rm Li}_{\frac{1+a+b p}{p}}(-\zeta)\;,
\eea 
we can check that the Eq. (\ref{power2}) is the normalization condition $\int d^d x \rho_N({\bf x}) =1$
of the density given by Eq. (\ref{power1}). Setting $p=2$ in the above formula gives the result for
the $d$-dimensional harmonic oscillator. In particular in $d=1$, using that ${\rm Li}_1(-\zeta)=-\ln(1+\zeta)$,
the implicit equation can be solved and one recovers the bulk scaling function for the density presented in Eqs. (\ref{densityT}) and (\ref{scaling_function}) in Section \ref{sec:1dTphys}.


The above result 
agrees with the LDA approximation which we briefly recall here for completeness. In the LDA, the local (unnormalized density) in position  and momentum is given by the 
Fermi-factor
\begin{equation}
n({\bf x},{\bf p}) = {1\over1+ \exp\left(\beta E({\bf x},{\bf p})-\beta \tilde \mu\right)}
\end{equation}
where $E({\bf x},{\bf p})$ is given by the classical energy 
\begin{equation}
E({\bf x},{\bf p}) = {{\bf p^2}\over 2m} + V({\bf x}).
\end{equation}
Integration over the momentum with the appropriate Planck volume normalization then gives
the local density in space as
\begin{equation}
N \rho_N({\bf x}) = K_{\tilde \mu}({\bf x},{\bf x}) = {1\over h^d}\int{d{\bf p}\over 1+ \exp\left(\beta E({\bf p},{\bf x})-\beta \tilde \mu\right)}.\label{kxxlda}
\end{equation}
Carrying out the angular integration in (\ref{kxxlda}) yields
\begin{equation}
 \rho_N({\bf x}) = {2\pi^{d\over 2}\over N \Gamma\left({d\over 2}\right) h^d}\int_0^\infty {dp \ p^{d-1}\over 1+ \exp\left(\beta E(p,{\bf x})-\beta \tilde \mu\right)} \;. \label{kxxxlda}
\end{equation}
Finally making the substitution $q= \beta p^2/2m$ shows that the LDA approximation Eq. (\ref{kxxxlda}) agrees with Eq. (\ref{gdenst}).

The formula for the bulk kernel at two unequal points is obtained by analyzing Eq. 
(\ref{205}) in the regime of separation $|{\bf x} - {\bf y}| \sim \lambda_T$, and of temperature such that $\beta V({\bf x}) = {\cal O}(1)$. In this limit, by expanding $V$ in the
integrand of \eqref{S1}, we can approximate $S_1({\bf x},{\bf y})$
by its leading term $V({\bf x})$, using that $\beta \lambda_T |\nabla V({\bf x})| \ll 1$
(always valid inside the bulk in this temperature regime, as can be checked explicitly). 
Replacing $S_1({\bf x},{\bf y})$ by $V({\bf x})$ in (\ref{205}) and using the change of
variable $u - \beta V({\bf x})=q$ and the integral representation of the Bessel function
in \eqref{eq:bessel}, we arrive at
\begin{equation}\label{KTd_bulk}
K_{\tilde \mu}({\bf x},{\bf y}) = {1\over \pi^{d-2\over 4}\lambda^d_T}\left(\lambda_T\over |{\bf x}-{\bf y}| \right)^{{d\over 2}-1}\int_0^\infty dq {q^{d-2\over 4}\over 1 + \zeta^{-1}\exp\left(q + \beta V({\bf x}) \right)}
J_{{d\over 2}-1}\left( 2\sqrt{q\pi}{|{\bf x}-{\bf y}|\over \lambda_T}\right) \;.
\end{equation}
This formula crosses over to the zero temperature kernel (\ref{eq:kernel_bulk2}) 
when $\lambda_T \gg \ell({\bf x})$ where $\ell({\bf x})$ is the typical separation
between particles defined in (\ref{lx}). This can be seen by performing a 
change of variable $q \to \beta q$, such that the Fermi factor becomes a Heaviside step function.
Thus in practice the above formula is valid in the range of separation $|{\bf x} - {\bf y}| \sim 
\min(\lambda_T,\ell({\bf x}))$. Beyond this scale the kernel decays to zero. 
Note that in $d=1$ and $d=3$ the formula simplifies slightly, since one has respectively $J_{-{1\over 2}}(x) = \sqrt{2/\pi x}\ \cos(x)$ and $J_{{1\over 2}}(x) = \sqrt{2/\pi x}\ \sin(x)$. 

%
%
%
%

\subsection{Results at the edge} \label{sec:6d}

In the edge region we expect the effects of fluctuations to be larger than in the bulk. To study the
bulk we had to scale the temperature as $T \sim T_F$. By contrast, in this section,
the relevant regime will involve lower temperature, $T \ll T_F$, hence everywhere
the variable $\beta \mu$ will be considered to be large. To estimate the 
chemical potential $\tilde \mu$ in this range of temperature, we can thus
use the Sommerfeld expansion valid for large $\beta \mu \gg 1$ \cite{Mahan}
\bea
\mu - \tilde \mu = \frac{\pi^2}{6} \frac{1}{\beta^2} \frac{\tilde \rho'(\mu)}{\tilde \rho(\mu)} 
+ {\cal O}\left(\frac{1}{(\beta \mu)^4}\right)
\eea
where we denote by $\tilde \rho(\epsilon) = \sum_{\bf k} \delta(\epsilon - \epsilon_{\bf k})$
the density of states (in energy). Note that for the harmonic oscillator in $d=1$ the
density of states is constant, and the corrections are exponentially small.

In this section we consider the kernel for a pair of points ${\bf x}, {\bf y}$ both located near
a point ${\bf r}_{\rm edge}$ at the edge, at finite temperature, in arbitrary $d$ and
for the large class of confining potential $V(r)$ described at the beginning of Section \ref{sec:Td}.
To proceed, we set
\bea \label{para} 
{\bf x} = {\bf r}_{\rm edge} + {\bf a}' \quad , \quad {\bf y} = {\bf r}_{\rm edge} + {\bf b}'
\eea
where $|{\bf r}_{\rm edge}|=r_{\rm edge}$ is defined in (\ref{redge2})
and we assume $|{\bf{a}}'|, |{\bf{b}}'| \ll {r}_{\rm edge}$. The goal of this section
is to show that the properly centered and scaled edge kernel becomes universal, i.e., 
independent of the details of the potential. 

Substituting (\ref{para}) in Eqs. (\ref{S1})-(\ref{S3}) and expanding the terms $S_1, S_2$ and $S_3$, 
in a gradient expansion,
we obtain
 \bea \label{exp1}
&& S_1  = V({\bf r}_{\rm edge}) + \frac{1}{2} {\bf \nabla} V({\bf r}_{\rm edge}) \cdot ({\bf a}'+{\bf b}') + \ldots \\
&& \label{exp2} S_2 = \frac{1}{6} \nabla^2 V({\bf r}_{\rm edge}) 
+ \frac{1}{12} {\bf \nabla} [\nabla^2 V({\bf r}_{\rm edge}) ] \cdot ({\bf a}'+{\bf b}') + \ldots  \\
&& S_3 = {1\over 12}\left[\nabla V({\bf r}_{\rm edge})\right]^2-{\hbar^2\over 120 m}  \nabla^2\nabla^2 V({\bf r}_{\rm edge})+ \ldots \label{exp3}
\eea 
Note that for potentials increasing as $V(r) \sim r^p$ for large $r$, each derivative brings 
an additional factor $1/r_{\rm edge}$ in the expansion, hence it is not necessary 
to keep higher order terms, at least for the potentials of this class. 

Let us start with Eq. (\ref{tkgen}) and substitute in it the short time expansion (\ref{S3}). We will justify a posteriori under what conditions the short time expansion can be stopped at order ${\cal O}(t^3)$. We make a 
change of variable $\beta(\mu'- \tilde \mu) = - b u$ in Eq. (\ref{tkgen}), where for convenience we have introduced 
a dimensionless parameter $b$ whose value will be chosen later. 
Setting ${\bf x}$ and ${\bf y}$ both close to ${\bf r}_{\rm edge}$, as in (\ref{para}), and using
(\ref{exp1})-(\ref{exp3}) 
we obtain
 \begin{eqnarray} \label{KedgeT} 
&& K_{\tilde \mu}({\bf x}, {\bf y})\simeq {b\over \beta} \int_{-\infty}^{\infty} {du\over 1+ \exp(-bu)}\int_\Gamma {dt\over 2\pi \hbar i} 
\left(m\over 2\pi \hbar t\right)^{d\over 2} \nonumber \\ 
&& \times  \exp\left[-{m\over 2\hbar t}\left( 
{\bf{a}}'-{\bf{b}}'\right)^2\right]\exp\left[   - \frac{u b t}{\beta \hbar} 
 -{t\over 2\hbar}|\nabla V({\bf r}_{\rm edge})|(a'_n+b'_n) +{t^3\over 24m\hbar}|\nabla V({\bf r}_{\rm edge})|^2\right], 
 \end{eqnarray}
where the upper bound $\tilde \mu \beta/b$ has been replaced by $+\infty$ since we are studying the
limit of large $\tilde \mu$ (large $N$). In deriving this equation we have kept both terms of $S_1$, but neglected 
$S_2$ and the second term in $S_3$ in Eqs. (\ref{exp1})-(\ref{exp3}), which will be justified later. Here $a'_n = {\bf a}' \cdot {\bf r}_{\rm edge}/r_{\rm edge}$ and $b'_n = {\bf b}' \cdot {\bf r}_{\rm edge}/{r}_{\rm edge}$ are projections of ${\bf a}'$ and ${\bf b}'$ in the radial direction as in Section \ref{sec:5c}. The term linear in time,
$\frac{t}{\hbar} (\tilde \mu - V({\bf r}_{\rm edge})) = \frac{t}{\hbar} (\tilde \mu - \mu)
\sim \frac{t}{\hbar \beta^2 \mu}$ 
has been set to zero, using that $\tilde \mu$ is very close to $\mu$ in this
temperature regime as discussed above (we also assume that $t$ is small, as justified below). 
 
 
 
Following the analysis of the harmonic oscillator at zero temperature, we then introduce the scaled dimensionless vectors $\bf{a}$ and ${\bf{b}}$ defined via ${\bf{a}}'= w_N\,{\bf{a}}$ and ${\bf{b}}'= w_N\,{\bf{b}}$, where the width $w_N$ has the dimension of length, and is determined as follows. We impose that both the second (of order ${\cal O}(t)$)
and third term (of order ${\cal O}(t^3)$) in (\ref{KedgeT}) are of order unity. This determines both $w_N$ and the typical time scale
$t_N$ as
 \begin{equation}\label{def_xi}
 w_N = {|\nabla V({\bf r}_{\rm edge})|^{-1/3}\hbar^{2\over 3}\over (2m)^{1\over 3}} 
 \quad , \quad t_N = |\nabla V({\bf r}_{\rm edge})|^{-2/3} (8m\hbar)^{1\over 3}
 \end{equation}
 where the amplitudes have been chosen for later convenience. With this choice
 it is clear that the diffusion term, 
 ${m\over 2\hbar t}\left({\bf{a}}'-{\bf{b}}'\right)^2 \sim w_N^2/t_N \sim {\cal O}(1)$ 
 is also of order unity. Finally, for the term containing temperature to be also of $\mathcal{O}(1)$ and for the parameter $b$ to be also of order unity, we must scale the temperature as a function of $N$ as follows
 \begin{equation} \label{bbeta} 
 \beta = {b t_N\over 2^{2\over 3}\hbar} =   {b (2m)^{1\over 3} \over \hbar^{2\over 3} |\nabla V({\bf r}_{\rm edge})|^{2\over 3}} \;.
 \end{equation}
Specializing to the harmonic oscillator $V(r) = \frac{1}{2} m \omega^2 r^2$, we obtain
\bea
 \beta = {b (2m)^{1\over 3} \over \hbar^{2\over 3} (m\omega^2 r_{\rm edge})^{2/3}} =  \frac{b}{\hbar\omega} \frac{1}{(\Gamma(1+d) N)^{1/(3 d)} } \;.
\eea 
In particular in $d=1$,
 \begin{equation}
 \beta = \frac{b}{\hbar\omega N^{1/3}} 
 \end{equation}
 in agreement with the result of Section \ref{shotedge} [see Eq. (\ref{def_b_1})]. In addition, the above formula (\ref{def_xi}) for the parameter
 $w_N$ when applied to the harmonic oscillator yields back the expression in Eq.~(\ref{wN}) in any $d$.
 Note that for more general potentials $V(r) \sim r^p$ one finds from \eqref{bbeta} 
that $\beta \sim b/T_{\rm edge}$ where the temperature scale which controls
the thermal fluctuations at the edge is 
\bea \label{Tedge} 
T_{\rm edge} \sim N^{\frac{4 (p-1)}{3 d (p+2)}}
\eea
which generalizes the scale $T_{\rm edge} \sim N^{1/3}$ for $p=2$ and $d=1$. 

Defining $\tau = t/t_N$ we can now rewrite
 \begin{equation}
 K_{\tilde \mu}({\bf x},{\bf y}) \simeq {1\over C_d w^d_N}\int_{-\infty}^\infty {du\over \exp(-bu)+1}\int_\Gamma {d\tau\over 2\pi i} {1\over \tau^{d\over 2}}
 \exp\left(-{({\bf a}-{\bf b})^2\over 2^{8\over 3}\tau}-\tau ({a_n+b_n+ 2u\over 2^{1\over 3}}) 
 + {\tau^3\over 3}\right) \;,
 \end{equation}
where $C_d = \pi^{d/2} 2^{(4 d-2)/3}$. Using the integral representation of the diffusive propagator in Eq.~(\ref{eq:diffusion}) and the one of the 
Airy function in Eq.~(\ref{eq:airy_integral}) we obtain the following scaling form for the edge
kernel at finite temperature in arbitrary dimension $d$ for two points near the edge
 \bea \label{kernelTVedge0} 
 && K_{\tilde \mu}({\bf x},{\bf y})  \simeq \frac{1}{w_N^d} {\cal K}^{{\rm edge}}_{d,b}\left(\frac{ {\bf x} - {\bf r}_{\rm edge} }{w_N}, 
 \frac{ {\bf y} - {\bf r}_{\rm edge} }{w_N}\right) 
 \eea
where we recall that $w_N = |\nabla V({\bf r}_{\rm edge})|^{-\frac{1}{3}}\hbar^{2\over 3} (2m)^{-{1\over 3}}$, and the scaling function
is given by
 \bea \label{scaling_kernelTVedge0} 
 && {\cal K}^{{\rm edge}}_{d,b}({\bf a}, {\bf b}) = 2^{2\over 3}\int_{-\infty}^\infty {du\over \exp(-bu)+1}
 \int \frac{d{\bf q}}{(2 \pi)^d}  e^{-i {\bf q} \cdot ({\bf a} - {\bf b})  } {\rm Ai}\left(2^{\frac{2}{3}} q^2 + \frac{a_n+b_n+2u}{2^{1/3}}\right) \;,
 \eea
which depends on a single dimensionless parameter 
$b =     \hbar^{2\over 3} |\nabla V({\bf r}_{\rm edge})|^{2\over 3} (2 m)^{-\frac{1}{3}}/T$ and the dimension of space $d$.
Note that this result is independent of the precise shape of the potential (within the broad
 class to be discussed below) and is thus universal. The dependence on
 the potential enters only in the width parameter $w_N$, and the
 dimensionless inverse temperature $b$. 

This representation can be expressed in an alternative form. We first
 split ${\bf q} = {\bf q}_t + q_n {\bf n}$ where ${\bf n}={\bf r}/r$ is the direction normal to the edge,
 and we similarly split ${\bf a} = {\bf a}_t + a_n {\bf n}$ and ${\bf b} = {\bf b}_t + b_n {\bf n}$. We then
 carry out the integral over $q_n$ using the identity (\ref{eq:identity_airy}). We obtain the 
 following alternative result for the scaling function of the edge kernel at finite temperature in space dimension $d$ 
 as 
 \bea
 {\cal K}^{{\rm edge}}_{d,b}({\bf a}, {\bf b}) &=& \int_{-\infty}^\infty {du\over \exp(-bu)+1}
 \int \frac{d{\bf q}_t}{(2 \pi)^{d-1}}  e^{-i {\bf q}_t \cdot ({\bf a}_t - {\bf b}_t)  } {\rm Ai}\left(
 a_n +{\bf q}_t^2 + u \right){\rm Ai}\left(
 b_n +{\bf q}_t^2 + u \right),  \label{kernelTVedge} \\
 & =&  \int \frac{d{\bf q}_t}{(2 \pi)^{d-1}}  e^{-i {\bf q}_t \cdot ({\bf a}_t - {\bf b}_t)  }  
 {\cal K}^{{\rm edge}}_b(a_n +{\bf q}_t^2,b_n +{\bf q}_t^2) \label{edgeKd2} 
 \eea
 where ${\cal K}^{{\rm edge}}_b$ is precisely the scaled edge kernel at finite temperature in
 $d=1$, given in Eq. (\ref{kff}). 
 
 
 {\it Average edge density at finite temperature}. The edge density is obtained simply by setting ${\bf a} = {\bf b}$ in Eq. (\ref{edgeKd2}).
This leads to
\bea
N \rho_N({\bf x}) = \frac{1}{w_N^d} F^{{\rm edge}}_{d,b}\left(\frac{r - r_{\rm edge} }{w_N}\right) 
\eea 
where the scaling function in dimension $d$ can be obtained in terms of the one 
in $d=1$ as
\bea \label{Fdb} 
F^{{\rm edge}}_{d,b}(z) = \int \frac{d{\bf q}_t}{(2 \pi)^{d-1}}  F^{{\rm edge}}_{1,b}(z +{\bf q}_t^2)
\eea
where the scaling function in $d=1$
is given in Eq. (\ref{F1bs}), which we recall here for convenience
 \bea
F_{1,b}(s) = \int_{-\infty}^{+\infty} du \frac{{\rm Ai}(s+u)^2}{1 + e^{-b u}} \;.
\eea 
Note that for $d=1$ there is no integration over ${\bf q}_t$ to perform and (\ref{Fdb}) reduces to an identity.
For $d>1$ one can perform an angular integration to obtain
\bea
F^{{\rm edge}}_{d,b}(z)  = \frac{2^{2-d} \pi^{\frac{1-d}{2}}}{\Gamma(\frac{d-1}{2})} \int_0^{+\infty} dq \, q^{d-2} \, F^{{\rm edge}}_{1,b}(z + q^2) \;.
\eea

 \medskip
 
{\it Zero temperature limit}. In the limit $T \to 0$, i.e. $b \to +\infty$, the Fermi factor converges to a
 Heaviside step function
 \bea
 {1\over \exp(-bu)+1} \to \theta(u) 
 \eea 
 and $\tilde \mu \to \mu$. The first representation of the finite temperature kernel in Eq.~(\ref{scaling_kernelTVedge0}) then
 recovers exactly the kernel obtained for the harmonic oscillator and given in 
(\ref{eq:edge_kernel3}). The second representation (\ref{kernelTVedge}) leads to the alternative form for
the zero temperature edge kernel in terms of the Airy kernel (\ref{airy_kernel.1}) as
  \begin{equation}
   {\cal K}^{\rm edge}_d({\bf a}, {\bf b}) = 
  \int \frac{d{\bf q}_t}{(2 \pi)^d}  e^{-i {\bf q}_t \cdot ({\bf a}_t - {\bf b}_t)  } K_{\rm Airy} \left(
 a_n +{\bf q}_t^2, 
 b_n +{\bf q}_t^2  \right) \;.  \label{kernel0Vedge} 
 \end{equation}
 This thus generalizes to any $d$ the result given in Eq.~(\ref{airy_kernel.1}) for $d=1$, i.e. the standard Airy kernel, and
 coincides, in an equivalent alternative form, with the harmonic oscillator result given in Eq. (\ref{eq:edge_kernel3}).


\medskip

{\it Validity of the method}. 
We now return to the question on the range of validity of this universal edge kernel at finite temperature.
In the derivation we essentially made two approximations: (i) a short time expansion keeping 
terms only up to order ${\cal O}(t^3)$, (ii) a gradient expansion of the potential, assuming that higher
order terms are subdominant for large $N$. To examine the validity of these two points, let us first
focus on the potentials of the form $V(r)\sim r^p$ (with $p>0$). In this case we know from 
\eqref{redgep} that $r_{\rm edge} \sim N^{\frac{2}{d(p+2)}}$.
It follows from (\ref{def_xi}), since $|\nabla V| \sim r_{\rm edge}^{p-1}$, that 
\bea
 w_N \sim N^{-\frac{2}{3d}(p-1)/(p+2)} \quad , \quad t_N \sim w_N^2 \;.
\eea 
Furthermore one can check that for $p>0$ all the neglected terms are
indeed subdominant for large $N$ (see Appendix~\ref{short_time}). For instance 
the leading term ${\cal O}(t^2)$, from Eq.~(\ref{exp2}), scales as $\sim \frac{t^2}{m} \nabla^2 V
\sim t_N^2 r_{\rm edge}^{p-2} \sim N^{-2/(3 d)}$ which is negligible compared to
the main ${\cal O}(1)$ terms at large $N$. Similarly it is easy to check that the neglected second term in $S_3$ in Eq.~(\ref{S3}) is indeed small since $\frac{\hbar^2 \nabla^4 V}{m |\nabla V|^2} \sim 
r_{\rm edge}^{-p-2} \sim N^{-2/d}$. Finally the gradient expansion is
controlled by the parameter $w_N/r_{\rm edge} \sim N^{- 2(p+4)/(3 d(p+2))}$ 
which is small for all fixed $p$ at large $N$. Our conclusion is thus that
all spherically symmetric polynomial potentials with leading degree $p > 0$ do satisfy the validity
criteria for our universal results at the edge. 

We expect that the class of such potentials is actually much broader. The precise conditions
are analyzed in the Appendix~\ref{short_time}. However there
are well identified cases where this universality breaks down, 
for instance for wall type potentials. One example is a box with an infinite hard wall. Another example is the
potential $V(r)= \frac{1}{r^2}+r^2$ potential. At zero temperature the latter is known to
be related to Wishart matrices, which have different edge properties (close the
origin) than the GUE ensemble. In this case the limiting kernel is the so-called Bessel kernel \cite{NM_interface,TW_Bessel}.}

In the case of the potential $V(r) \sim r^p$ with $0< p < 1$, some additional peculiarities
arise. First one finds, from the above
estimates, that the typical width of the edge region, $w_N$, {\it increases} with increasing $N$.
In addition, the temperature scale 
$T_{\rm edge}$ defined in \eqref{Tedge} {\it decreases} with increasing $N$
which is consistent with the fact that the potential is rather shallow so even a
small temperature is sufficient to excite the system. Nevertheless, as discussed in the
Appendix~\ref{short_time}, the conditions for the universality class of the $p=2$ case studied in this paper, seem to still
hold, at least at a perturbative level. To assess more precisely the validity of 
this statement for $0<p<1$ would require further studies.



\section{Discussion and open problems}\label{sec:conclusion}

\subsection{Bosonization in the bulk and interactions: beyond the Sine kernel} 


In the bulk of the Fermi gas, the density $N \rho_N(x)$ varies very slowly compared
to the typical spacing between fermions and other, more standard, methods can be
applied. As discussed in the text, see around formula (\ref{ftker}), the universal sine-kernel correlations (in $d=1$)
and its high dimensional generalizations can also be derived using the LDA method. 

In $d=1$ another tool can be applied: the bosonization technique. We will briefly
recall here the results which can be obtained from this method. One motivation
is that it also allows to treat the case of interacting fermions, hence to address, 
to some extent, the question of universality in presence of interactions. Furthermore,
although well-known in the condensed matter community, these results do not seem
widely known to mathematicians working on random matrices.

The bosonization method allows one to represent fermions in $d=1$ with uniform local
density $\rho(x) := N \rho_N(x) \approx \rho_0$, as some "exponentials"
of two conjugated bosonic fields, described by a quadratic Hamiltonian
(see e.g. \cite{BookGiam} for a pedagogical introduction). 
This representation is exact for free fermions with a
linear dispersion relation. For more general dispersion relations, and for 
interacting fermions, it remains an accurate effective description in the hydrodynamic
limit, i.e., on spatial scales $(x-x') \rho_0 \gg 1$: this is the so called Luttinger liquid (LL). The effective quadratic Hamiltonian
is parametrized by the (renormalized) Luttinger parameter $K$, which contains all
the information about the large scales. The special value $K=1$ corresponds to non-interacting fermions,
while attraction leads to $K>1$ and repulsion to $K<1$. Using these methods
it was shown \cite{haldane81,BookGiam} that the correlation function of the density at $T=0$ is given by \cite{foot_LL}
\bea
\langle \rho(x) \rho(0) \rangle_0 \simeq \rho_0^2 \big( 1 - \frac{2 K}{(2 \pi \rho_0 x)^2} 
+ \sum_{m=1}^{+\infty} A_m (\rho_0 x)^{- 2 K m^2} \cos( 2 \pi m \rho_0 x) \big) \;, \label{corrrho}
\eea 
while the correlation function of the fermionic field (which is the analogue of the kernel) is
\bea
\langle \Psi^\dagger(x) \Psi(0) \rangle_0 \simeq \rho_0 \sum_{m=0}^{+\infty} 
C_m (\rho_0 x)^{- \frac{1}{2 K} - 2 K (m+ \frac{1}{2})^2} \sin \left( 2 \pi \left(m+ \frac{1}{2}\right) \rho_0 x\right) \;.
\label{corrPsi} 
\eea 
These formulae (\ref{corrrho}) and (\ref{corrPsi}) are valid for $\rho_0 x \gtrsim 1$. 
Here $A_1$ in (\ref{corrrho}) and $C_0$ in (\ref{corrPsi}) represent the leading behaviors
at large $\rho_0 x$, while the terms $A_m$, $m \geq 2$ and $C_m$, $m \geq 1$ 
represent the contributions of
higher harmonics (often neglected in LL studies). 

For non-interacting fermions, $K=1$, $C_0=1$ and all $C_{m}=0$ for $m \geq 1$, and the
expression in (\ref{corrPsi}) becomes exact. It is precisely the sine kernel
$\sin(\pi \rho_0 x)/x = \ell(x)^{-1} {\cal K}^{\rm bulk}(x/\ell(x))$ with $\ell(x)=2/(\pi \rho_0)$ 
and ${\cal K}^{\rm bulk}(y)=\sin(2 y)/\pi y$, proved in this paper to further hold for
fermions in a trap (in the bulk), using the mapping to RMT. In presence of interactions we see that the correlation function of the fermionic field in (\ref{corrPsi}) in the ground state now decays at large $x$ as
\bea
\langle \Psi^\dagger(x) \Psi(0) \rangle_0
 \sim  \sin(\pi \rho_0 x)/x^\eta \quad , \quad \eta = \frac{1}{2} (K + K^{-1})
\eea 
with a non-universal prefactor. The exponent $\eta$ is thus always larger than for
non-interacting fermions, and its precise value depends on the Luttinger parameter
$K$, hence on the strength of the interactions. 

Unfortunately, the standard bosonization methods fail near the edge where 
$k_F = \pi \rho_0$ vanishes. We have found in this paper that for non-interacting fermions
in a trap, the correlation function of the fermion field is described by the Airy kernel.
Deriving that result using bosonization techniques seems at present out of reach.
Nevertheless, it may be possible, at least qualitatively, to recover the leading asymptotics of the left tail 
of that kernel, i.e. the limit where both points enter into the bulk. Indeed, in the non-interacting case, 
we observe that by changing $\pi \rho_0 x \to \int_x^{x_{\rm edge}} dy k_F(y)$ and inserting $k_F(y) \sim \sqrt{x_{\rm edge}-y}$ in \eqref{corrPsi}, we obtain the
Airy function asymptotics at large negative arguments. 
Note that since the edge regime is diluted, it is possible that the effect of (at least short-range) interactions is less important than in the bulk, and the edge universality is more robust. This however remains to be studied in detail.

\subsection{Connection to the KPZ equation} \label{sec:KPZ}


In a recent paper \cite{us_prl} we have unveiled a remarkable connection between the problem of
non-interacting fermions in a one-dimensional trap at finite temperature 
and the continuum $1d$ Kardar-Parisi-Zhang (KPZ) growth equation at finite time, 
with the so-called droplet initial condition (also called curved geometry).
The KPZ equation \cite{kpz} describes the stochastic time evolution 
of the height field $h(x,t)$ of an interface, at point $x \in \mathbb{R}$ and time $t$
\begin{eqnarray}\label{eq:KPZ}
\partial_t h(x,t) = \nu \partial_x^2 h(x,t) + \frac{\lambda_0}{2}\, (\partial_x h(x,t))^2 + \sqrt{D} \eta(x,t) \;,
\end{eqnarray}
where $\nu > 0$ is the coefficient of diffusive relaxation, $\lambda_0 > 0$ is the strength of the 
non-linearity and $\eta(x,t)$ is a centered Gaussian white noise with  
correlator $\langle \eta(x,t) \eta(x',t')\rangle = \delta(x-x')\delta(t-t')$. 
From now on we express height and time in the natural units \cite{footnote1}
\bea 
t^*=2(2 \nu)^5/(D^2 \lambda_0^4) \quad , \quad 
x^* =  (2 \nu)^3/(D \lambda_0^2) \quad , \quad 
h^* = 2 \nu/\lambda_0 \;.
\eea

Here we start from the ``narrow wedge'' initial condition, $h(x,0) = - w |x| + \ln(w/2)$, with $w \gg 1$,
which gives rise to a curved (or {\em droplet}) {mean profile} 
$\langle h(x,t) \rangle=\langle h(0,t) \rangle - \frac{x^2}{4 t}$ 
as time evolves.  Equivalently to (\ref{eq:KPZ}) the Cole-Hopf transformed field $Z(x,t)=e^{h(x,t)}$ 
satisfies the stochastic heat equation (with multiplicative noise) 
\bea \label{Z} 
\partial_t Z(x,t) = \partial_x^2 Z(x,t) + \sqrt{2} \eta(x,t) Z(x,t) \;,
\eea 
with initial condition $Z(x,t=0)=\delta(x)$. The continuum KPZ equation \cite{footnote0} 
is usually defined by the equation (\ref{Z}) with the Ito convention, 
implying that the first moment $\langle Z(x,t) \rangle =Z_0(x,t)$ where 
$Z_0(x,t):=\frac{1}{\sqrt{4 \pi t}} e^{-x^2/(4 t)}$ is the free diffusion propagator
\cite{footnote2}. This is the definition we use here. 
In the natural units defined above, one can define the scaled height at position $x=0$
\bea
\tilde h(0,t) = \frac{ h(0,t) + \frac{t}{12} }{t^{1/3} } \;.
\eea 
The exact results of Refs. \cite{SS10,CLR10,DOT10,ACQ11} can then
be expressed as follows. Let us define the time-dependent generating
function 
\begin{eqnarray} \label{gen}
g_{t}(s) = \langle \exp(- e^{t^{1/3} (\tilde h(0,t) - s)})\rangle \;
\end{eqnarray}
of the rescaled height at $x=0$. It is expressed for all time $t >0$ 
as a Fredholm determinant:
\begin{eqnarray}
&&g_{t}(s) = \det(I - P_s K^{\rm KPZ}_{t} P_s) \label{GF}\\
&&K^{\rm KPZ}_{t}(x,y) = \int_{-\infty}^\infty \frac{{\rm Ai}(z+x) {\rm Ai}(z+y)}{e^{- t^{1/3} z} + 1} 
\,{dz}  \label{eq:KPZKernel} \;,
\end{eqnarray}
where $P_s$ is the projector on the interval $[s,+\infty)$. We can now compare the finite time kernel for KPZ, $K^{\rm KPZ}_{t}$ in Eq.~(\ref{eq:KPZKernel}) with 
the finite temperature kernel ${\cal K}_b^{\rm edge}$ for the fermions at the edge, given in Eq. (\ref{kff}), and see
that they are {\it identical} provided one identifies
\bea \label{rel2} 
b := \frac{\hbar \omega N^{1/3}}{T}  \equiv  t^{1/3} \;,
\eea 
i.e., large time in KPZ corresponds to zero temperature of the fermions, and 
high temperature for the fermions to small time in KPZ [in original KPZ units the r.h.s. should be replaced
by $(t/t^*)^{1/3}$]. The correspondence appears more direct if one introduces an additional
random variable $\gamma$, {\it independent from $h(0,t)$} and distributed according to the Gumbel distribution $P(\gamma)=e^{-\gamma - e^{-\gamma}}$. Using that $ \langle \theta(x- \gamma) \rangle_\gamma = e^{-e^{-x}}$ we can rewrite the generating function (\ref{gen}) and obtain the identity in law
\bea \label{inlaw} 
\lim_{N \to + \infty, ~ {\rm fixed} ~ b} ~ \xi:= \frac{x_{\max}(T) - x_{\rm edge}}{w_N} \equiv_{\rm in \; law} \frac{ h(0,t) + \frac{t}{12} + \gamma }{t^{1/3} }
\eea 
where the random variables on the left and right hand sides have identical PDF. On the left hand side the limit of large $N$ is taken at fixed $b$, and the identity in law then holds for arbitrary $t$ and $b$ related by (\ref{rel2}). In the limit of large time (respectively low $T$) the Gumbel variable can be ignored in the r.h.s. and one recovers that both the scaled variable $\xi = \frac{x_{\max}(T) - x_{\rm edge}}{w_N}$, and the scaled height $\tilde h(0,t)$ tend
to the Tracy-Widom GUE distribution [of CDF $F_2(s)$ given in Eq. (\ref{fredholm_F2})].

In the limit of small time one can use the small $t$ expansion obtained in \cite{CLR10}.
Using the formulae (11) and (12) therein (and below the formula for the skewness)
with the correspondence $\ln z \to h(0,t) - \ln Z_0(0,t) + t/12$ and
$\lambda^3 = t/4= b^3/4$ one first obtains the first three cumulants for the KPZ height
field $h(0,t)$ at small $t$. From this, using (\ref{inlaw}), together with the cumulants of the Gumbel distribution
$\langle \gamma^n \rangle^c_\gamma = (n-1)! \zeta(n)$ for $n \geq 2$ and 
$\langle \gamma \rangle_\gamma = \gamma_E$, as well as the independence of
the two random variables, 
we obtain the high temperature expansion of the cumulants of
the variable $\xi$, i.e. the scaled position of the rightmost fermion. 
The corresponding formula are displayed in Section~\ref{sec:crossover}. 

It is interesting to note that the relation (\ref{inlaw}) between random variables can
be "inverted" to express the PDF of the KPZ field. Indeed one can also write
\cite{SS10,CLR10,ACQ11}
\bea \label{second1} 
\tilde h(0,t) = u -  \frac{\gamma'}{t^{1/3}} 
\eea 
where $\gamma'$ is (yet another) Gumbel random variable, and 
$u$ is a random variable, independent of $\gamma'$ and 
of time-dependent PDF $p_t(u)$ obtained as
\bea \label{second2} 
&& p_t(u) = {\rm Det}[ I - P_u (B_t - Ai Ai^\dagger) P_u] - {\rm Det}[ I - P_u B_t P_u] \\
&& B_t(r,r') = {\rm P. V.} \int_{-\infty}^{+\infty}  dv \frac{{\rm Ai}(r+v) {\rm Ai}(r'+v)}{1 - e^{- t^{1/3} v}} 
\eea 
in terms of the kernel $B_t$ and of the projector $Ai Ai^\dagger(r,r')={\rm Ai}(r) {\rm Ai}(r')$ and where ${\rm P. V.}$ denotes the Cauchy
principal value. The large time expansion, i.e., the corrections around the TW distribution
(or low temperature expansion for the fermions) is studied in
Appendix \ref{sec:expansion} by using equivalently (\ref{GF}) or (\ref{second2}).

Our results establish a precise connection between free fermions at finite temperature and the KPZ equation~(\ref{eq:KPZ}) at finite time $t$. At this stage this connection exists only for the droplet initial condition of KPZ and also for the one-point distribution of $h(0,t)$ (the one-point distribution of $h(x,t)$ being identical
up to a shift by its average $-x^2/(4 t)$). 
Let us mention two other examples where fermions appear in the context of models in the 
KPZ universality class. One is the polynuclear growth model, also related to a zero temperature
model of directed paths~\cite{praeho_spohn_prl, praeho_spohn_jsp}. This problem is related to fermions
on a $1d$ lattice in a linear potential, with a time-dependent slope. A second example
was studied recently in a model of semi-discrete directed polymers~\cite{sasamoto_fermion}. 
It is at present unclear whether the connection between fermions and the KPZ equation hides a 
deeper correspondence, e.g. extending to many point correlations or different initial conditions.

\subsection{Conclusion}

In this paper, we have developed a unified framework to study the statistical mechanics of $N$ non-interacting
fermions trapped by a confining potential, in any dimension $d$ and at any finite temperature $T$. 
The trapping potential gives rise to an edge, i.e., a distance from the center of the trap  
above which the average density vanishes. Consequently, when $N$ is large, 
there are two distinct scaling regimes in the local correlations: the ``bulk regime'', i.e., near the center of the trap where the density of fermions varies smoothly, and the ``edge regime'', i.e., close to the edge where the density is vanishing and where the (quantum and thermal) fluctuations are thus large.    
In the ``bulk regime'', our method recovers and puts on firmer basis the results of the standard Local Density
(or Thomas-Fermi) Approximation (LDA) \cite{Castin,butts}. On the other hand, at the edge, where the LDA fails, we have
obtained a detailed description of the correlations. Indeed, we have shown that, even at finite temperature $T>0$, 
the system, in the limit of a large number of fermions, is a determinantal point process characterized by a kernel
which depends on both the dimension $d$ and the temperature $T$ [see Eq.~(\ref{kernel0Vedge})], which generalizes the Airy 
kernel (\ref{airy_kernel.1}), which is a fundamental object in RMT. Remarkably, we have shown, using a path integral representation
of this kernel, that is universal, i.e., independent of the details of the trapping potential, for a wide class of spherically symmetric 
potentials $V(|{\bf x}|)$ which behaves at large distance as $V(|{\bf x}|) \sim |{\bf x}|^p$, with $p>0$.
In the special case $d=1$, we have studied in detail the cumulative distribution function of the position of the rightmost fermion $x_{\max}(T)$ at $T>0$ and we have shown that this CDF in Eq. (\ref{gap_proba}) (i) generalizes the well known Tracy-Widom distribution, which describes the fluctuations of $x_{\max}(T=0)$ and (ii) displays a remarkable connection with the (1+1)-dimensional Kardar-Parisi-Zhang equation in a droplet geometry~(\ref{inlaw}).   

Therefore, we believe that the results presented here open the way for interesting bridges between the techniques of many-body physics and of random matrix theory. We hope that our results will stimulate further works at this interface. In addition, one of the main outcome of our paper is a precise set of predictions for systems of non-interacting cold fermions at finite temperature. It would be exciting if our theoretical predictions could be verified in cold atom experiments, for example using the state of the art quantum microscopes~\cite{Cheuk:2015,Haller:2015,Parsons:2015} .

%
%
%
%

\newpage 
 
\appendix

\section{General short time expansion of the imaginary time Schr\"odinger propagator}\label{short_time}

Here we derive, using probabilistic methods, the general short time expansion for the imaginary time propagator of a one-body Hamiltonian with an arbitrary soft potential. The semi-classical expansion of the path integral for the quantum mechanical propagator,  an expansion in $\hbar$,  has been extensively studied in the literature \cite{klein}. However the short-time expansion has received less attention \cite{MM88}. As shown in the text, and detailed again in the next Appendix, this expansion allows the calculation of  the kernel of trapped non-interacting fermions in the bulk and at the edge. We first show the following formula for the propagator: \\

{\bf Proposition:}
The solution of the imaginary time Schr\"odinger equation in Eq. (\ref{itshro}) can be written, for 
an arbitrary potential $V(\bf x)$, as
\begin{equation}
G({\bf x},{\bf y};t) = \left({m\over 2\pi \hbar t}\right)^{d\over 2}
\exp\left(-{m\over 2 t \hbar}({\bf x}-{\bf y})^2\right) \left\langle \exp\left(-{t\over \hbar}\int_0^1 du\  V( {\bf x}{(1-u)} + {\bf y}{u} + \sqrt{D_0 t}{\B}_u)\right) \right\rangle_{{\B}} ,\label{repbb}
\end{equation}
where $D_0=\hbar/m$ is the diffusion coefficient and ${\B}_u=\{B_{iu}\}_{i=1}^d$ is a $d$-dimensional Brownian bridge on the interval $[0,1]$, i.e. a Gaussian
process with mean zero and correlation function
\begin{equation}
\left\langle B_{iu} B_{ju'}\right\rangle_{{\B}} = \delta_{ij} \ g(u,u') \quad , \quad g(u,u') = \min(u,u') - u u'
\end{equation}
hence with $\B_0=\B_1=0$.\\

{\it Proof.} To show this we use the Feynman-Kac formula \cite{oks} to express the propagator as a path integral
\begin{equation}
G({\bf x},{\bf y};t) = \int_{{\bf X}(0)={\bf x}} d[{\bf X}] \delta({\bf X}_t -{\bf y})\exp\left(-{1\over \hbar}\int_0^t ds\left[{1\over 2}m \left({d{\bf X_s}\over ds}\right)^2 + V({\bf X}_s)\right]\right),
\end{equation}
over all paths staring at ${\bf x}$ at time $t=0$, with a delta function weight so that only paths which end at ${\bf y}$ at time $t$ contribute. We now make a shift of variables in the path integral
\begin{equation}
{\bf X}_s = {\bf x}{(t-s)\over t} + {\bf y}{s\over t} + {\bf Z}_s, 
\end{equation}
under which the expression for the propagator becomes
\begin{equation}
G({\bf x},{\bf y};t) = \exp\left(-{m\over 2 t \hbar}({\bf x}-{\bf y})^2\right) \left\langle \exp\left[-{1\over \hbar}\int_0^t ds V \left( {\bf x}{(t-s)\over t} + {\bf y}{s\over t} + {\bf Z}_t\right)\right] \right\rangle_{\bf Z} \times {\cal N}_t
\end{equation}
where ${\cal N}_t = \int_{{\bf Z}(0)={\bf 0}} d[{\bf Z}] \delta\left[{\bf Z}_t \right]\exp\left(-{m\over 2\hbar}\int_0^t ds \left({d{\bf Z}_s\over ds}\right)^2\right)$ is a normalization. In simplifying the kinetic energy contribution we have used that only paths where ${\bf Z}_t={\bf 0}$ contribute to the path integral, and we denote by $\langle \ldots \rangle_{\bf Z}$ an average with respect to the normalized measure 
on the paths ${\bf Z}$ given by
\begin{equation}
P({\bf Z}) = \frac{1}{{\cal N}_t} \delta\left[{\bf Z}_t \right]\exp\left(-{m\over 2\hbar}\int_0^t ds \left({d{\bf Z}_s \over ds}\right)^2\right)
.\label{eqpz}
\end{equation}
To evaluate the normalization ${\cal N}_t$ we apply the Feynman-Kac formula again and see that the solution is given by the solution of the $d$-dimensional (free) diffusion equation,
with a delta function initial condition. This gives ${\cal N}_t = \left({m\over 2\pi \hbar t}\right)^{d\over 2}$.
The process ${\bf Z}$ with the measure given  in Eq. (\ref{eqpz}) is in fact a $d$-dimensional Brownian bridge, a Gaussian process with zero mean and temporal correlation function
\begin{equation}
\left\langle Z_{is} Z_{js'}\right\rangle_{\bf Z} = \delta_{ij}D_0\left(\min(s,s') -{ss'\over t}\right) \;. \label{cfz}
\end{equation} 
To show this, we compute the 
generating functional
\begin{equation}
g[\boldsymbol{\lambda}] := \left\langle \exp\left(-i\int_0^t {\boldsymbol \lambda(s)}\cdot{\bf Z}_s\right) \right\rangle_{\bf Z}
= {\left\langle \delta({\bf W}_t)(\exp\left(-i\int_0^t {\boldsymbol \lambda(s)}\cdot{\bf W}_s\right) 
\right\rangle_{\bf W}\over\left\langle \delta({\bf W}_t)\right\rangle_{\bf W}} \label{eqglambda}
\end{equation}
where $\langle \ldots \rangle_{\bf W}$ denotes an average with respect to an unconstrained Brownian motion with correlation function
\begin{equation}
\left\langle W_{is}W_{js'}\right\rangle_{\bf W} = D_0 \min(s,s') \;.
\end{equation}
In the second equality in \eqref{eqglambda} we enforced ${\bf W}_t=0$ by inserting a delta function. 
The averaging over ${\bf W}$ can be carried out by using a Fourier representation of the delta function
$\delta({\bf W}_t) = \int {d{\bf q}\over (2\pi)^d}\  \exp(-i{\bf q}\cdot{\bf W}_t)$. Inserting this representation, both in the denominator and in numerator of Eq. (\ref{eqglambda}), allows the Gaussian averaging to be carried out yielding
\begin{equation}
g[\boldsymbol{\lambda}] ={ \int {d{\bf q}\over (2\pi)^d}\exp\left(-{D_0\over 2} \int_0^t\int_0^t dsds'\  \min(s,s')\boldsymbol{\lambda}(s)\cdot\boldsymbol{\lambda}(s')
-{D_0t\over 2} {\bf q}^2 -\int_0^t ds\  D_0 s{\bf q}\cdot\boldsymbol{\lambda}(s)\right)
\over \int {d{\bf q}\over (2\pi)^d}\exp\left(-{D_0t\over 2}{\bf q}^2\right)} \;.
\end{equation}
Now carrying out the integral over ${\bf q}$ in both the numerator and denominator gives
\begin{equation}
g[\boldsymbol{\lambda}] = \exp\left(-{D_0\over 2} \left[\int_0^t\int_0^t dsds'\  \min(s,s')\boldsymbol{\lambda}(s)\cdot\boldsymbol{\lambda}(s')  -{1\over t}\left(\int_0^t ds \ s\boldsymbol{\lambda}(s)\right)^2\right]\right).
\end{equation}
From this expression for the generating functional of the Gaussian process we immediately see that the temporal correlation function is given by Eq. (\ref{cfz}). Making now the change of variables $s=ut$ and use Brownian scaling leads to the proof of the Proposition in (\ref{repbb}). \\

The representation of the propagator in Eq. (\ref{repbb}) identifies explicitly the occurrence of the variable $t$. 
Let us now derive the small $t$ expansion up to $\mathcal{O}(t^3)$, and recover Eq. (\ref{expt3}).
To this purpose it is sufficient to Taylor expand the integral in the exponential on the right hand side of Eq. (\ref{repbb}) to $\mathcal{O}(t^3)$, giving
\begin{equation}
\int_0^1 du\  V\left( {\bf x}{(1-u)} + {\bf y}{u} + \sqrt{D_0 t}{\B}_u)\right) 
= A_0 + t^{1\over 2} A_1 + t A_2 + A_3 t^{3\over 2} + A_4 t^2 + o(t^2) \;.
\end{equation}
The various terms are explicitly given by (using the Einstein summation convention)
\be
A_0 = \frac{1}{\hbar} \int_0^1 du\ V({\bf X}(u)) \quad , \quad 
A_p = {D_0^{p/2}\over p!\hbar}  \int_0^1 du\ \nabla_{i_1} \cdots \nabla_{i_p} V({\bf X}(u)) B_{i_1 u} \cdots B_{i_p u} \quad , \quad {\bf X}(u) = {\bf x} + u ({\bf y}-{\bf x})
\ee
where ${\bf X}(u)$ denotes the straight line path between ${\bf x}$ and ${\bf y}$ for $u\in[0,1]$. The resulting expansion can then be averaged using the cumulant expansion which, keeping only terms of ${\mathcal O}(t^3)$,
gives, 
\begin{eqnarray}
&&\left\langle \exp\left(-{t\over \hbar}\int_0^1 du\  V( {\bf x}{(1-u)} + {\bf y}{u} + \sqrt{D_0 t}{\B}_u)\right) \right\rangle_{{\B}}=\nonumber \\
&&\approx \exp\left(-  tA_0  - \langle t^{3\over 2} A_1 + t^2 A_2 + t^{5\over 2} A_3 + t^3 A_4 \rangle_{{\B}} + {1\over 2} t^3[ \langle A_1^2\rangle_{{\B}}-\langle A_1\rangle^2_{{\B}}]\right)
\end{eqnarray}
where we have used the fact that $A_0$ does not depend on ${\B}$. Now 
we note that odd moments of ${\B}$ are zero so $\langle A_1\rangle_{{\B}} = \langle A_3\rangle_{{\B}} =0$ and consequently we have
\begin{eqnarray}
\left\langle \exp\left(-{t\over \hbar}\int_0^1 du\  V( {\bf x}{(1-u)} + {\bf y}{u} + \sqrt{D_0 t}{\B}_u)\right) \right\rangle_{{\B}}\approx
\exp\left(-  tA_0  - t^2\langle A_2 \rangle_{{\B}}+  t^3[{1\over 2}  \langle A_1^2\rangle_{{\B}}-\langle A_4\rangle_{{\B}}]\right) \;.
\end{eqnarray}
Explicit computation of the moments then leads to the result given in Eq. (\ref{expt3}). It is interesting to note that the short time expansion as formulated here is an expansion about the, constant  velocity, direct straight-line path between ${\bf x}$ and ${\bf y}$ (in contrast to the semi-classical expansion in powers of $\hbar$ for the real time path integral, which is given about classical paths obeying Newton's equation).\\

{\it An alternative derivation using diagrammatic techniques}. To address the calculation of the kernel in a more systematic way, we now develop a diagrammatic method. 
Let us consider the formula \eqref{laplace_inverse} for the kernel (for simplicity at $T=0$)
and use the expression of the propagator in \eqref{repbb}.
Let us further set $d=1$ and work in the natural units where $m=\hbar=1$, for simplicity. The 
generalization to higher dimension is straightforward. One has 
\begin{equation}
K_\mu(x,y) = \int_{\Gamma} {dt\over (2\pi t)^{3/2} i} 
\exp\left(-{ (x-y)^2 \over 2 t} + \mu t - S(x,y;t)\right) \quad , \quad S(x,y;t)=- \ln 
\left\langle e^{- t \int_0^1 du\  V(x (1-u) + y u + \sqrt{t}{\B}_u) } \right\rangle_{{\B}} \;. \label{Kmunew}
\end{equation}
Making a standard cumulant expansion of \eqref{Kmunew} and Wick contractions
for the Gaussian Brownian bridge, one can calculate term by term using an elegant
diagrammatic expansion, as shown below. Let
us denote the vertex at position $u$, and the Brownian bridge correlator 
$g(u,u')=\left\langle {\B}_u \B_{u'}  \right\rangle_{{\B}} $ respectively by
\bea
 \diagram{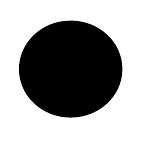} = - t \ V(x (1-u) + y u) \quad , \quad 
\diagram{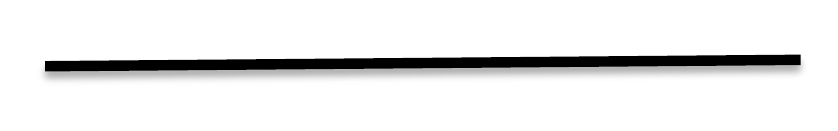}  = t \ g(u,u')  \label{rule} \;.
\eea 
Then the function $- S(x,y;t)$ is the sum of all connected diagrams
\bea \label{seriest}
&& - S = \diagram{graph0} (t) + \diagram{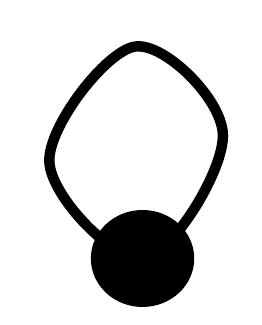} (t^2) +\diagram{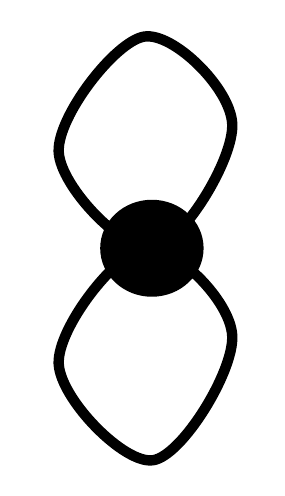} (t^3) + \cdots  + \diagram{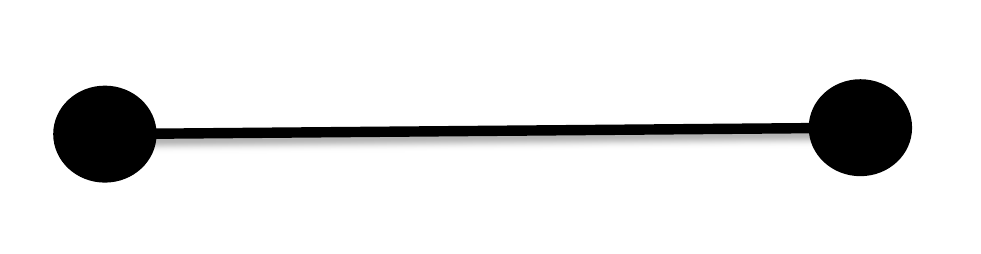}  (t^3) + \diagram{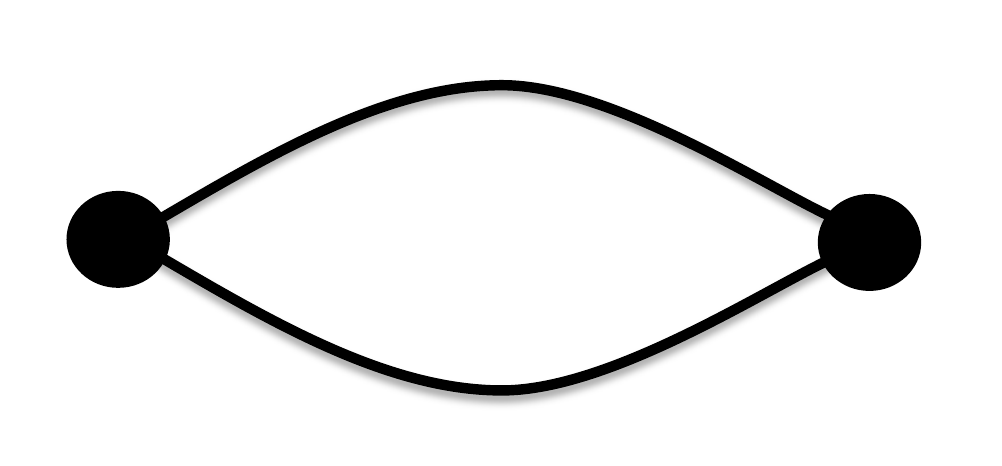} (t^4)  \\
&& + \diagram{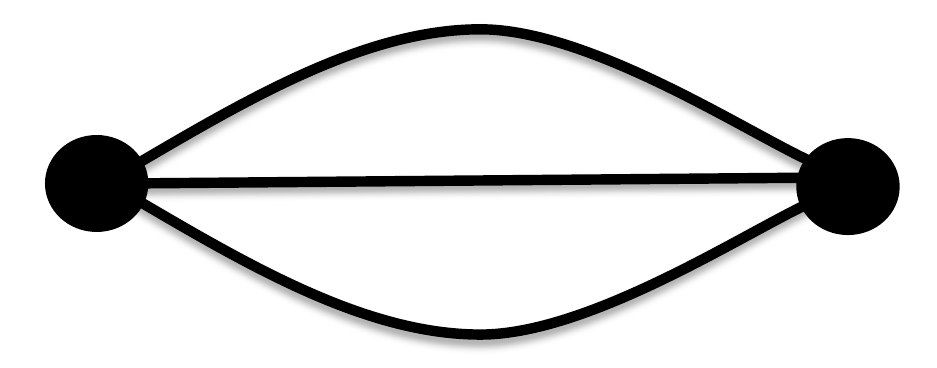} (t^5)  +  \cdots +  \diagram{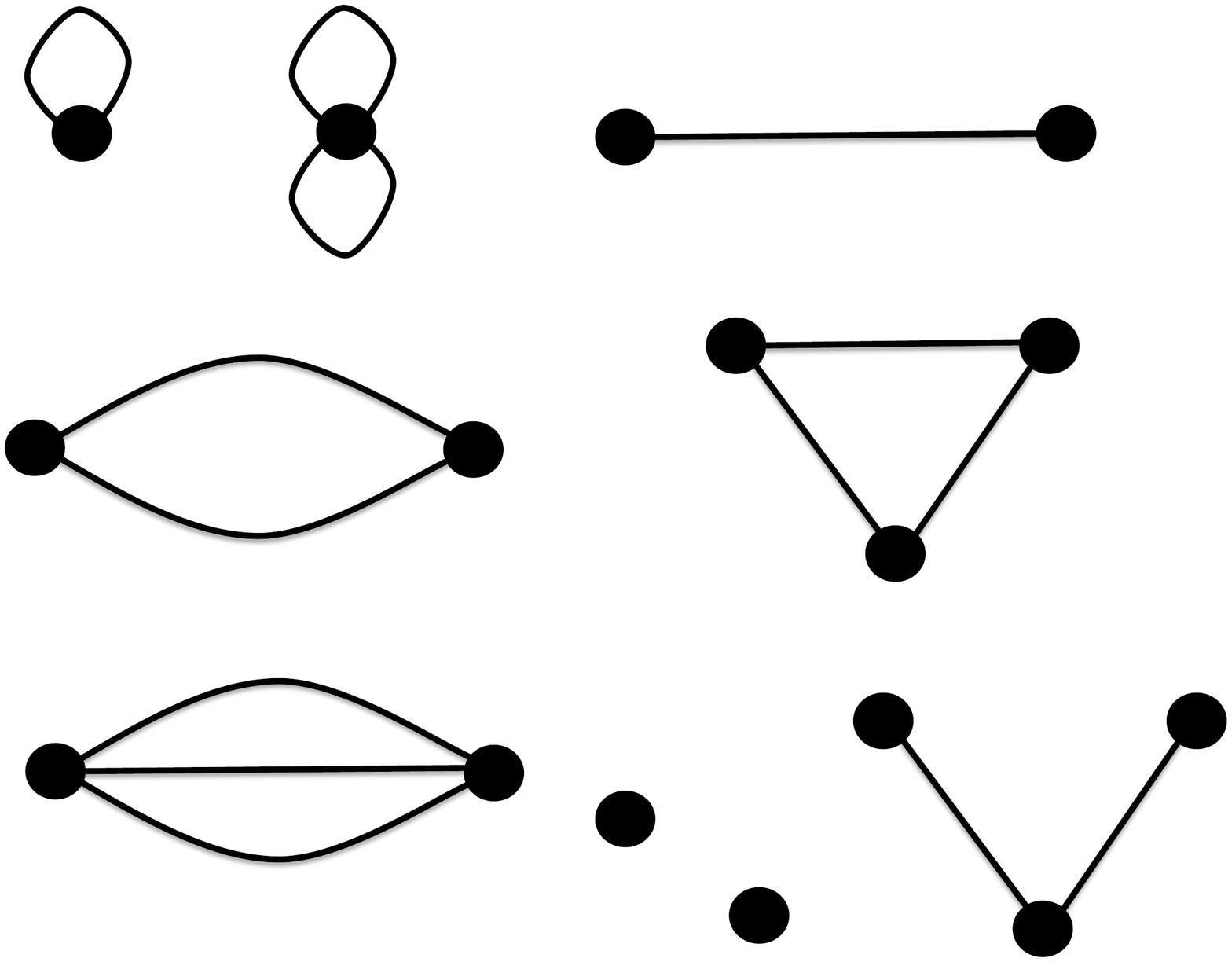} (t^5) +  \diagram{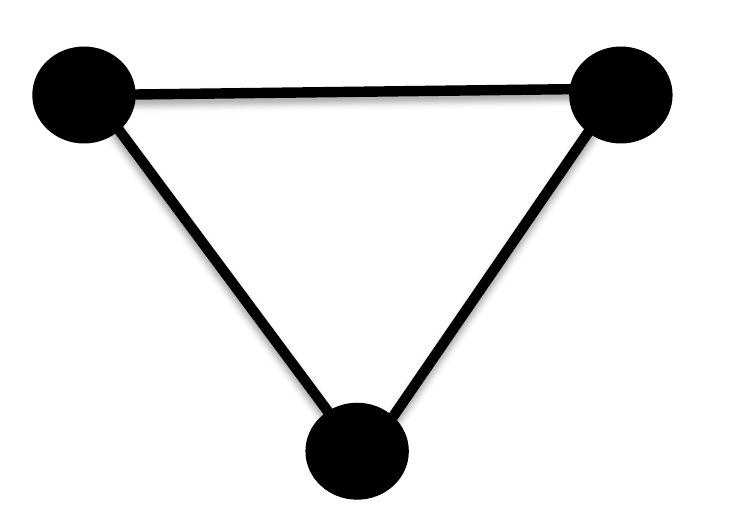}  (t^6)+ \cdots \nn
\eea
The first family of diagrams is ${\cal O}(V)$, the second ${\cal O}(V^2)$, the third ${\cal O}(V^3)$ and so on. 
The power of $t$ is indicated next to the graph in parenthesis, the rule being, from \eqref{rule}, that each vertex
and each propagator brings a power of $t$. There is one distinct index $u_i$ per vertex, each line emerging from a vertex leads to a derivative acting on the vertex, and all the $u_i$'s are integrated at the end, each in $[0,1]$.
For instance (omitting the combinatorial factor)
\bea
\diagram{graph6}  = \int_0^1 du_1 \int_0^1 du_2 g(u_1,u_2) V'(x (1-u_1) + y u_1) V'(x (1-u_2) + y u_2) \;. \label{g6}
\eea 
Till now, all terms in this systematic expansion in powers of $t$ have been written exactly,
and are valid for arbitrary potentials (sufficiently differentiable, although some extensions
are possible). At this stage it is so general that it is not clear what is the true expansion parameter 
of the above series \eqref{seriest}, in other words, for the series to be under control,
$t$ must be small compared to what? The answer
to that question will depend on the class of physical situations studied.

Now we assume a soft potential and discuss the gradient expansion near the edge. 
By definition of the edge, $\mu=V(x_e)$ and here, for compactness we
denote $x_e=x_{\rm edge}$. We set $x=x_{e} + w_N a$, $y=x_{e} + w_N b$
where $a,b$ are assumed to be of order ${\cal O}(1)$ in the large $N$ limit. Hence
$X(u)=x (1-u) + y u= x_{e} + w_N (a (1-u) + b u)$ and the scale factor $w_N$ is
to be fixed later. For instance \eqref{g6} becomes
\bea
\! \! \! && \diagram{graph6}  = c_{00} V'(x_{e})^2  + 2 c_{12} V'(x_{e}) V''(x_{e}) w_N
+ {\cal O}(w_N^2) \\
&& c_{pq} = \int_0^1 du_1 \int_0^1  g(u_1,u_2) (a (1-u_1) + b u_1)^p (a (1-u_2) + b u_2)^q \;.
\eea 
We can now enumerate the various terms, and keep the leading and sub-leading ones in each family. 
In the exponential in \eqref{Kmunew}, we have $(a-b)^2 \frac{w_N^2}{2 t} - S$, where 
the terms in $S$ are [after cancellation of $\mu$ and $V(x_e)$] (without showing the combinatorial factors)
\bea
\! \! \! && \diagram{graph0} V'(x_{e}) w_N  t + V''(x_{e}) w_N^2 t  + \cdots \quad , \quad  \diagram{graph4} V''(x_{e}) \ t^2 + \cdots  \label{te1} \\
\! \! \! && \! \! \! \diagram{graph6} V'(x_{e})^2 t^3 + V'(x_{e}) V''(x_{e}) w_N t^3 + \cdots 
 ,   \diagram{graph2} V''(x_{e})^2 t^4 + V''(x_{e}) V'''(x_{e}) w_N t^4 + \cdots \nn \\
&&  \diagram{graph9} V'(x_e)^2 V''(x_e) t^5 + V''(x_e)^3 w_N t^5 + \cdots , 
\diagram{graph7} V''(x_e)^3 t^6 + V''(x_e)^2 V'''(x_e) w_N t^6 + \cdots \;. \nn
\eea 

To classify the possible behaviors we start with the diffusion term $(a-b)^2 \frac{w_N^2}{2 t}$.
Assume that the exponent $(a-b)^2 \frac{w_N^2}{2 t} - S$ is dominated 
by some typical value of $t$ (called $t_N$ in the
text). There are three possibilities: (i) the diffusion term blows up, i.e., $t_N \ll w_N^2$ (ii) it vanishes, i.e., $t_N \gg w_N^2$
or (iii) it is of order unity $t_N \sim w_N^2$. The cases (i) and (ii) being more exotic, 
we now consider the case (iii).

Now let us first discuss the standard class (leading to the Airy kernel). 
It is such that the first term in \eqref{te1}, using $t_N \sim w_N^2$, is $V'(x_{e}) w_N  t_N \sim V'(x_{e}) w_N^3 = {\cal O}(1)$
which means $w_N \sim |V'(x_{e})|^{-1/3}$, as found in the text. Interestingly the first term of
the first diagram in the second line of \eqref{te1}, $V'(x_{e})^2 t_N^3 = [ V'(x_{e}) w_N^3 ]^2={\cal O}(1)$
is also of order unity, and that leads to the Airy kernel (\ref{airy_kernel.1}), if we can neglect all other terms. 
Let us examine now the conditions for neglecting all the other terms.

The diagrams of the first line (tadpoles, ${\cal O}(V)$ diagrams) can all be
written as $\sim t_N (w_N \partial_{x_e})^q (t_N \partial^2_{x_e})^\ell V(x_e)$
with $\ell \geq 0$ number of loops, and $q \geq 0$ the degree in the gradient expansion,
with $(\ell,q)=(0,0)$ being excluded. Now we want all these diagrams to be subdominant, excluding
$(\ell,q)=(0,1)$ which is ${\cal O}(1)$. Hence it requires
$|V^{(2 \ell+q)}(x_e)| \ll |V'(x_e)|^{(2 \ell + 2 + q)/3}$ for $\ell=0,q \geq 2$ and for $\ell \geq 1, q \geq 0$,
which is equivalent to the set of conditions
\bea \label{cond0} 
|V^{(2+n)}(x_e)| \ll |V'(x_e)|^{(4+n)/3} \quad , \quad n=0,1,2,...
\eea 
with $n=2 \ell+q-2$, the first of these conditions being
\bea
|V''(x_e)| \ll |V'(x_e)|^{4/3} \;.  \label{cond1} 
\eea 

Similarly, the diagrams on the second line (${\cal O}(V^2)$ diagrams) can all be written as  
$\sim t_N^{\ell+3} (w_N \partial_{x_e})^q [ \partial_{x_e}^{\ell+1} V(x_e) ]^2$ with
$\ell \geq 0$ and $q \geq 0$. Now we want all to be subdominant, excluding
$(\ell,q)=(0,0)$ which is ${\cal O}(1)$. Hence it requires
\be
| \partial_{x_e}^q [ \partial_{x_e}^{\ell+1} V(x_e) ]^2 |
\sim \left | \sum_{p=0}^q V^{(\ell+1+p)}(x_e) V^{(\ell+1+q-p)}(x_e) \right |  \ll |V'(x_e)|^{(2 \ell + 6 + q)/3} |
 , \quad \ell=0,q \geq 1 \quad \text{and} \quad 
\ell \geq 1, q \geq 0
\ee
Remarkably, it is a simple exercise to check that if the conditions (\ref{cond0}) are verified, then 
this new set of conditions is also verified. 

It is a tedious exercise to continue this process, and check that if the conditions (\ref{cond0}) are verified
then {\it all} conditions for all graphs ${\cal O}(V^k)$, $k \geq 1$ are satisfied. 
As an illustration, let us
point out for instance the conditions from the leading terms 
of the first two diagrams ${\cal O}(V^3)$ on the next line. One finds simply (\ref{cond1}). Hence it means that the set of conditions (\ref{cond1}) defines the {\it exact} basin of
attraction of the Airy kernel edge statistics at $T=0$. 

At this stage we have specified very little about $V(x)$, apart from being smooth. All we know is that
\bea
&& V(x_e) = \mu \quad , \quad N = \frac{\sqrt{2}}{\pi} \int dx \sqrt{\mu-V(x)} \, \theta(\mu-V(x))  \;.
\eea
The only large parameter being $N$, this equation determines $\mu$ and $x_e$. From there
the conditions (\ref{cond1}) can be checked, the width of the edge region being $w_N \sim |V'(x_e)|^{-1/3}$.
There are many possible cases to consider, especially if one also considers $N$-dependent
trap potentials. There are however two large classes, with either $x_e \to \pm \infty$, or $x_e \to x_e^*$ as
$N \to +\infty$. 

For the ($N$-independent) power law potentials $V(x) \sim |x|^p$, the condition \eqref{cond0} becomes
\bea \label{condsimp} 
|x_e|^{p-2-n} \ll |x_e|^{(p-1) \frac{4+n}{3}}  \Rightarrow 1 \ll |x_e|^{(p+2) \frac{1+n}{3}} 
\eea 
The conditions are thus verified for all $p>0$, since in all these cases $x_e \to +\infty$. 
This case was studied in the text. It also includes limiting cases such as
$V(x)\propto \ln |x|$. Note that the width is $w_N \sim x_e^{(1-p)/3}$. 


The case $p<0$ occurs in two situations, the first one is a (bounded from below) confining potential with
a power law tail at large $|x|$
\bea \label{binding} 
V(x) \simeq_{|x| \to +\infty} V_0 - c |x|^{-\alpha}
\eea 
with $\alpha=-p>0$. It is known that such a potential has an infinite number of
bound states if $\alpha < 2$ 
\cite{friedrich}. Since then $|x_e| \to +\infty$, the conditions \eqref{condsimp} are
satisfied which indicates that it belongs to the Airy class at $T=0$
with a width scaling as $w_N \sim x_e^{(1+\alpha)/3} \ll x_e$.
In the case $\alpha >2$ there is only a finite number of bound states,
which does not enter in the class studied here (since the limit of large $N$ then cannot
be studied). The case $\alpha=2$ is marginal and deserves a separate
discussion, but there clearly the conditions \eqref{condsimp} fail. 

The second situation with $p=-\alpha<0$ corresponds to {\it wall potentials} $V(x) \sim 1/x^{\alpha}$ near $x=0$. In this case the average density will have
a left edge $x_e \to 0^+$ as $N \to +\infty$ and
one finds that the condition \eqref{cond0} is valid only if $p<-2$, i.e. $\alpha > 2$. 
In addition the width now scales as $w_N \sim x_e^{(1+\alpha)/3}$. Hence
for $\alpha > 2$ one has $w_N \ll x_e$, which is completely physical, while for $2 < \alpha <0$ one finds
$w_N \gg x_e$ which is in contradiction with the assumptions (because of the wall). 
Hence both criteria
are valid for $\alpha>2$, and break down for $0 < \alpha \leq 2$. 


The above analysis concerns the Airy universality at $T=0$. While the class $V(x) \sim |x|^p$
leads to the universal finite temperature kernel studied in this paper (same as the
harmonic oscillator), the class of potential of type (\ref{binding}) is much more
sensitive to temperature since its spectrum has a continuum part and the fermions can
unbind. Hence we expect there a different behavior which remains to be studied. 
Note that for potentials $V(x) \sim |x|^p$ with $p<1$ the temperature scale 
$T_{\rm edge}$ defined in \eqref{Tedge} {\it decreases} with increasing $N$.

\section{Large time/low temperature expansions for KPZ/fermions around the Tracy Widom
distribution}  \label{sec:expansion} 


We start from the solution for droplet initial conditions recalled in the text, expressed as
\bea \label{appfd} 
\Pr(\xi<s) = \Pr\left(h(0,t) + \frac{t}{12} + \gamma < s t^{1/3} \right) = {\rm Det}[ I - P_s K_{t}^{\rm KPZ} P_s] \;,
\eea 
where $\gamma$ is a Gumbel variable, independent of $h(0,t)$. In this form it describes
both the fermions, via the variable $\xi$ defined in the text, and the KPZ height, and we recall that
\be \label{Kt} 
 K_{t}^{\rm KPZ}(r,r')  = {\cal K}^{\rm edge}_{b=t^{1/3}}(r,r')  = \int_{-\infty}^{+\infty} du \frac{{\rm Ai}(r+u) {\rm Ai}(r'+u)}{1+e^{- t^{1/3}  u} } \;.
\ee
Using that $\frac{1}{1+e^{-u}}=1 - \frac{1}{1+e^{u}}$ and the same 
manipulations as to derive the Sommerfeld expansion of the Fermi factor
(see e.g. \cite{wiki_sommer}) we can rewrite
\bea
&& K_{t}^{\rm KPZ}(r,r') = K_{{\rm Ai}}(r,r') - t^{-1/3} \int_{-\infty}^{+\infty} du 
\frac{{\rm Ai}(r+\frac{u}{t^{1/3}}) {\rm Ai}(r'+\frac{u}{t^{1/3}}) {\rm sgn}(u) }{1+e^{|u|} } \:,
\eea 
which, until now, is an exact expression for all $t$. 

Let us now perform an expansion for large $t$, by first expanding the kernel. 
Expanding in powers of the factor $\frac{u}{t^{1/3}}$, and performing the 
integrals over $u$, one finds that only odd terms $\sim u^{1+2 q}$ survive, leading to
\bea
K_{t}^{\rm KPZ}(r,r') = K_{{\rm Ai}}(r,r') - t^{-2/3} \sum_{q \geq 0} t^{- \frac{2 q}{3}}  2(1- 2^{-1-2 q})  \zeta(2+2 q) 
\partial_s^{1+2 q}[ {\rm Ai}(r+s) {\rm Ai}(r'+s) ]  |_{s=0} \;.
\eea 
Replacing $t=b^3$, where $b=N^{1/3} \hbar \omega/T$, we simultaneously obtain the low $T$ 
expansion of the edge kernel for the fermions. For instance, let us give the complete
expansion for the universal scaling function of the density near the edge
\bea
F_{1,b}(r) = {\cal K}^{\rm edge}_{b}(r,r)
= F_1(r) - b^{-2} \sum_{q \geq 0} b^{-2 q}  2(1- 2^{-1-2 q})  \zeta(2+2 q) 
\partial_r^{1+2 q} [ {\rm Ai}(r)^2] \;.
\eea 

Now we would like to obtain the large $t$ expansion of the PDF of the 
KPZ height, equivalently the low $T$ expansion of the PDF of the position
of the rightmost fermion, around the Tracy-Widom distribution. For this we need
to expand the Fredholm determinant in \eqref{appfd}.
Let us rewrite the FD as follows
\bea
{\rm Det}[ I - P_s K_{t}^{\rm KPZ} P_s]  = {\rm Det}[ I - P_0 K_{t,s} P_0]  \quad , \quad 
K_{t,s}(r,r')=K_t^{\rm KPZ}(r+s,r'+s)  \;.
\eea 
Let us write the two leading corrections
\bea
K_{t,s}(r,r') = K_{{\rm Ai},s}(r,r') - \frac{\pi^2}{6 t^{2/3}} \partial_s [{\rm Ai}(r+s) {\rm Ai}(r'+s)] 
- \frac{7 \pi ^4}{360 t^{4/3}} \partial^3_s [{\rm Ai}(r+s) {\rm Ai}(r'+s)] +{\cal O}(t^{-2}) \;,
\eea 
where we denote 
$K_{{\rm Ai},s}(r,r') = K_{{\rm Airy}}(r+s,r'+s)$. Recalling that the CDF of the
Tracy-Widom distribution is $F_2(s) = {\rm Det}[ I - P_0 K_{{\rm Ai},s} P_0]$, we
can expand the FD to second order as
\bea\label{B8}
{\rm Det}[ I - P_0 K_{t,s} P_0]  = F_2(s) + t^{-2/3} \frac{\pi^2}{3}  {\rm Tr} [ P_0 (I - P_0 K_{{\rm Ai},s} P_0)^{-1}  
{Ai}'_s {Ai}^\dagger_s ] F_2(s) +  Q_4(s) t^{-4/3} + {\cal O}(t^{-2})  \;,
\eea 
where $Ai'_s Ai^\dagger_s(r,r')={\rm Ai}'(r+s) {\rm Ai}(r'+s)$ and 
$Q_4(s)$ can in principle be computed. We can now use the identity (see a derivation 
in Ref. \cite{jacopo}) 
\bea
F_2''(s) = 2 F_2(s)  {\rm Tr} [ P_0 (I - P_0 K_{{\rm Ai},s} P_0)^{-1}  
Ai'_s Ai^\dagger_s ] 
\eea 
to simplify the expression in (\ref{B8}) and obtain
\bea
{\rm Det}[ I - P_0 K_{t,s} P_0]  
= F_2(s) + t^{-2/3} \frac{\pi^2}{6} F_2''(s) + Q_4(s) t^{-4/3} + {\cal O}(t^{-2}) \;.
\eea 

Taking a derivative we thus obtain the PDF, noted $p_t(s)$ of the variable $\xi$ for large $t$ as
\bea \label{pdfexp} 
p_t(s) = f_2(s) + t^{-2/3} \frac{\pi^2}{6} f_2''(s) + q_4(s) t^{-4/3} + {\cal O}(t^{-2}) \;,
\eea 
where $f_2(s)=F_2'(s)$ and $q_4(s)=Q_4'(s)$. Let us denote $m_p=\langle \chi^p \rangle_{GUE}$ and
$\kappa_p=\langle \chi^p \rangle^c_{GUE}$ the moments and cumulants 
of the TW distribution for GUE, and $a_p$ the moments associated to $q_4(s)$. From (\ref{pdfexp}) we 
obtain
\bea
\langle \xi^p \rangle = m_p +  t^{-2/3} \frac{\pi^2}{6} p(p-1) m_{p-2} 
+ a_p t^{-4/3} + {\cal O}(t^{-2}) \;.
\eea 
Let us give the first three cumulants of the variable $\xi$ for the fermions
\bea
&& \langle \xi \rangle = m_1 + a_1 t^{-4/3} \;, \\
&& \langle \xi^2 \rangle^c = \kappa_2 + \frac{\pi^2}{3} t^{-2/3} + (a_2 - 2 a_1 m_1) t^{-4/3} \;, \\
&& \langle \xi^3 \rangle^c = \langle \xi^3 \rangle - 3 \langle \xi^2 \rangle
\langle \xi \rangle+ 2 \langle \xi \rangle^3 =
\kappa_3 + (a_3 - 3 a_2 m_1 - 3 a_1 m_2 + 6 a_1) t^{-4/3} \;.
\eea 
Since the cumulants are additive for uncorrelated variables one immediately
obtains the large time expansion of the cumulants for the scaled KPZ field
$\tilde h \equiv \tilde h(0,t)$ as
\bea
\langle \tilde h^p \rangle^c = \langle \xi^p \rangle^c - \gamma_p 
\eea 
where the cumulants of the Gumbel distribution are $\gamma_p=\langle \gamma^p \rangle^c_\gamma = (p-1)! \zeta(p)$ for $p \geq 2$ and $\gamma_1=\gamma_E$. Note that for the first three cumulants
the leading correction to TW is equal to the corresponding cumulant of {\it minus}
a Gumbel variable
\bea
\langle \tilde h^p \rangle^c = \kappa_p + (-1)^p \gamma_p t^{-p/3} + {\cal O}(t^{-4/3}) \quad , \quad p=1,2,3 \;,
\eea 
the leading correction to all higher cumulants being {\it a priori} of order ${\cal O}(t^{-4/3})$. 
This fact is actually known \cite{takeuchi}. It can be obtained equivalently from 
(\ref{second1})-(\ref{second2}), showing that $p_t(u)=f_2(u)+ {\cal O}(t^{-4/3})$.
Indeed, manipulations of the integrals similar as above allow to show that 
\bea
&& B_t(r,r')=K_{{\rm Airy}}(r,r') + t^{-1/3} \int_{-\infty}^{+\infty} dv \frac{{\rm sgn}(v)}{e^{|v|}-1} 
\left({\rm Ai}\left(r + \frac{v}{t^{1/3}}\right) {\rm Ai}\left(r' + \frac{v}{t^{1/3}}\right) - {\rm Ai}(r) {\rm Ai}(r')\right) \\
&& = K_{{\rm Airy}}(r,r') + t^{-2/3} \frac{\pi^2}{3} \left( {\rm Ai}(r) {\rm Ai}'(r) + {\rm Ai}'(r) {\rm Ai}(r) \right) + {\cal O}(t^{-4/3}) \label{expB} \;.
\eea 
Now, since $Ai Ai^\dagger$ is a rank one projector, one can rewrite (\ref{second2}) as
\bea
p_t(u) = \frac{1}{2} {\rm Det}[ I - P_u (B_t - Ai Ai^\dagger) P_u] -  \frac{1}{2} {\rm Det}[ I - P_u (B_t + Ai Ai^\dagger) P_u] \;,
\eea 
and inserting (\ref{expB}), it is clear that the ${\cal O}(t^{-2/3})$ term cancels between the two terms.
Thus the (non-trivial) leading correction to $p_t(u)$ around the TW distribution is ${\cal O}(t^{-4/3})$,
hence the first three cumulants are corrected first by (minus) the Gumbel
variable in (\ref{second1}). The method that we used here, however, allows
one to obtain  the cumulants for the fermion problem in a more direct manner.



\begin{thebibliography}{10}


\bibitem{BDZ08}
I. Bloch, J. Dalibard, W. Zwerger, Rev. Mod. Phys. {\bf 80}, 885 (2008).

\bibitem{GPS08}
S. Giorgini, L. P. Pitaevski, S. Stringari, Rev. Mod. Phys. {\bf 80}, 1215 (2008).


\bibitem{Mahan}
G.~D. Mahan, {\it Many particle physics}, Plenum, NY (1981). 


\bibitem{Castin} Y. Castin, {\it Basic theory tools for degenerate Fermi 
gases}, in Proceedings of the International School of Physics Enrico
Fermi, Vol. 164: {\it Ultra-cold Fermi Gases}, edited by M. Inguscio, W.   
Ketterle, and C. Salomon, Varenna Summer School Enrico Fermi (IOS Press,
Amsterdam, 2006), arXiv:0612613.





\bibitem{Castin2}Y. Castin,
in {\it Quantum gases in low dimensions}, 
J. Phys. IV France, {\bf 116} 89 (2004), arXiv:0407118.




\bibitem{Cheuk:2015}
L.W. Cheuk, M.A. Nichols, M. Okan, T. Gersdorf, R.Vinay, W. Bakr, T. Lompe, M. Zwierlein, Phys. Rev. Lett. {\bf 114},  193001, (2015). 


\bibitem{Haller:2015}
E. Haller, J. Hudson, A. Kelly, D. A. Cotta, B. Peaudecerf, G. D. Bruce, S. Kuhr, Nature Physics {\bf 11}, 738 (2015).

 \bibitem{Parsons:2015} 
 M. F. Parsons, F. Huber, A. Mazurenko, C. S. Chiu, W. Setiawan, K. Wooley-Brown, S. Blatt, M. Greiner, Phys. Rev. Lett. {\bf 114}, 213002 (2015).

\bibitem{butts}
D. A. Butts, D. S. Rokshar, Phys. Rev. A {\bf 55}, 4346 (1997).





\bibitem{Kohn}
W. Kohn, A. E. Mattsson, Phys. Rev. Lett. {\bf 81} 3487 (1998). 





\bibitem{us_prl}
D. S. Dean, P. Le Doussal, S. N. Majumdar, G. Schehr, Phys. Rev. Lett. {\bf 114}, 110402 (2015). 


\bibitem{us_epl}
 D. S. Dean, P. Le Doussal, S. N. Majumdar, G. Schehr, Europhys. Lett. {\bf 112}, 60001 (2015).




\bibitem{mehta}{M.~L. Mehta, {\it Random Matrices} 
(Academic Press, Boston, 1991).}

\bibitem{forrester}{ P.~J. Forrester, 
{\it Log-Gases and Random Matrices} 
(London Mathematical Society monographs, 2010).  }

\bibitem{johansson} 
See e.g. K. Johansson, {\it Random matrices and determinantal processes}, in Lecture Notes of the Les
Houches Summer School 2005 (A. Bovier, F. Dunlop, A. van Enter,
F. den Hollander, and J. Dalibard, eds.), Elsevier Science, (2006); arXiv:math-ph/0510038.

\bibitem{borodin_determinantal}
A. Borodin, {\it Determinantal point processes}, in {\it The Oxford Handbook of Random Matrix Theory}, 
G. Akemann, J. Baik, P. Di Francesco (Eds.), Oxford University Press, Oxford (2011).


\bibitem{tracy_widom_determinantal}
C. A. Tracy, H. Widom, J. Stat. Phys. {\bf 92}, 809 (1998).


\bibitem{eisler_prl}
V. Eisler, Phys. Rev. Lett. {\bf 111}, 080402 (2013).

\bibitem{marino_prl} R. Marino, S. N. Majumdar, G. Schehr, P. Vivo, Phys.
Rev. Lett. {\bf 112}, 254101 (2014).


\bibitem{castillo}
I.~P\'erez-Castillo, Phys. Rev. E {\bf 90}, 040102(R) (2014).

\bibitem{CDM14} P. Calabrese, P. Le Doussal, S. N. Majumdar, Phys. Rev. A
{\bf 91}, 012303 (2015).

\bibitem{marino_pre}
R. Marino, S. N. Majumdar, G. Schehr, P. Vivo, Phys. Rev. E {\bf 94}, 032115 (2016).  


\bibitem{TW}
C.~A. Tracy, H.~Widom, Commun. Math. Phys. {\bf 159}, 151 (1994).

\bibitem{baik} J. Baik, P. Deift, K. Johansson, {J. Am. Math. Soc.} {\bf 12}, 1119 (1999). 

\bibitem{johann} K. Johansson, {Commun. Math. Phys.} {\bf 209}, 437 (2000).



\bibitem{poli} J. Baik, E. M. Rains, {J. Stat. Phys.} {\bf 100}, 523 (2000). 

\bibitem{growth} M. Pr\"ahofer, H. Spohn, {Phys. Rev. Lett.} {\bf 84}, 4882 
(2000); J. Gravner, C. A. Tracy, H. Widom, {J.~Stat. Phys.} {\bf 102}, 
1085 (2001); S. N. Majumdar, S. Nechaev, {Phys. Rev. E} {\bf 69}, 011103 (2004); 
T.~Imamura, T. Sasamoto, {Nucl. Phys. 
B} {\bf 699}, 503 (2004). 


\bibitem{SS10} T. Sasamoto, H. Spohn, Phys. Rev. Lett. {\bf 104}, 230602 
(2010).

\bibitem{CLR10} P. Calabrese, P. Le Doussal, A. Rosso, Europhys. Lett. {\bf 
90}, 20002 (2010).

\bibitem{DOT10} V. Dotsenko, Europhys. Lett. {\bf 90}, 20003 (2010).

\bibitem{ACQ11} G. Amir, I. Corwin, J. Quastel, Comm. Pure and Appl. Math.
{\bf 64}, 466 (2011).

\bibitem{sequence} S. N. Majumdar, S. K. Nechaev, {Phys. Rev. E} {\bf 72}, 020901(R) (2005). 

\bibitem{dots} M. G. Vavilov, P. W. Brouwer, V. Ambegaokar, C. W. J. Beenakker, {Phys. Rev. Lett.} {\bf 86}, 874 (2001); A. Lamacraft, B. D. Simons, {Phys. Rev. B} {\bf 64}, 014514 (2001); P. M. Ostrovsky, M. A. Skvortsov, M. V. Feigel'man, {Phys. Rev. Lett.}
{\bf 87}, 027002 (2001); J. S. Meyer, B. D. Simons, {Phys. Rev. B} {\bf 64}, 134516 (2001); A. Silva, L. B. Ioffe, {Phys. Rev. B}
{\bf 71}, 104502 (2005). 

\bibitem{FMS11}
P. J. Forrester, S. N. Majumdar, G. Schehr, Nucl. Phys. B {\bf 844}, 500 (2011).

\bibitem{Lie12}
K. Liechty, J. Stat. Phys. {\bf 147}, 582 (2012). 

\bibitem{NM_interface} C. Nadal, S.~N. Majumdar, Phys. Rev. E, {\bf 79}, 
061117 (2009).

\bibitem{biroli} G. Biroli, J.-P. Bouchaud, M. Potters, {Eur. Phys. Lett.} {\bf 78}, 10001 (2007). 

\bibitem{maj}
S. N. Majumdar, Les Houches lecture notes on {\it Complex Systems} (2006), ed. by J.-P. Bouchaud, M. M\'ezard and J. Dalibard 
[arXiv: cond-mat/0701193]. 

\bibitem{MS_thirdorder}
S. N. Majumdar, G. Schehr, J. Stat. Mech. P01012 (2014) .

\bibitem{takeuchi}
K . A. Takeuchi, M. Sano, Phys. Rev. Lett. {\bf 104}, 230601 (2010); K . A. Takeuchi, M. Sano, T. Sasamoto, H. Spohn, Sci. Rep. (Nature) {\bf 1}, 34 (2011); K . A. Takeuchi, M. Sano,  J. Stat. Phys. {\bf 147}, 853 (2012).

\bibitem{davidson} M. Fridman, R. Pugatch, M. Nixon, A. A. Friesem, N. 
Davidson, {Phys. Rev. E} {\bf 85}, R020101 (2012).











\bibitem{gleisberg}
F. Gleisberg, W. Wonneberger,
U. Schl\"oder, C. Zimmermann, Phys. Rev. A {\bf 63}, 602 (2000).

\bibitem{calabrese_prl} P. Calabrese, M. Mintchev, E. Vicari, Phys.
Rev. Lett. {\bf 107}, 020601 (2011); J. Stat. Mech. P09028 (2011).

\bibitem{vicari_pra} E. Vicari, Phys. Rev. A {\bf 85}, 062104 (2012).

\bibitem{vicari_pra2} M. Campostrini, E. Vicari, Phys. Rev. A {\bf 82},
063636 (2010).

\bibitem{vicari_pra3} A. Angelone, M. Campostrini, E. Vicari, Phys.
Rev. A {\bf 89}, 023635 (2014).

\bibitem{einstein}
T. L. Einstein, Ann. Henri Poincar\'e {\bf 4}, Suppl. 2, S811 (2003).





\bibitem{BB91}
M. Bowick, E. Br\'ezin, Phys. Lett. B {\bf 268}, 21 (1991). 

\bibitem{For93}
P. J. Forrester, Nucl. Phys. B {\bf 402}(3), 709 (1993).



\bibitem{BTW92}
E. L. Basor, C. A. Tracy, H. Widom, Phys. Rev. Lett. {\bf 69}, 5 (1992).

\bibitem{Meh92}
M. L. Mehta, Z. Phys. B {\bf 86}, 285 (1992).

\bibitem{Grim2004}
U. Grimm, Phys. Stat. Sol. B {\bf 241}, 2139 (2004).

\bibitem{fredholm}
We recall that, for a trace-class operator $K(x,y)$ such that ${\rm Tr} K = \int dx K(x,x)$ is well defined, 
$\det(I - K) = \exp{[-\sum_{n=1}^\infty{{\rm Tr \,} K^n}/{n}]}$, where ${\rm Tr}\, K^n = \int dx_1 \cdots \int dx_n K(x_1,x_2) K(x_2,x_3)\cdots K(x_n,x_1)$. The effect of the projector $P_s$ in (\ref{fredholm_F2}) is simply to restrict the integrals over $x_i$ to the interval $[s,+\infty )$.


\bibitem{BBD08}
J. Baik, R. Buckingham, J. DiFranco, Commun. Math. Phys. {\bf 280}, 463 (2008). 


\bibitem{Dys1962}
F. J. Dyson, J. Math. Phys. {\bf 3}, 140 (1962).


\bibitem{CL1995}
O. Costin, J. L. Lebowitz, Phys. Rev. Lett. {\bf 75}, 69 (1995).

\bibitem{FS1995}
M. M. Fogler, B. I. Shklovskii, Phys. Rev. Lett. {\bf 74}, 3312 (1995).


\bibitem{EP2014}
V. Eisler, I. Peschel, J. Stat. Mech. P04005 (2014).


\bibitem{Johansson_Lambert}
K. Johansson, G. Lambert, preprint arXiv:1504.06455.

\bibitem{MNS94} M. Moshe, H. Neuberger, B. Shapiro, Phys. Rev. Lett. {\bf 73}, 1497 (1994).

\bibitem{Andreief}
C. Andreief, M\'em. de la Soc. Sci., Bordeaux, (3) {\bf 2}, 1 (1883).  

\bibitem{hough} J.~B. Hough, M.~Krishnapur, Y.~Peres, B.~Vir\'ag, Probability Surveys {\bf 3}, 206 (2006).


\bibitem{Gaudin}
M. Gaudin, Nucl. Phys. {\bf 15}, 89 (1960).

\bibitem{Joh07}
{K. Johansson, Probab. Theory Rel. {\bf 138}, 75 (2007).}

\bibitem{FFG06}
P. J. Forrester, N. E. Frankel, T. M. Garoni, J. Math. Phys. {\bf 47}, 023301 (2006).

\bibitem{JP1}
{R. Allez, J.~-P. Bouchaud, A. Guionnet, Phys. Rev. Lett. {\bf 109}, 094102 (2012).}

\bibitem{JP2}
{R. Allez, J.~-P. Bouchaud, S. N. Majumdar, P. Vivo, J. Phys. A: Math. Theor. {\bf 46}, 015001 (2013).}

\bibitem{Verba}
A. M. Garc{\' i}a-Garc{\' i}a, J.~J.~M.~Verbaarschot, Phys. Rev. E {\bf 67}, 046104 (2003). 


\bibitem{LargeDev_KPZ}
P. Le Doussal, S. N. Majumdar, G. Schehr, EPL {\bf 113}, 60004 (2016). We use formula (55) and (57) of the Supp. Matt. of the arXiv version, with the correspondence that $t \to b^3$ and $y \to \tilde s$.

\bibitem{SpohnProlhac}
S. Prolhac, H. Spohn, Phys. Rev. E {\bf 84}, 011119 (2011).

\bibitem{borneman}
F. Bornemann, Math. Comp. {\bf 79}, 871 (2010).

\bibitem{Short_time_PRL}
P. Le Doussal, S. N. Majumdar, A. Rosso, G. Schehr, 
Phys. Rev. Lett. {\bf 117}, 070403 (2016).


\bibitem{Ferrari_Frings}
P. L. Ferrari, R. Frings, J. Stat. Phys. {\bf 144}, 1123 (2011).


\bibitem{Gumbel} 
E. J. Gumbel, {\it Statistics of extremes}, NY Dover (1958). 

\bibitem{BoGo}
A. Borodin, V. Gorin, preprint arXiv:1608.01557.

\bibitem{feynman_hibbs}
R.~P. Feynman, A.~R. Hibbs, {\it Quantum Mechanics and Path Integrals}, (McGraw-
Hill, New York, 1965).

\bibitem{Grad} I.~S. Gradshteyn, I.~M. Ryzhik, {\it Table of Integrals, 
Series and Products}, edited A. Jeffrey and D. Zwilinger (Academic Press, 
Elsevier, 2007), 7th ed.


\bibitem{scardicchio}
A.~Scardicchio, C.~E. Zachary, S. Torquato, Phys. Rev. E {\bf 79} 041108 (2009); 
S. Torquato, A. Scardicchio, C.~E. Zachary, J. Stat. Mech P 11019 (2008).

\bibitem{vallee}O. Vall\'ee, M. Soares, {\it Airy Functions and 
Applications to Physics}, (Imperial College Press, London, 2004).



\bibitem{MM88}N. Makri, W.~H. Miller, Chem. Phys. Lett. {\bf 151}, 1 (1988).

\bibitem{TW_Bessel}
C. A. Tracy, H. Widom, Commun. Math. Phys., 161(2), 289-309 (1994).



\bibitem{BookGiam} T. Giamarchi, {\it Quantum physics in one dimension}, Oxford Clarendon Press (2004). 


\bibitem{haldane81}
F. D. M. Haldane, Phys. Rev. Lett. {\bf 47}, 1840 (1981). 


\bibitem{foot_LL}
Note a misprint in the sign of the second term in Ref. \cite{haldane81} that was corrected later in Eq. (3.40) of \cite{BookGiam}.


\bibitem{kpz}
M. Kardar, G. Parisi, Y-C. Zhang, Phys. Rev. Lett. {\bf 56}, 889 (1986).


\bibitem{footnote1} 
These units correspond to the choice $\bar c=1$ and 
$\lambda=(t/4)^{1/3}$ in the notations of \cite{CLR10,DOT10}.
The dimensionless unit choice in \cite{SS10,ACQ11} corresponds to the choice 
$t^*=2$, and relates as $\gamma_t=2^{2/3} \lambda$ with the notations of
\cite{CLR10,DOT10}.



\bibitem{footnote0} 
As is well known, a proper definition of (\ref{eq:KPZ}) in the continuum 
requires a smooth noise at microscopic scale. The limit of white noise leads to
non-trivial renormalizations (since e.g. $\langle (\partial_x h)^2 \rangle$ diverges in that
limit). As shown by Hairer \cite{Hairer} this limit can still be defined
rigorously, and its proper definition is consistent
with the (more naive) one given via the Hopf-Cole transform $Z(x,t)$, presented here. 

\bibitem{Hairer}
M. Hairer, Proceedings of the ICM (2014), arXiv:1403.6353; 
I. Corwin, Rand. Mat. Theo. Appl. {\bf 1}, 1130001 (2012).

\bibitem{footnote2}
The continuum solution is related to the physical solution up to a non-universal
shift (i.e. renormalization). Let us call $h^{phys}(x,t)$ the solution to 
a physical KPZ problem with a regularized, i.e. 
smooth noise at small scale. Then, above a correspondingly small time and length scale,
the continuum and physical solutions are related as follows
$h(x,t)= \ln Z^{phys}(x,t)/\langle Z^{phys}(x,t) \rangle + \ln Z_0(x,t) =
h^{phys}(x,t) - \ln( \langle \exp(h^{phys}(x,t)) \rangle ) + \ln Z_0(x,t)$.
It holds for an arbitrary initial condition provided $Z_0(x,t)$ is chosen as the solution of
the free diffusion equation with that initial condition. See \cite{FlatShortTime}
for a more detailed discussion in case of flat initial conditions. 





\bibitem{FlatShortTime}
T. Gueudr\'e, P. Le Doussal, A. Rosso, A. Henry, P. Calabrese, Phys. Rev. E {\bf 86}, 041151 (2012).





\bibitem{praeho_spohn_prl}
M. Praehofer, H. Spohn,  Phys. Rev. Lett. {\bf 84}, 4882 (2000).


\bibitem{praeho_spohn_jsp}
M. Praehofer, H. Spohn,  J. Stat. Phys. {\bf 108}, 1071 (2002).

\bibitem{sasamoto_fermion}
T. Imamura, T. Sasamoto,  preprint arXiv:1506.05548.



\bibitem{klein}H. Kleinert, {\it Path Integrals in Quantum Mechanics, Statistics, Polymer 
Physics and Financial Markets}, (World Scientific, Singapore, 2004) 3rd ed.


\bibitem{oks}B. Oksendal, {\it Stochastic Differential Equations}, (Springer-Verlag, Berlin Heidelberg, 1992), 3rd ed.



\bibitem{friedrich}
H. Friedrich, in {\it Proceedings of the Fourth International Conference on Dynamical Systems and Differential Equations},
Wilmington, USA, 288 (2002).


\bibitem{wiki_sommer}
See \url{https://en.wikipedia.org/wiki/Sommerfeld_expansion.}


\bibitem{jacopo}
J. de Nardis, P. Le Doussal, in preparation. 




















































































 











\end{thebibliography}
\end{document}